\documentclass[twocolumn,twocolappendix]{aastex631} % ,linenumbers
\usepackage[mathlines]{lineno} % for arXiv

\usepackage{newtxtext,newtxmath}
\usepackage[OT2,T1]{fontenc}
\usepackage[russian,english]{babel}
\usepackage{CJKutf8}

\DeclareRobustCommand{\VAN}[3]{#2}
\let\VANthebibliography\thebibliography
\def\thebibliography{\DeclareRobustCommand{\VAN}[3]{##3}\VANthebibliography}

\usepackage{graphicx}	% Including figure files
\usepackage{amsmath}	% Advanced maths commands
\usepackage{amssymb}	% Extra maths symbols
\usepackage{color}
\usepackage{booktabs}
\usepackage{tcolorbox}
\usepackage[normalem]{ulem}
\usepackage{orcidlink}

\usepackage{multirow}
\usepackage{url}
\usepackage[frozencache,cachedir=.]{minted}

\definecolor{F184}{rgb}{0.65,0.1,0.0}
\definecolor{H158}{rgb}{0.35,0.4,0.0}
\definecolor{J129}{rgb}{0.0,0.4,0.35}
\definecolor{Y106}{rgb}{0.0,0.1,0.65}

\newcommand{\papone}{Paper I}
\newcommand{\paptwo}{Paper II}
\newcommand{\papfour}{Paper IV}

\newcommand{\NewEdit}[1]{#1} % {\bfseries{#1}}

\shorttitle{Image combination for {\slshape Roman} III}
\shortauthors{K. Cao et al.}

\begin{document}

\title{Simulating image coaddition with the {\textit{\textbf{Nancy Grace Roman Space Telescope}}}:\\III. Software improvements and new linear algebra strategies}

\correspondingauthor{Kaili Cao}
\email{cao.1191@osu.edu}

\author[0000-0002-1699-6944]{Kaili Cao (\begin{CJK*}{UTF8}{gbsn}曹开力\end{CJK*}$\!\!$)}
\affiliation{Center for Cosmology and AstroParticle Physics (CCAPP), The Ohio State University, 191 West Woodruff Ave, Columbus, OH 43210, USA}
\affiliation{Department of Physics, The Ohio State University, 191 West Woodruff Ave, Columbus, OH 43210, USA}

\author{Christopher M. Hirata}
\affiliation{Center for Cosmology and AstroParticle Physics (CCAPP), The Ohio State University, 191 West Woodruff Ave, Columbus, OH 43210, USA}
\affiliation{Department of Physics, The Ohio State University, 191 West Woodruff Ave, Columbus, OH 43210, USA}
\affiliation{Department of Astronomy, The Ohio State University, 140 West 18th Avenue, Columbus, OH 43210, USA}

\author{Katherine Laliotis}
\affiliation{Center for Cosmology and AstroParticle Physics (CCAPP), The Ohio State University, 191 West Woodruff Ave, Columbus, OH 43210, USA}
\affiliation{Department of Physics, The Ohio State University, 191 West Woodruff Ave, Columbus, OH 43210, USA}

\author{Masaya Yamamoto}
\affiliation{Department of Physics, Duke University, Box 90305, Durham, NC 27708, USA}
\affiliation{Department of Astrophysical Sciences, Princeton University, Princeton, NJ 08544, USA}

\author{Emily Macbeth}
\affiliation{Center for Cosmology and AstroParticle Physics (CCAPP), The Ohio State University, 191 West Woodruff Ave, Columbus, OH 43210, USA}
\affiliation{Department of Physics, The Ohio State University, 191 West Woodruff Ave, Columbus, OH 43210, USA}
\affiliation{Department of Astronomy, The Ohio State University, 140 West 18th Avenue, Columbus, OH 43210, USA}

\author{M.~A.~Troxel}
\affiliation{Department of Physics, Duke University, Box 90305, Durham, NC 27708, USA}

\begin{abstract}

The {\slshape Nancy Grace Roman Space Telescope} will implement a devoted weak gravitational lensing program with its High Latitude Wide Area Survey. For cosmological purposes, a critical step in {\slshape Roman} image processing is to combine dithered undersampled images into unified oversampled images and thus enable high-precision shape measurements.
{\sc Imcom} is an image coaddition algorithm which offers control over point spread functions in output images. This paper presents the refactored {\sc Imcom} software, featuring full object-oriented programming, improved data structures, and alternative linear algebra strategies for determining coaddition weights. Combining these improvements and other acceleration measures, to produce almost equivalent coadded images, the consumption of core-hours has been reduced by about an order of magnitude.
We then re-coadd a $16 \times 16 \,{\rm arcmin}^2$ region of our previous image simulations with three linear algebra kernels in four bands, and compare the results in terms of {\sc Imcom} optimization goals, properties of coadded noise frames, and measurements of simulated stars. The Cholesky kernel is efficient and relatively accurate, yet its irregular windows for input pixels slightly bias coaddition results.
The iterative kernel avoids this issue by tailoring input pixel selection for each output pixel; it yields better noise control, but can be limited by random errors due to finite tolerance. The empirical kernel coadds images using an empirical relation based on geometry; it is inaccurate, but being much faster, it provides a valid option for ``quick look'' purposes. We fine-tune {\sc Imcom} hyperparameters in a companion paper.

\end{abstract}

\keywords{Astronomy image processing (2306) --- Weak gravitational lensing (1797)} %% https://astrothesaurus.org

\section{Introduction} \label{sec:intro}

Gravitational lensing is the bending of optical paths due to gravity of massive celestial objects. In the case of weak gravitational lensing (hereafter weak lensing), such bending is strong enough to distort the shape of background sources to an observable degree, but not so that several images of the same objects are produced, as in the case of strong lensing.
Weak lensing is directly sensitive to the mass distribution in the Universe, and is thus a powerful probe of the growth of cosmic structure \citep[e.g.][]{2001PhR...340..291B, 2013PhR...530...87W, 2015RPPh...78h6901K}. However, the great potential of weak lensing can only be revealed if astronomers are able to measure galaxy shapes precisely and accurately.

Weak gravitational lensing surveys have grown steadily, leading up to few percent level constraints from the three large programs of the past decade using wide-field optical imaging cameras: the Dark Energy Survey \citep{2022PhRvD.105b3514A, 2022PhRvD.105b3515S}, the Hyper Suprime Cam \citep{2019PASJ...71...43H, 2020PASJ...72...16H}, and the Kilo Degree Survey \citep{2022A&A...664A.170V, 2023A&A...679A.133L}. The next generation of surveys aims for sub-percent precision. These include the recently-launched {\slshape Euclid} space telescope \citep{2011arXiv1110.3193L, 2022A&A...662A.112E, 2024arXiv240513491E}; the upcoming Legacy Survey of Space and Time at the Vera Rubin Observatory (hereafter ``Rubin;'' \citealt{2012arXiv1211.0310L, 2019ApJ...873..111I}); and the infrared survey to be conducted with the {\slshape Nancy Grace Roman Space Telescope} (hereafter {\slshape Roman}; \citealt{2019arXiv190205569A}).

{\slshape Roman} is planned to launch in late 2026 or early 2027 and start a five-year mission, and implement a devoted weak lensing program with its High Latitude Wide Area Survey (HLWAS).
Given {\slshape Roman}'s vantage point in outer space, specifically at Sun-Earth Lagrange Point 2 (L2), its point spread function (PSF) will not be affected by Earth's atmosphere, and will be narrower than those of ground-based instruments. A wide-field infrared imager with a $2.4$-meter telescope is possible due to advances in detector array technology: the {\slshape Roman} sensor chip assemblies (SCAs) have $4088\times 4088$ active pixels each, and 18 will fly in the {\slshape Roman} focal plane for a total of 300 Mpix \citep{2020JATIS...6d6001M}. But even with these large-format imagers, an efficient survey requires the pixels to be undersampled relative to the diffraction-limited PSF (in the sense that the pixel scale is larger than $\lambda/2D$, where $\lambda$ is the wavelength of observation and $D$ is the entrance pupil diameter). Given the native pixel size of $0.11 \,{\rm arcsec}$, raw {\slshape Roman} exposures are unable to fully resolve the PSFs and permit high-quality shape measurements.
Fortunately, image coaddition allows us to take groups of dithered undersampled images and combine them into unified oversampled images.

Several image coaddition algorithms have been developed; these are usually formulated as a linear transformation from input pixels to output pixels. (Linearity is necessary for the output image to have a well-defined PSF; see \citealt{2023OJAp....6E...5M} for a detailed discussion.)
Most of these algorithms, e.g. {\sc Drizzle} \citep{2002PASP..114..144F, 2012drzp.book.....G}, simply assign weights by computing geometric overlaps between input and output pixels; the resulting output PSF varies with sub-pixel position, and there is not a fundamental understanding of how to calibrate weak lensing shear estimators with such images. There are Fourier-domain techniques for recovering sampling that allow control over the output PSF \citep{1999PASP..111..227L}, but there are challenges adopting these to fast surveys that have varying geometric distortions, rolls, and missing samples.
The {\sc Imcom} technique \citep{2011ApJ...741...46R} can accept arbitrary rolls, distortions, missing pixels, and dithering patterns, provides users with the freedom to specify target output PSFs, and reports deviations between actual reconstructed PSFs and their ideal counterparts as a measure of quality control.
\citet[][hereafter \papone]{2024MNRAS.528.2533H} re-implemented {\sc Imcom} as a Python program with a C back end, extended it to enable coadding larger areas of the sky, and tested it using synthetic {\slshape Roman} images produced by \citet{2023MNRAS.522.2801T}.
\citet[][hereafter \paptwo]{2024MNRAS.528.6680Y} further diagnosed the output images in terms of resulting noise power spectra and shape parameters of simulated point sources. Systematic errors in the shape of the output PSF caused by {\sc Imcom} itself were found to meet {\slshape Roman} requirements, although several areas such as noise-induced bias and higher-order PSF moments warrant further attention (and of course the calibration of PSF and linearity in the input images themselves).

Much work on {\sc Imcom} remains to be done before its application to real {\slshape Roman} data. Therefore, a better software architecture is desirable for maintenance and extension purposes.
In the discussion section of \paptwo, we included a preliminary to-do list, in which the foremost item was to improve computing efficiency, as it would be prohibitively expensive to apply the original implementation of {\sc Imcom} to the entire HLWAS region, which is $\sim 2000 \,{\rm deg}^2$ in 4 bands in the {\slshape Roman} Reference Survey \citep{2015arXiv150303757S, 2023MNRAS.522.2801T}.
In addition, boundary effects around {\sc Imcom} postage stamps reported in \papone\ may confuse source identification or shape measurement algorithms, and a ${\cal O}(n^3)$ complexity ($n$ is the number of selected input pixels) is unaffordable for deep fields. We seek to address these issues via alternative strategies to determine coaddition weights.
These are topics of the current paper, which is structured as follows. In Section~\ref{sec:framework}, we present the refactored {\sc Imcom} framework, featuring full object-oriented programming (OOP), improved data structures, and additional output maps for diagnostic purposes\NewEdit{. A more in-depth description of the new software is included in Appendix~\ref{app:pyimcom}, and} some additional acceleration measures are detailed in Appendix~\ref{app:accel}.
Then we introduce alternative linear algebra strategies and the philosophy behind them in Section~\ref{sec:lakernel}. In Section~\ref{sec:sims}, we describe how we configure simulations in this work and present some ``preview'' results, as well as {\sc Imcom} diagnostics.
We then compare the Cholesky, iterative, and empirical kernels in terms of noise power spectra and measurements of simulated point sources in Sections~\ref{sec:noiseps} and \ref{sec:gsstar14}, respectively. Finally \NewEdit{we} wrap up this work by summarizing it in Section~\ref{sec:summary}. % We
A set of holistic and objective evaluation criteria for coaddition results will be applied to fine-tune {\sc Imcom} hyperparameters in a companion paper (Cao et al. in prep, hereafter \papfour).

\section{Improvements to the {\sc Imcom} framework} \label{sec:framework}

Our previous implementation of the {\sc Imcom} software consists of two GitHub repositories:
\begin{itemize}
\item {\sc furry-parakeet}, which contains utilities to coadd individual postage stamps (usually $1.25 \times 1.25 \,{\rm arcsec}^2$);
\item {\sc fluffy-garbanzo}, a driver to coadd blocks (large arrays of postage stamps; $48 \times 48$ per block in \papone\ and this paper, not including padding on its boundaries).
\end{itemize}
To facilitate improvements and extensions, we have refactored {\sc Imcom} into a single repository, {\sc PyImcom}.\footnote{Links to all these repositories can be found in the Data Availability section at the end of this paper.}
After recapping the {\sc Imcom} formalism in Section~\ref{ss:formalism}, we describe the new framework in Section~\ref{ss:pyimcom} \NewEdit{(for a more in-depth version, see Appendix~\ref{app:pyimcom})}, and then discuss additional output maps to diagnose the results in Section~\ref{ss:outmaps}.
Unless otherwise noticed, known issues of {\sc Imcom} {\slshape per se} mentioned in \papone\ Section~4.4 have been addressed; for additional acceleration measures in the current implementation, see Appendix~\ref{app:accel}.

\subsection{Recap of {\sc Imcom}} \label{ss:formalism}

We briefly recap the aspects of {\sc Imcom} here relevant to the optimization; the reader is referred to \citet{2011ApJ...741...46R} for full details on the mathematics and \papone\ for details of the previous implementation.
We recall that {\sc Imcom} attempts to make an output image 
\begin{equation}
H_\alpha = \sum_{i=0}^{n-1} T_{\alpha i} I_i,
\label{eq:coadd}
\end{equation}
where $I_i$ represents the input images (flattened and concatenated, with input pixel indexed by $i=0... n-1$), and $H_\alpha$ represents the output image (with output pixel indexed by $\alpha=0... m-1$).\footnote{Since the implementation is in Python, we follow the Python indexing scheme in this paper, and start arrays with 0.}
The input images are assumed to have PSF $G_i$ (which may be different for each image or at each position) and pixels centered at ${\boldsymbol r}_i$. We aim to have an output with a uniform ``target'' PSF $\Gamma$ at pixels centered at ${\boldsymbol R}_\alpha$.

{\sc Imcom} finds the matrix ${\mathbf T}$ that attempts to minimize
\begin{equation}
U_\alpha = \left\Vert {\rm PSF}_{\alpha,\rm out} - \Gamma \right\Vert^2
~~{\rm and}~~
\Sigma_\alpha = \sum_{i,j} N_{ij} T_{\alpha i}T_{\alpha j},
\label{eq:U_Sigma}
\end{equation}
where $\Vert\cdot\Vert$ represents the $L^2$ norm, and $N_{ij}$ is the input noise covariance.
Here ${\rm PSF}_{\alpha,\rm out}$ is the as-realized coadded PSF in pixel $\alpha$, consisting of the appropriately translated input PSFs:
\begin{equation}
{\rm PSF}_{\alpha,\rm out}({\boldsymbol R}_\alpha -{\boldsymbol s})
= \sum_{i=0}^{n-1} T_{\alpha i} G_i({\boldsymbol r}_i - {\boldsymbol s}).
\end{equation}
We identify $U_\alpha$ as a ``PSF leakage'' metric. Since the output PSF is linear in $T_{\alpha i}$, we may write
\begin{equation}
U_\alpha =\sum_{i,j} A_{ij} T_{\alpha i}T_{\alpha j} + \sum_i B_{\alpha i}T_{\alpha i} + C,
\label{eq:U_mat}
\end{equation}
where $C = \Vert \Gamma \Vert^2$ is the square norm of the target output PSF, and the matrices ${\mathbf A}$ and ${\mathbf B}$ can be described by
\begin{equation}
A_{ij} = [G_j\otimes G_i]({\boldsymbol r}_i-{\boldsymbol r}_j)
~{\rm and}~
-\frac12 B_{\alpha i} = [\Gamma\otimes G_i]({\boldsymbol r}_i-{\boldsymbol R}_\alpha),
\label{eq:Aij_Bi}
\end{equation}
where $\otimes$ denotes the correlation. As for $\Sigma_\alpha$, identified as a ``noise amplification'' metric, we assume $N_{ij} = \delta_{ij}$ (the Kronecker delta), so that its expression can be simplified to
\begin{equation}
\Sigma_\alpha = \sum_{i=0}^{n-1} T_{\alpha i}^2.
\label{eq:Sigma_mat}
\end{equation}
This is the simplest form of the input noise covariance, assuming no correlation. (Note that this is being used for optimal weighting; if the true noise is correlated, as it will be for {\slshape Roman}, no bias is introduced but the output noise variance may be different from $\Sigma_\alpha$.) 

Since the minimum of $U_\alpha$ is generally not also the minimum of $\Sigma_\alpha$, a trade-off must be made between the two. This is described by a Lagrange multiplier $\kappa_\alpha$: we minimize the combination $U_\alpha + \kappa_\alpha \Sigma_\alpha$, resulting in
\begin{equation}
T_{\alpha i} =\sum_j [({\mathbf A} + \kappa_\alpha {\mathbb I}_{n\times n})^{-1}]_{ij} \left(-\frac12B_{\alpha j}\right),
\label{eq:T-AB}
\end{equation}
where ${\mathbb I}_{n\times n}$ denotes the $n\times n$ identity matrix. An algorithmic choice must always be made to determine $\kappa_\alpha$: larger $\kappa_\alpha$ places more priority on lowering the noise metric at the expense of worse PSF leakage, while smaller $\kappa_\alpha$ places more priority on lowering the PSF leakage metric at the expense of more noise. In \papone, the choice of $\kappa_\alpha$ was based on the following step algorithm: first, we determine whether it is possible to have $\Sigma_\alpha\le \frac12$ (per-pixel noise a factor of at least 2 in variance lower than the input images) and $U_\alpha/C = 10^{-6}$ (i.e., 0.1\% PSF leakage in a root-sum-square sense).\footnote{The choice of thresholds may vary according to the application and will likely be tuned prior to {\slshape Roman} launch. But as an initial guess for weak lensing applications, it is reasonable to aim for $<0.1\%$ leakage in all aspects of the PSF combined and for some reduction of noise by combining images.} If this is possible, \papone\ fixes $U_\alpha/C = 10^{-6}$ and tries to minimize the noise; otherwise, \papone\ fixes $\Sigma_\alpha=\frac12$ and minimizes $U_\alpha/C$.

The implementation proceeds through the following steps on each postage stamp:
\begin{itemize}
    \item Read input data, principally input signals ($I_i$), pixel positions (${\boldsymbol r}_i$), and PSFs ($G_i$); parse configuration to get output pixel positions (${\boldsymbol R}_\alpha$) and the target PSF ($\Gamma$).
    \item Perform fast Fourier transform (FFT) and inverse FFT to compute PSF overlaps ($G_j\otimes G_i$, $\Gamma\otimes G_i$, and $C$).
    \item Perform interpolations (see \papone\ Appendix A for details) using pixel positions to obtain system matrices (${\mathbf A}$ and ${\mathbf B}$).
    \item Solve linear systems to get the Lagrange multiplier $\kappa_\alpha$ and coaddition weights $T_{\alpha i}$ for each output pixel $\alpha$.
    \item Compute the output map ($H_\alpha$), and report diagnostics for its quality ($U_\alpha/C$ and $\Sigma_\alpha$).
\end{itemize}
Since transformation matrices (${\mathbf T}$) only depend on input and target PSFs, not input signals, it is economical to coadd multiple versions of the images, referred to as layers, simultaneously.
These \NewEdit{include science images, injected sources (using {\sc GalSim} \citep{Rowe2015A&C} utilities or {\sc IMCOM} routines), and noise fields. They} are read from files or made according to user-specified parameters as {\sc Imcom} prepares the input data. See \papone\ Section~3 for further details.
Provided that the complexity of linear system solving is usually ${\mathcal O}(n^3)$, it is necessary to limit $n$ by selecting only input pixels relatively close to the region being coadded.
To this end, {\sc Imcom} divides each block ($1.0 \times 1.0 \,{\rm arcmin}^2$ in \papone\ and this paper) into a two-dimensional (2D) array of postage stamps, and constructs system matrices for each of them.
Below we detail some of the technical choices regarding the above procedure in our new implementation.

\subsection{New framework: {\sc PyImcom}} \label{ss:pyimcom}

\begin{table*}
    \centering
    \caption{\label{tab:pyimcom}Layout of {\sc PyImcom} core modules.}
    \begin{tabular}{ccll}
        \hline
        Module & Class & Description & Rough mapping to the previous implementation \\
        \hline
        \multirow{2}{*}{\tt config} & {\tt Settings} & {\sc PyImcom} background settings & {\sc fluffy-garbanzo}: first part of {\tt coadd\_utils.py} \\
        & {\tt Config} & {\sc PyImcom} configuration, with JSON file interface & {\sc fluffy-garbanzo}: first part of {\tt run\_coadd.py} \\
        \hline
        \multirow{4}{*}{\tt coadd} & {\tt InImage} & Input image customized for each {\tt Block} instance &  \\
        & {\tt InStamp} & Data structure for input pixel positions and signals & {\sc fluffy-garbanzo}: second part of {\tt run\_coadd.py}, \\
        & {\tt OutStamp} & Driver for postage stamp coaddition & \quad second part of {\tt coadd\_utils.py}, and {\tt psf\_utils.py} \\
        & {\tt Block} & Driver for block coaddition &  \\
        \hline
        \multirow{5}{*}{\tt layer} & {\tt GalSimInject} & Utilities to inject objects using {\sc GalSim} & {\sc fluffy-garbanzo}: {\tt inject\_galsim\_obj.py} \\
        & {\tt GridInject} & Utilities to inject stars using {\sc Imcom} C routine & {\sc fluffy-garbanzo}: {\tt grid\_inject.py} \\
        & {\tt Noise} & Utilities to generate $1/f$ noise frames & {\sc fluffy-garbanzo}: {\tt inject\_complex\_noise.py} \\
        & \multirow{2}{*}{\tt Mask} & \multirow{2}{*}{Utilities for permanent and cosmic ray masks} & {\sc fluffy-garbanzo}: segments of {\tt run\_coadd.py} \\
        &  &  & \quad and {\tt coadd\_utils.py} \\
        \hline
        \multirow{5}{*}{\tt psfutil} & {\tt OutPSF} & Simple target output PSF models & {\sc furry-parakeet}: first part of {\tt pyimcom\_interface.py} \\
        & {\tt PSFGrp} & Group of either input or output PSFs & {\sc furry-parakeet}: {\tt pyimcom\_interface.py}, \\
        & {\tt PSFOvl} & Overlap (correlation) between PSF arrays & \quad class {\tt PSF\_Overlap} \\
        & {\tt SysMatA} & System matrix ${\mathbf A}$ attached to a {\tt Block} instance & {\sc furry-parakeet}: {\tt pyimcom\_interface.py}, \\
        & {\tt SysMatB} & System matrix ${\mathbf B}$ attached to a {\tt Block} instance & \quad function {\tt get\_coadd\_matrix} \\
        \hline
        \multirow{5}{*}{\tt lakernel} & {\tt \_LAKernel} & Abstract base class of linear algebra kernels & New feature, see \NewEdit{Section~\ref{sec:lakernel}} of this paper \\ % Section~\ref{ss:pyimcom}
        & {\tt EigenKernel} & LA kernel using eigendecomposition & {\sc furry-parakeet}: {\tt pyimcom\_lakernel.py} \\
        & {\tt CholKernel} & LA kernel using Cholesky decomposition & New feature, see Section~\ref{ss:kappa-search} of this paper \\
        & {\tt IterKernel} & LA kernel using conjugate gradient method & New feature, see Section~\ref{ss:iter-kernel} of this paper \\
        & {\tt EmpirKernel} & Mock LA kernel using empirical relation & New feature, see Section~\ref{ss:empir-kernel} of this paper \\
        \hline
        \multirow{6}{*}{\tt analysis} & {\tt OutImage} & Wrapper for coadded images (blocks) & New feature, see Section~\ref{ss:outmaps} of this paper \\
        & {\tt NoiseAnal} & Utilities to analyze noise frames & Previously not included in either repository \\
        & {\tt StarsAnal} & Utilities to analyze point sources & {\sc fluffy-garbanzo}: {\tt starcube.py} \\
        & {\tt \_BlkGrp} & Abstract base class for groups of coadded images & New feature, developed for analysis in this paper \\
        & {\tt Mosaic} & Wrapper for 2D arrays of coadded images & Previously not included in either repository \\
        & {\tt Suite} & Wrapper for hashed arrays of coadded images & New feature, developed for analysis in \papfour \\
        \hline
    \end{tabular}
\end{table*}

The refactored version of the {\sc Imcom} software, also known as the {\sc PyImcom} package, reorganizes the entire functionality of the \papone\ implementation into a fully object-oriented framework.
Table \ref{tab:pyimcom} lists principal {\sc PyImcom} modules and classes, describes their respective roles, and states rough mapping between them and {\sc fluffy-garbanzo} and {\sc furry-parakeet} modules.\footnote{In addition, we have translated the C back end of our previous implementation ({\sc furry-parakeet}: {\tt pyimcom\_croutines.c}, based on {\sc NumPy} C-API) into Python (the {\tt routine.py} module, based on {\sc Numba}) to make {\sc PyImcom} a standalone package. As {\tt routine.py} functions are slightly slower than their {\tt pyimcom\_croutines.c} counterparts, installing {\sc furry-parakeet} is encouraged for performance purposes; {\sc PyImcom} automatically detects and uses the C back end if available.}
The rest of this section \NewEdit{presents some important intermediate results of the {\sc PyImcom} software; for an in-depth description of its workflow, see Appendix~\ref{app:pyimcom}.\footnote{We thank the anonymous reviewer for this insightful suggestion.}}

\begin{figure}
    \centering
    \includegraphics[width=\columnwidth]{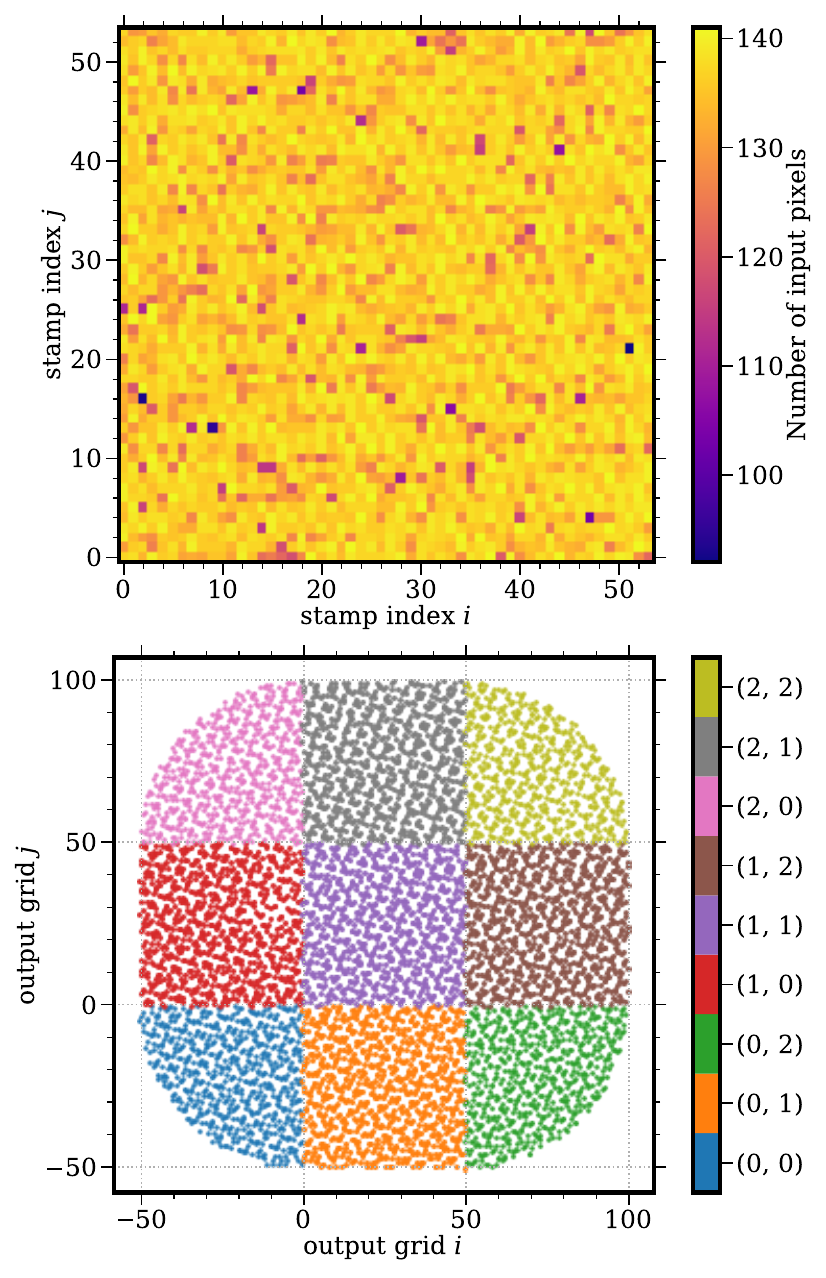}
    \caption{\label{fig:inpix} Diagrams for partitioning and selection of input pixels.
    {\it Upper} panel: Diagram showing how pixels from input exposures ({\tt InImage} instances) are partitioned into $54 \times 54$ input postage stamps ({\tt InStamp} instances). In this particular example, \papone\ block (0, 0) in Y106 band, $394338$ pixels are selected from input exposure (95060, 12), and the most populous postage stamp contains $141$ of them.
    {\it Lower} panel: Diagram showing how input pixels are selected for output postage stamps ({\tt OutStamp} instances). Pixels from different input postage stamps (labeled as $(j_{\rm stamp}, i_{\rm stamp})$) are shown in different colors; the output postage stamp being coadded overlaps with the one in purple. Note that the dotted grid lines are offset by $-0.5$ output pixels in both directions relative to postage stamp boundaries due to the finite output pixel size.}
\end{figure}

\NewEdit{Each {\sc PyImcom} run corresponds to the coaddition of a block. It starts by parsing the configuration and preparing the input data (see Section~\ref{ss:app-config}). Since a {\slshape Roman} image ($7.5 \times 7.5 \,{\rm arcmin}^2$) is much larger than an {\sc Imcom} block ($1.0 \times 1.0 \,{\rm arcmin}^2$ in \papone\ and this paper), {\sc PyImcom} partitions input pixels and only stores information about relevant ones in order to save memory space (or avoid using virtual memory for this purpose).}
An example of partitioning results is shown in the upper panel of Fig.~\ref{fig:inpix}; the expected maximum number of input pixels per exposure per postage stamp is $\lceil n_2\Delta\theta / s_{\rm in}\rceil^2 = \lceil 1.25/0.11\rceil^2 = 144$, where $\lceil \cdot\rceil$ is the ceiling function and $s_{\rm in} = 0.11 \,{\rm arcsec}$ is the native pixel size of {\slshape Roman}. {\tt InImage} instances request slightly more storage than this estimate to account for plate distortions.

For each output postage stamp, {\sc PyImcom} selects all input pixels within an acceptance radius $\rho_{\rm acc}$ (the {\tt INPAD} configuration entry) of its output pixels (not including transition pixels), and the reorganization is to facilitate this process.
An example of such selection is shown in the lower panel of Fig.~\ref{fig:inpix}, where $\rho_{\rm acc}$ is set to $1.25 \,{\rm arcsec}$ following \papone, coincidentally equal to the postage stamp size $n_2\Delta\theta$.
Given this layout, the expected number of input pixels relevant to an output stamp is
\begin{equation}
\langle n\rangle_{\rm rdsq} = \bar{n}_{\rm image} \frac{ (n_2\Delta\theta)^2 + 4(n_2\Delta\theta)\rho_{\rm acc} + \pi\rho_{\rm acc}^2}{ s_{\rm in}^2},
\label{eq:n-rdsq}
\end{equation}
where the subscript ``rdsq'' stands for rounded square, and $\bar{n}_{\rm image}$ is the mean coverage, which can be strictly defined as the number of unmasked input pixels per unit area. For the \papone\ acceptance radius, this yields $\langle n\rangle_{\rm rdsq} \approx 1051 \;\bar{n}_{\rm image}$.
See Section~\ref{ss:iter-kernel} for an alternative layout; the choice of $\rho_{\rm acc}$ will be examined in \papfour.

\begin{figure*}
    \centering
    \includegraphics[width=0.75\textwidth]{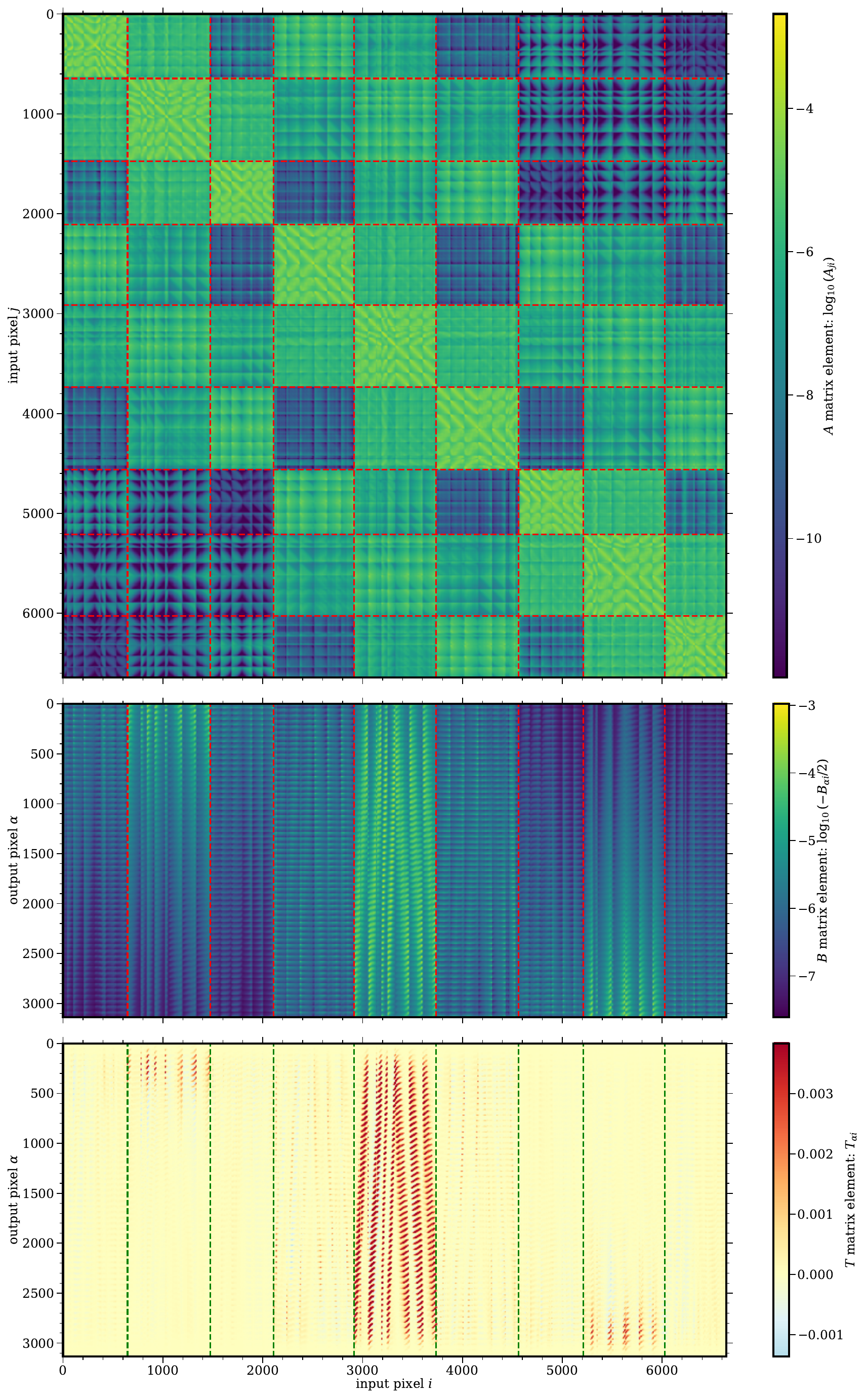}
    \caption{\label{fig:matrix}Example system matrices ({\it upper}: ${\mathbf A}$ matrix; {\it middle}: $-{\mathbf B}/2$ matrix) in logarithmic scale and the resulting transformation matrix ({\it lower}: ${\mathbf T}$ matrix) in linear scale. Boundaries between input postage stamps are shown as red or green dashed lines.
    For the ${\mathbf A}$ matrix, a threshold is set so that values close to zero (some of them are negative due to numerical errors) are uniformly shown in dark purple; the $-{\mathbf B}/2$ matrix does not have this issue. For the ${\mathbf T}$ matrix, the coloring scheme was chosen to better display structures, not to highlight significant weights (both positive and negative).}
\end{figure*}

\NewEdit{{\sc PyImcom} then computes PSF overlaps and matrix elements using Eq.~(\ref{eq:Aij_Bi}) and managed them in specially designed ways (see Section~\ref{ss:app-linsys}).}
Each {\tt OutStamp} instance stitches relevant system submatrices into their own ${\mathbf A}$ and ${\mathbf B}$ system matrices, which it solves using a linear algebra kernel (an instance of a concrete derived class of {\tt \_LAKernel}, see Section~\ref{sec:lakernel}).
Following the layout in the lower panel of Fig.~\ref{fig:inpix}, each ${\mathbf A}$ matrix has $9 \times 9$ submatrices, and each ${\mathbf B}$ matrix has $9$ submatrices; a pair of example system matrices, together with the resulting ${\mathbf T}$ matrix, are shown in Fig.~\ref{fig:matrix}.
Because of the different ordering of input pixels, these system matrices seem fragmental compared to Figs.~5 and 6 of \citet{2011ApJ...741...46R}; nevertheless, thanks to the properties of linear systems, the coaddition results are supposed to be the same, as long as the mapping between rows of ${\mathbf T}$ matrices and input signals is unaffected.\footnote{Each input pixel $i$ corresponds to a row of the ${\mathbf T}$ matrix in the sense that we rewrite Eq.~(\ref{eq:T-AB}) in matrix form: $\vec T_{\alpha} = [({\mathbf A} + \kappa_\alpha {\mathbb I}_{n\times n})^{-1}] \left(-\frac12\vec B_{\alpha}\right)$, where we consider $\vec T_{\alpha}$ and $\vec B_{\alpha}$ as column vectors. The middle and lower panels of Fig.~\ref{fig:matrix} plot the transposed version of ${\mathbf B}$ and ${\mathbf T}$ matrices for a better layout.}
As one can see from Fig.~\ref{fig:matrix}, the submatrices themselves also have structures, since each of them involves several input images; some of the substructures of submatrices seem flipped or broken, due to the ordering induced by our partitioning process.
In the ${\mathbf A}$ matrix (upper panel), block-diagonal submatrices \NewEdit{are} computed for the same group of input pixels from an {\tt InStamp} instance, while others \NewEdit{are} computed for a pair of different {\tt InStamp} instances; some of the columns and rows are narrower than others, because only part of the pixels are selected from the corresponding {\tt InStamp} instances. % were
Since the ${\mathbf A}$ matrices are symmetric, below-diagonal submatrices are simply their above-diagonal counterparts transposed.

\NewEdit{The} {\tt Block} instance loops over output postage stamps, with each {\tt OutStamp} building (with the help of {\tt SysMatA} and {\tt SysMatB}) and solving (with a concrete derived class of {\tt \_LAKernel}, which we detail in Section~\ref{sec:lakernel}) its own linear system.
Finally, the {\tt Block} finishes its job by writing the final output file, after all postage stamps have been coadded and integrated to its output arrays.\footnote{To save the partial progress, the {\tt Block} also writes an intermediate output file each time a row of $2 \times 2$ postage stamp groups is completed. When ${\tt fade\_kernel} > 0$ (see \NewEdit{Section~\ref{ss:app-linsys}}), $2 \times {\tt fade\_kernel}$ rows or columns on block boundaries in intermediate outputs are not directly usable, since the ``fading'' is only recovered for the final output.} % the previous footnote

\subsection{Additional output maps} \label{ss:outmaps}

In addition to coadded versions of all input layers in primary HDU of the FITS file, {\sc Imcom} can write additional output maps in extension HDUs to help diagnose image quality. % \NewEdit{\st{several}}
\papone\ defines the ``fidelity'' as $-10\log_{10} (U_\alpha/C)$, where $U_\alpha$ is the PSF leakage defined in Eq.~(\ref{eq:U_Sigma}) and $C = \Vert \Gamma \Vert^2$ is the square norm of the target output PSF. {\sc fluffy-garbanzo} stores this map as unsigned 8-bit integers ({\tt numpy.uint8}); for transition pixels between postage stamps, it records the minimum fidelity.
It does not include the noise amplification ($\Sigma_\alpha$) map; it stores both maximum and minimum values (which can be different for transition pixels) of the Lagrange multiplier ($\kappa_\alpha$) as 32-bit floating point numbers ({\tt numpy.float32}) in a separate FITS file if requested.

{\sc PyImcom} supports all these three quantities, and incorporates two additional diagnostics for each output pixel.
As its name indicates, the total weight is defined as a summation of coaddition weights over all input pixels
\begin{equation}
T_{{\rm tot}, \alpha} = \sum_i T_{\alpha i}.
\label{eq:Tsum}
\end{equation}
Since {\sc Imcom} tries to conserve surface brightness rather than flux, $T_{\rm tot}$ is expected to be $\sim 1$ for each output pixel; however, since input PSFs $G_i$ and the target output PSF $\Gamma$ have different levels of concentration, it should not be surprising if $T_{\rm tot}$ is slightly off. Compared to specific values of the total weight, its uniformity among all output pixels is arguably more important for shape measurements.

To test whether {\sc Imcom} takes full advantage of all available input exposures, we define the effective coverage as
\begin{equation}
\bar{n}_{{\rm eff}, \alpha} = \frac{
\left( \sum_{\bar i} |t_{\alpha\bar i}| \right)^2
}{ \sum_{\bar i} t_{\alpha\bar i}^2},
~~~~ t_{\alpha\bar i} = \sum_{i\in\bar{i}} T_{\alpha i},
\label{eq:Neff}
\end{equation}
where $\bar{i}$ denotes the set of input pixel indices corresponding to each input image. This quantity is normalized so that its maximum value is equal to the number of selected exposures; it is trivial to prove that this maximum is reached if and only if the absolute total contribution, $|t_{\alpha\bar i}|$, is equal for all these exposures.
We emphasize that all these five output maps can be computed for any strategy determining the coaddition weights, regardless of whether it actually solves {\sc Imcom} linear systems, so they are useful for \NewEdit{comparison} purposes. % comparisons

In order to reduce storage usage while maintaining reasonable precision, {\sc PyImcom} compresses these additional maps into 16-bit integers, either unsigned ({\tt numpy.uint16}) or signed ({\tt numpy.int16}), by multiplying the logarithm of the five quantities (note that they are all non-negative by definition, and practically almost always positive) by appropriate coefficients.
Below we summarize extension HDU options, data types, and adopted coefficients of the five supported output maps:\footnote{Which map or maps to include in the output FITS file can be specified via the {\tt OUTMAPS} configuration entry, which should be a string containing uppercase initial letter(s) of quantity name(s): {\tt \textquotesingle U\textquotesingle} for $U_\alpha/C$, {\tt \textquotesingle S\textquotesingle} for $\Sigma_\alpha$, {\tt \textquotesingle K\textquotesingle} for $\kappa_\alpha$, {\tt \textquotesingle N\textquotesingle} for $T_{{\rm tot}, \alpha}$, and {\tt \textquotesingle N\textquotesingle} for $\bar{n}_{{\rm eff}, \alpha}$; the order does not matter. If a map is supposed to be uniform, for example when a single $\kappa/C$ value is specified, {\sc PyImcom} automatically excludes it.}
\begin{itemize}
    \item {\tt \textquotesingle FIDELITY\textquotesingle} ({\tt numpy.uint16}): $-5000\log_{10} (U_\alpha/C)$;
    \item {\tt \textquotesingle SIGMA\textquotesingle} ({\tt numpy.int16}): $-10000\log_{10} \Sigma_\alpha$;
    \item {\tt \textquotesingle KAPPA\textquotesingle} ({\tt numpy.uint16}): $-5000\log_{10} \kappa_\alpha$;
    \item {\tt \textquotesingle INWTSUM\textquotesingle} ({\tt numpy.int16}): $200000\log_{10} T_{{\rm tot}, \alpha}$;
    \item {\tt \textquotesingle EFFCOVER\textquotesingle} ({\tt numpy.uint16}): $50000\log_{10} \bar{n}_{{\rm eff}, \alpha}$.
\end{itemize}
The {\tt OutImage} class of the {\tt analysis.py} module provides utilities to automatically recover floating point numbers from these ({\tt u}){\tt int16} maps by removing coefficients (parsed from header comments) and undoing the logarithm.

\subsection{Fixes to known issues} \label{ss:fixes}

The following issue identified in Paper II has been fixed in this work and in the {\sc PyImcom} toolkit:
\begin{list}{$\bullet$}{}
\item The script that collected the injected stars in Paper II did not cut enough pixels around the block padding regions, resulting in $\approx 8\%$ of the objects being duplicates because they are near block boundaries (and $\sim 0.7\%$ are repeated 4$\times$ because they are near the block corners). This resulted in non-uniform weighting of the 1-point statistics histograms and the 2-point correlation functions of the injected star moments (including likely $\sim 15\%$ underestimate of the error bars in the latter).
We note that \paptwo\ results were still unbiased, and our main conclusions were not affected by this issue.
\end{list}

\section{Linear algebra strategies} \label{sec:lakernel}

This section introduces alternative linear algebra (LA) strategies of {\sc Imcom}, including three newly developed LA kernels and different approaches to determine the Lagrange multiplier $\kappa$.
{\sc PyImcom} implements {\tt \_LAKernel} as an abstract base class for all LA kernels, which are supposed to have the same forms of input and output.
To enable efficient bisection search on optimal $\kappa$ values, {\sc furry-parakeet} performs eigendecomposition on the ${\mathbf A}$ matrix of each postage stamp; this has become the first LA kernel of {\sc PyImcom} (the first concrete derived class of {\tt \_LAKernel}), the eigendecomposition kernel ({\tt EigenKernel}).

In Section~\ref{ss:kappa-search}, we demonstrate that searching $\kappa$ in a reduced space provides almost equivalent results, and allows us to use the more efficient Cholesky decomposition instead; this is our first new kernel, the Cholesky kernel ({\tt CholKernel}).
\papone\ has shown that solving linear systems as they are formulated in {\sc furry-parakeet} lead to significant postage stamp boundary effects, which may confuse source identification algorithms and cause biases in shape measurements.
Therefore in Section~\ref{ss:iter-kernel}, we develop an additional LA kernel: the iterative kernel ({\tt IterKernel}), which tries to avoid these effects by tailoring input pixel selection for each output pixel and solving linear systems with an iterative method.
In Section~\ref{ss:empir-kernel}, we introduce a mock LA kernel, the empirical kernel ({\tt EmpirKernel}), which employs an empirical relation based on geometry to quickly mimic results produces by other kernels.
These are the four LA kernels currently supported by {\sc PyImcom}.\footnote{To switch linear algebra kernel, one sets the {\tt LAKERNEL} configuration entry to {\tt \textquotesingle Eigen\textquotesingle} for {\tt EigenKernel}, {\tt \textquotesingle Cholesky\textquotesingle} for {\tt CholKernel}, {\tt \textquotesingle Iterative\textquotesingle} for {\tt IterKernel}, or {\tt \textquotesingle Empirical\textquotesingle} for {\tt EmpirKernel}. The current default is {\tt \textquotesingle Cholesky\textquotesingle}.}

Regarding the search strategy for optimal $\kappa_\alpha$ values, each ``true'' LA kernel supports two modes: either bisection search within an interval ({\tt EigenKernel}), or interpolation between a series of nodes ({\tt CholKernel} and {\tt IterKernel}), and adoption of a single value.\footnote{The $\kappa/C$ values, where $C = \Vert \Gamma \Vert^2$, are set specified by the {\tt KAPPAC} configuration entry. This entry should always be an iterable (typically a Python {\tt list}); if it contains multiple elements, they must be sorted in increasing order. When there is only one element, no searching is performed; if there are more than one, {\tt EigenKernel} uses the first and last elements to set the interval for bisection search, while {\tt CholKernel} and {\tt IterKernel} consider all elements as $\kappa/C$ nodes for the reduced space. {\tt EmpirKernel} simply ignores this entry.}

\subsection{Searching the space of $\kappa$} \label{ss:kappa-search}

A key challenge in implementing the {\sc Imcom} algorithm is that Eq.~(\ref{eq:T-AB}) has a different matrix inversion for each possible value of $\kappa_\alpha$. \citet{2011ApJ...741...46R} solved this by computing the eigendecomposition of ${\mathbf A}$. But eigendecomposition is still expensive (order several$\times n^3$ since it is an iterative process) compared to other matrix-matrix operations (e.g., Cholesky decomposition, which requires $n^3/6$ multiply-adds). Therefore it is worth examining whether one can get almost the same performance in the $(U_\alpha,\Sigma_\alpha)$-plane by using a few Cholesky decompositions rather than an eigendecomposition.

To implement this idea, we define a sequence of $N_{\rm v}$ ``nodes'' in $\kappa$-space: $\tilde\kappa^{(1)}<\tilde\kappa^{(2)}<... < \tilde\kappa^{(N_{\rm v})}$. We restrict our search space to linear combinations of solutions at the nodes:
\begin{equation}
T_{\alpha i} = \sum_{p=0}^{N_{\rm v}-1} \omega_\alpha^{(p)} T_{\alpha i}^{(p)},
\label{eq:T-LC}
\end{equation}
where the node matrices
\begin{equation}
T_{\alpha i}^{(p)} =
\sum_j [({\mathbf A} + \tilde\kappa^{(p)} {\mathbf I}_{n\times n})^{-1}]_{ij} \left(-\frac12B_{\alpha j}\right)
\label{eq:T-AB-L}
\end{equation}
are Eq.~(\ref{eq:T-AB}) evaluated at each node. The matrix system solutions are obtained using {\tt scipy.linalg.cholesky} ($N_{\rm v}n^3/6$ operations) and {\tt scipy.linalg.cho\_solve} ($N_{\rm v}n^2m$ operations).

Then the leakage metric and noise metric can be written as quadratic functions of the weights $w_\alpha^{(p)}$:
\begin{equation}
U_\alpha = \sum_{p,q} E_{\alpha}^{(pq)}\omega_\alpha^{(p)}\omega_\alpha^{(q)} - 2\sum_p D_\alpha^{(p)}\omega_\alpha^{(p)} + C
\label{eq:U_reduced}
\end{equation}
and
\begin{equation}
\Sigma_\alpha = \sum_{p,q} N_{\alpha}^{(pq)}\omega_\alpha^{(p)}\omega_\alpha^{(q)},
\label{eq:Sigma_reduced}
\end{equation}
where ${\mathbf E}_\alpha$ and ${\mathbf N}_\alpha$ are $N_{\rm v}\times N_{\rm v}$ symmetric positive semidefinite (positive definite in practical cases) matrices if viewed in terms of the $(pq)$ indices at fixed $\alpha$. We may explicitly write
\begin{align}
D_\alpha^{(p)} &= \sum_i \left(-\frac12B_{\alpha i}\right) T_{\alpha i}^{(p)}
,
\nonumber \\
N_\alpha^{(pq)} &= \sum_i T_{\alpha i}^{(p)} T_{\alpha i}^{(q)},
~~{\rm and}
\nonumber \\
E_\alpha^{(pq)} &= \sum_{ij} A_{ij} T_{\alpha i}^{(p)} T_{\alpha j}^{(q)}
= D_\alpha^{(q)} - \tilde\kappa^{(p)} N_\alpha^{(pq)}.
\label{eq:EDN}
\end{align}
As long as the final expression is used to evaluate ${\mathbf E}_\alpha$, the computation of ${\mathbf N}_\alpha$ dominates at ${\cal O}(N_{\rm v}^2 mn)$ operations, and is still fast compared to the construction of the $T^{(p)}_{\alpha i}$.

One can then find the weights $\omega_\alpha^{(p)}$ that minimize the combination $U_\alpha + \kappa_\alpha\Sigma_\alpha$ for any Lagrange multiplier $\kappa_\alpha$:
\begin{equation}
\omega_\alpha^{(p)} = \sum_q [({\mathbf E}_\alpha + \kappa_\alpha {\mathbf N}_\alpha)^{-1}]^{(pq)} D_\alpha^{(q)}.
\label{eq:omega}
\end{equation}
This is fast, involving only $N_\kappa mN_{\rm v}^3/6$ operations, where $N_\kappa$ is the number of trials of $\kappa_\alpha$ used for each output pixel $\alpha$. We perform a bisection search in $\log \kappa_\alpha$ space with $N_\kappa=12$ iterations over the space $\tilde\kappa^{(1)} < \kappa_\alpha < \tilde\kappa^{(N_{\rm v})}$ to meet the targets in \papone.

\begin{figure}
\includegraphics[width=\columnwidth]{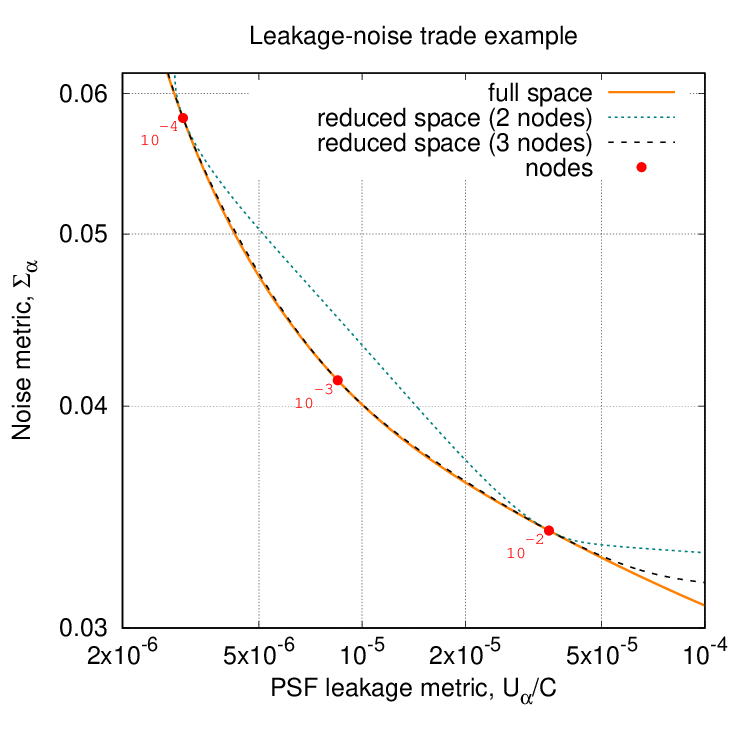}
\caption{\label{fig:trade}An example of the trade between the PSF leakage metric $U_\alpha/C$ and noise metric $\Sigma_\alpha$, for pixel $(504,504)$ in block $(0,0)$, Y band. The solid orange curve shows the locus in $(U_\alpha/C,\Sigma_\alpha)$-space traced out as one varies the Lagrange multiplier $\kappa_\alpha$ using the ``full space'' search (Eq.~\ref{eq:T-AB}). The red points labeled by $\log_{10}(\kappa_\alpha/C)$ indicate the $N_{\rm v}=3$ nodes used for this example. The dotted green curve shows the locus of linear combinations (Eq.~\ref{eq:T-LC}) obtained with the weights of Eq.~(\ref{eq:omega}) for 2 nodes (at $\tilde\kappa^{(p)}/C=10^{-4}$ and $10^{-2}$) and the dashed blue curve shows the results for 3 nodes (at $\tilde\kappa^{(p)}/C=10^{-4}$, $10^{-3}$, and $10^{-2}$). One sees that the 3-node curve achieves almost as good PSF leakage and noise performance as the full space search, but with reduced computational cost since only 3 Cholesky decompositions are used instead of an eigendecomposition.} % \NewEdit{\papone?}
\end{figure}

An example of this approach to searching the noise vs.\ PSF leakage space is shown in Fig.~\ref{fig:trade}. One sees that excellent performance can be obtained with 3 nodes.
Based on this ``shortcut,'' we have developed the Cholesky kernel ({\tt CholKernel}), which is supposed to produce coadded images almost equivalent to those yielded by the eigendecomposition kernel ({\tt EigenKernel}), but is faster when $N_{\rm v} < 6$.

\subsection{Iterative kernel ({\tt IterKernel})} \label{ss:iter-kernel}

The underlying philosophy of eigendecomposition \citep[][\papone]{2011ApJ...741...46R} and Cholesky decomposition (Section~\ref{ss:kappa-search}) is the same in a sense: the expensive ${\mathcal O}(n^3)$ decomposition only needs to be performed once or a few ($N_{\rm v}$) times for each postage stamp, and the results can be shared among all its output pixels.
The philosophy of the iterative scheme is completely different: nothing but ${\mathbf A}$ matrix elements is shared among output pixels, yet each row of the ${\mathbf T}$ matrix can be computed in a less complex (${\mathcal O}(n^2)$ with a smaller $n$) way.
Thanks to this decoupling, instead of a rounded square selection of input pixels (see the lower panel of Fig.~\ref{fig:inpix} and Eq.~\ref{eq:n-rdsq}), each output pixel only needs to select input pixels within an acceptance radius $\rho_{\rm acc}$ of itself, of which the expected number is
\begin{equation}
\langle n\rangle_{\rm circ} = \bar{n}_{\rm image} \frac{\pi \rho_{\rm acc}^2}{s_{\rm in}^2},
\label{eq:n-circ}
\end{equation}
where the subscript ``circ'' stands for circle; for $\rho_{\rm acc} = n_2\Delta\theta = 1.25 \,{\rm arcsec}$ (\papone), this yields $\langle n\rangle_{\rm circ} \approx 405.7 \;\bar{n}_{\rm image}$, or $\pi / (5+\pi) \approx 38.6\%$ of $\langle n\rangle_{\rm rdsq}$.
For each output pixel $\alpha$, it is expected to take $\langle n\rangle_{\rm rdsq}$ (number of pre-selected input pixels for each output postage stamp) distance calculations to select $\langle n\rangle_{\rm circ}$ relevant input pixels within $\rho_{\rm acc}$, and then $\langle n\rangle_{\rm circ}^2 + \langle n\rangle_{\rm circ}$ operations to extract its own system matrix ${\mathbf A}_\alpha$ and right vector ${\mathbf b}_\alpha$.
It turns out that the time consumption of distance-based selection is almost negligible, but the overhead due to array extractions and assignments is very significant, limiting the overall efficiency of the iterative kernel.\footnote{An performance improvement has been implemented during \papfour\ experiments, accelerating the iterative kernel by about a quarter (depending on the configuration).}

As for a specific iterative method, the conjugate gradient method \citep[CG;][]{hestenes1952methods} is specifically designed for real, symmetric, and positive definite system matrices, like ${\mathbf A}$ matrices in {\sc Imcom}. Thanks to the definition of ${\mathbf A}$ matrix elements (Eq.~\ref{eq:Aij_Bi}), reality, symmetry, and positive definiteness are all maintained when we extract elements corresponding to a given selection of input pixels.
For a given ${\mathbf A}_\alpha$, CG takes $\langle n\rangle_{\rm circ}^2$ multiply-adds to initialize, and then $\langle n\rangle_{\rm circ}^2$ additional multiply-adds per iteration. Denoting the average number of iterations as $\bar{n}_{\rm iter}$, the total complexity for an entire postage stamp is $N_{\rm v}(1+\bar{n}_{\rm iter}) \langle n\rangle_{\rm circ}^2 m$, where $N_{\rm v}$ is the number of ``nodes'' in $\kappa$-space (see Section~\ref{ss:kappa-search}).
For a coverage of $\bar{n}_{\rm image} = 8$, acceptance radius $\rho_{\rm acc} = n_2\Delta\theta = 1.25 \,{\rm arcsec}$, $m = n_2^2 = 2500$ output pixels, and $N_{\rm v} = 3$, equating the above expression with $N_{\rm v} (\langle n\rangle_{\rm rdsq}^3/6 + \langle n\rangle_{\rm rdsq}^2m)$, theoretically CG should perform as good as Cholesky decomposition when $\bar{n}_{\rm iter} \approx 9.48$.\footnote{Since the iterative kernel is not subject to boundary effects by definition, transition pixels between postage stamps are not needed. If we use $m = 2500$ for the iterative kernel and $m = 3136$ (corresponding to ${\tt fade\_kernel} = 3$, used in both \papone\ and this work) for the Cholesky kernel, this answer becomes $\bar{n}_{\rm iter} \approx 11.19$, which is not significantly different.}
Although in reality this LA strategy is limited by overhead described above, replacing ${\cal O}(n^3)$ with ${\cal O}(mn^2)$ naturally provides a possibility to attenuate the deep-field difficulty (same $m$ but much larger $n$; see \paptwo\ Section~6).

For efficiency, we implement our own conjugate gradient based on {\tt scipy.sparse.linalg.cg}\footnote{\url{https://docs.scipy.org/doc/scipy/reference/generated/scipy.sparse.linalg.cg.html}} and accelerate it with {\sc Numba}.
For a general linear system $A\vec{x} = \vec{b}$, the balance between precision and performance of an iterative method is set by the tolerance, usually defines as the maximum allowed error $\varepsilon \equiv |\vec{b} - A\vec{x}|$.
In {\tt scipy.sparse.linalg.cg}, this value is set to $\varepsilon = \max(\{{\tt rtol} \cdot |\vec{b}|, {\tt atol}\})$, where {\tt rtol} ({\tt atol}) is the relative (absolute) tolerance parameter. For simplicity, our implementation only supports {\tt rtol}, of which the default value is tentatively set to $1.5 \times 10^{-3}$; according to some initial tests, this leads to a typical number of iterations of $14$.\footnote{The relative tolerance {\tt rtol} is set by the {\tt ITERRTOL} configuration entry; the maximum number of iterations {\tt maxiter} is set by the {\tt ITERMAX} configuration entry, which currently defaults to $30$.}
With a smaller tolerance, CG can yield more exact results, but requires a larger number of iterations; fine-tuning this hyperparameter is a part of \papfour. Due to the finite tolerance, $U_\alpha$ and $\Sigma_\alpha$ computed using Eqs.~(\ref{eq:U_reduced}) and (\ref{eq:Sigma_reduced}) are not exact, by they are good approximations --- after being compressed to 16-bit integers (see Section~\ref{ss:outmaps}), the absolute discrepancy is usually $1$ if not $0$ --- hence we adopt these results instead of full calculations based on Eqs.~(\ref{eq:U_mat}) and (\ref{eq:Sigma_mat}) to save time.

\subsection{Empirical kernel ({\tt EmpirKernel})} \label{ss:empir-kernel}

The {\sc Drizzle} algorithm can coadd images in a much more efficient manner, as it assigns coaddition weights using only geometric relationship between input and output pixels. If {\sc Imcom} solutions could be approximated by a relatively simple pattern, we would be able to benefit from existing experience and obtain ``quick look'' coadds in drastically less time.
Since linear system solutions usually do not obey any simple empirical rules, this approximation is not expected to give the quality of standard {\sc Imcom}. But by being much faster, it can be useful in testing: it allows us to find larger issues with the input data or downstream products before we go to precision measurements.

\begin{figure*}
    \centering
    \includegraphics[width=0.9\textwidth]{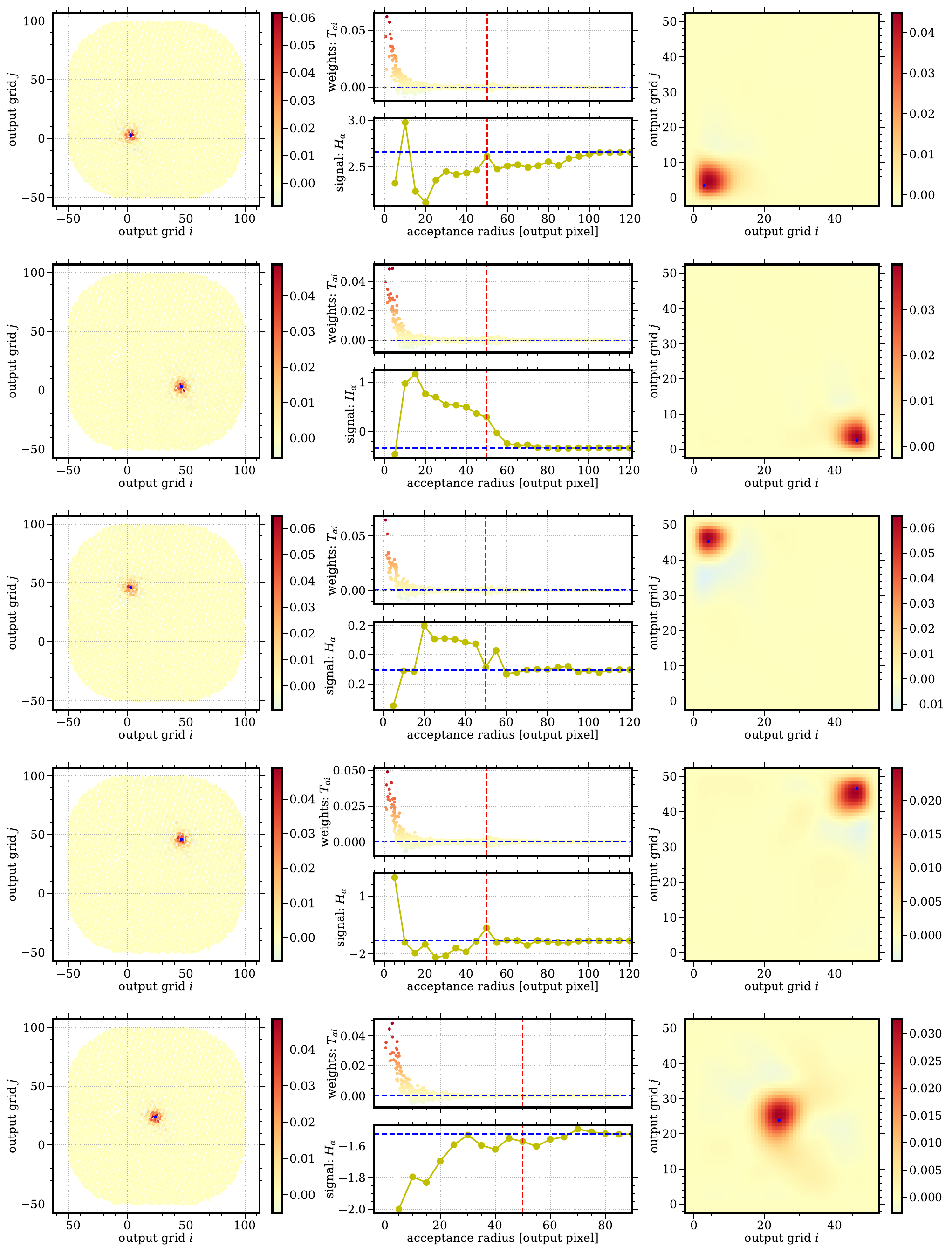}
    \caption{\label{fig:locality}Locality of the transformation matrix ${\mathbf T}$. {\it Left} column: ${\mathbf T}$ entries for five specific {\bf output} pixels (shown as blue dots) near the corners and the center of the postage stamp.
    {\it Middle} column, {\it upper} panels: ${\mathbf T}$ entries for the same output pixels, shown as a function of the distance between them and the input pixels, in units of output pixels; {\it lower} panels: output signal if we discard the contribution from input pixels outside a certain acceptance radius (but use the same ${\mathbf T}$ matrix elements for those inside), also as a function of distance.
    {\it Right} column: ${\mathbf T}$ entries for five specific {\bf input} pixels (shown as blue dots) near the corners and the center of the postage stamp; note that these maps have a smaller scale than those in the left column.}
\end{figure*}

To this end, we study correlation between ${\mathbf T}$ matrix elements (given by the Cholesky kernel) and pixel positions in Fig.~\ref{fig:locality}. The left column shows that input pixels with significant weights form circles surrounding our selected output pixels.
By plotting $T_{\alpha i}$ versus $d_{\alpha i} \equiv |{\bf r}_i - {\bf R}_\alpha|$, distance between input pixel $i$ and output pixel $\alpha$, in the upper panels of the middle column, we see clearly that weights outside a certain radius, $\lesssim 20$ output pixels or $0.5 \,{\rm arcsec}$, are at least an order of magnitude smaller than the maximum weight.
In the lower panels of the middle column, we discard the contribution from input pixels outside a gradually increasing acceptance radius, and plot the resulting signal in each output pixel versus this radius. Although the signal fluctuates, the dynamic range of such fluctuation is relatively small, indicating that distant input pixels are insignificant.
Finally in the right column, we display weights of same input pixels as maps of this example postage stamp. Again, we see that significant weights, positive and negative, can only be found within a relatively small distance. \NewEdit{Note that output pixels on the boundaries appear to receive smaller weights because of the transition between postage stamps (see \papone\ Eq.~4).}

To summarize, we find that the transformation matrix has a locality feature: input pixels with significant weights, usually positive but sometimes negative, are consistently those surrounding the output pixels we are looking at; discarding contribution from distant input pixels does not change the output signal to a large degree. Inspired by weight-versus-distance panels of this figure, here we add the empirical kernel ({\tt EmpirKernel}) to the list of kernels.
Instead of solving linear systems,\footnote{Although the empirical kernel does not need to solve linear systems, we still need to compute the system matrices if we want to obtain PSF leakage and noise amplification (Eq.~\ref{eq:U_Sigma}); this is time consuming, as we have to use Eqs.~(\ref{eq:U_mat}) and (\ref{eq:Sigma_mat}). For practical purposes, it is non-conditionally much faster than any kernel that solves linear systems.} this kernel employs an empirical relation to set ${\mathbf T}$ matrix elements:
\begin{equation}
T_{\alpha i} (d_{\alpha i}) = \mathcal{N}_\alpha \theta(\rho_{\rm acc} - d_{\alpha i}) (1 - d_{\alpha i} / \rho_{\rm acc}),
\label{eq:empir}
\end{equation}
where $d_{\alpha i} \equiv |{\bf r}_i - {\bf R}_\alpha|$, $\mathcal{N}_\alpha$ is a normalization factor so that $T_{{\rm tot}, \alpha} = \sum_\alpha T_{\alpha i} = 1$, and $\theta(\cdot)$ presents the Heaviside step function. Note that the acceptance radius $\rho_{\rm acc}$ plays a significantly different role here: for the empirical kernel, it directly affects the assignment of coaddition weights.

A raw exposure is a discrete sampling of the convolution between the true sky signal $f$ and the input PSF $G_i$ plus noise, while our desired output image is a continuous sampling (though represented by a discrete array) of the convolution between $f$ and the target output PSF $\Gamma$.
From this perspective, our empirical kernel can be considered as an almost fixed transformation kernel which we convolve with $G_i$ to approximate $\Gamma$. Thus it is similar to {\sc Drizzle}; the difference is that the choice of $\rho_{\rm acc}$ is informed by {\sc Imcom} results produced by other kernels.

In the following sections, we conduct simulations and compare coaddition results produced by different kernels.

\section{Simulations in this work} \label{sec:sims}

To compare the three new linear algebra kernels introduced in Section~\ref{sec:lakernel}, we re-coadd with each of them one ninth ($16 \times 16$ blocks, $1.0 \times 1.0 \,{\rm arcmin}^2$ each) of the \papone\ mosaic in all four bands (Y106, J129, H158, and F184).
For each combination of band and LA kernel, we use its ``benchmark'' configuration, detailed in Section~\ref{ss:config}.
Before going into thorough analysis of coadded layers, we discuss performance of each kernel and present some example images in Section~\ref{ss:preview}, and describe statistics of {\sc Imcom} diagnostics in Section~\ref{ss:diagnostics}.

\subsection{Configuration technical details} \label{ss:config}

Simulations in this work use the same input as \papone: the {\slshape Roman} arm \citep{2023MNRAS.522.2801T} of the Legacy Survey of Space and Time (LSST) Dark Energy Science Collaboration (DESC) Data Challenge 2 (DC2) simulations \citep{2019ApJS..245...26K, 2021arXiv210104855L, 2021ApJS..253...31L, 2022OJAp....5E...1K}.\footnote{The input image (133729, 13) is missing as of we run these simulations. This only affects $13$ blocks in F184 band and does not change our main conclusions.} We direct interested readers to \papone\ Section~3.1 for a thorough description in the context of {\sc Imcom}.
In addition to simulated {\tt \textquotesingle SCI\textquotesingle} and {\tt \textquotesingle truth\textquotesingle} images, we incorporate simulated white and $1/f$ noise frames ({\tt \textquotesingle whitenoise1\textquotesingle} and {\tt \textquotesingle 1fnoise2\textquotesingle}) as well as an injected point source grid drawn by {\sc GalSim} ({\tt \textquotesingle gsstar14\textquotesingle}) from \papone\ Table~1.
Quantitative analysis in this work is mainly performed on the noise frames and point sources, but simulated images are included to make sure that {\sc PyImcom} largely replicates the behavior of our previous implementation of {\sc Imcom}. More details about noise frames and point sources will be presented in Sections~\ref{sec:noiseps} and \ref{sec:gsstar14}, respectively.

In preparation for coaddition of the \NewEdit{OpenUniverse2024 simulations \citep{OpenUniverse2025arXiv}}, we have found that a simple Gaussian function may outperform an obscured Airy disk smoothed by Gaussian (adopted by \papone) as target output PSF.\footnote{The form and width of target output PSF is set by the {\tt OUTPSF} and {\tt EXTRASMOOTH} configuration entries, respectively. For the form, currently three options are supported: simple 2D Gaussian ({\tt \textquotesingle GAUSSIAN\textquotesingle}) and Airy disk, either obscured by $0.31$ (same as {\sl Roman}; {\tt \textquotesingle AIRYOBSC\textquotesingle}) or unobscured ({\tt \textquotesingle AIRYUNOBSC\textquotesingle}), convolved with a 2D Gaussian. The characteristic width of an Airy disk is set by $\lambda/D$, where $\lambda$ is the central wavelength of the filter, and $D$ is the diameter of {\sl Roman} aperture; {\tt EXTRASMOOTH} always corresponds to the width of the Gaussian component, whether or not it is the only component. Following our previous implementation, {\sc PyImcom} also supports multiple target output PSFs in the same run. This can be done by adding pairs of configuration entries like {\tt OUTPSF2} and {\tt EXTRASMOOTH2}, etc.; we do not use this feature in this work.} % Open Universe 2024 (Troxel et al. in prep) simulations
Due to the inclusion of charge diffusion in our new image simulations, the PSF width is supposed to be different for the two versions of synthetic images. Therefore, in each band, we adopt as target PSF a Gaussian function with the same full width at half maximum (FWHM) as in \papone\ Table~4.
Note that \papone\ adopted monotonically larger width for bluer band to better mitigate undersampling of input images; fine-tuning the target PSF width will be a part of \papfour.
Basically, other configuration entries are set following \papone, except for the strategy for determining the Lagrange multiplier $\kappa$, which we discuss next.

As mentioned in Section~\ref{ss:formalism}, the Lagrange multiplier $\kappa_\alpha$ (subscript $\alpha$ is included here to emphasize that $\kappa$ can be different for each output pixel) is how the two optimization goals of {\sc Imcom}, PSF leakage $U_\alpha$ and noise amplification $\Sigma_\alpha$, are balanced. As shown in the appendix of \citet{2011ApJ...741...46R}, as $\kappa_\alpha$ increases, $U_\alpha$ always increases, while $\Sigma_\alpha$ always decreases.
Both \citet{2011ApJ...741...46R} and \papone\ performed a bisection search within a specific range $(\kappa_{\min}, \kappa_{\max})$ with $n_{\rm bis} = 53$ steps, and Fig.~\ref{fig:trade} shows that $N_{\rm v} = 3$ nodes in the $\kappa$ space are sufficient to yield almost the same $\Sigma_\alpha$--$U_\alpha$ curve.
This largely facilitates the search for optimal $\kappa_\alpha$ values, enabling the usage of Cholesky decomposition or iterative method (these would be much slower than the original eigendecomposition if we still use bisection search).
Using $\tilde\kappa$ nodes does not change the general strategy at all: {\sc Imcom} optimizes $\kappa$ for each output pixel $\alpha$ separately, and even neighboring output pixels can have drastically different values (in logarithmic space).
However, during the development of {\sc PyImcom}, we have noticed that the $\kappa$ map produced in this way is usually close to \NewEdit{being} uniform, and non-uniformity may aggravate postage stamp boundary effects, suggesting that a fixed $\kappa$ value for all output pixels is a possibly better choice, despite its simplicity.
If this works, the number of systems to solve using Cholesky decomposition or iterative method can be divided by $N_{\rm v}$, leading to another desirable speed-up.

Through testing, we find that for the iterative kernel, impact of the Lagrange multiplier $\kappa$ only becomes significant when it reaches the level of $10^{-3} C$, which is arguably large.
Since $\kappa$ was induced to improve the noise control, and the iterative kernel already controls the noise reasonably well (see Section~\ref{sec:noiseps}), we can set $\kappa = 0$ and focus on the minimization of PSF leakage. Note that for the iterative kernel, neither $U/C$ nor $\Sigma$ is a strictly monotonic function of $\kappa$, as it does not yield exact solutions to linear systems.
In the case of the Cholesky kernel, zero or small $\kappa/C$ values make linear systems unstable (i.e., not positive definite due to numerical errors); we adopt $\kappa/C = 2\times 10^{-4}$ for the Cholesky kernel in all four bands.

Some further adjustments are needed for the new linear algebra strategies.
As for the acceptance radius, we adopt $\rho_{\rm acc} = n_2\Delta\theta = 1.25 \,{\rm arcsec}$ following \papone\ for the Cholesky kernel, but smaller values for the other two kernels.
In principle, the same acceptance radius could be used for the iterative kernel. However, since this kernel is slow due to significant overhead and large number of iterations, we halve the \papone\ value and adopt $\rho_{\rm acc} = 0.625 \,{\rm arcsec}$. How the quality of the iterative kernel output scales with $\rho_{\rm acc}$ is another topic of \papfour.
As mentioned in Section~\ref{ss:empir-kernel}, the acceptance radius directly affects the assignment of coaddition weights yielded by the empirical kernel. Based on a visual inspection of Fig.~\ref{fig:locality}, $\rho_{\rm acc} = 0.25 \,{\rm arcsec}$ seems reasonable for Y106 band; since the main purpose of this work is to demonstrate new capabilities of the {\sc Imcom} software, we adopt the same value for the other three bands without fine-tuning.

All $12$ configuration files used for simulations in this work can be found in the {\tt configs/paper3\_configs/} subdirectory of the {\sc PyImcom} repository (v1.0.2).

\subsection{Performance and example images} \label{ss:preview}

\begin{table}[]
    \centering
    \caption{\label{tab:time_consump}Number of cores and average time consumption (together with standard deviation) per block ($1.0 \times 1.0 \,{\rm arcmin}^2$) for each linear algebra strategy. Note that this work uses the same machine, namely the Pitzer cluster at the Ohio Supercomputer Center \citep{Pitzer2018}, as most of \papone\ simulations.}
    \begin{tabular}{cccc}
    \hline
    LA strategy & \papone & Cholesky & Iterative \\
    \hline
    No. of cores & $2$ or $3$ & $1$ or $1.25$ & $1$ or $1.25$ \\
    Y106 (hours) & $41.54$ & $7.87 \pm 2.07$ & $24.88 \pm 7.05$ \\
    J129 (hours) & $47.53$ & $13.11 \pm 6.21$ & $38.46 \pm 9.16$ \\
    H158 (hours) & $61.02$ & $11.74 \pm 5.37$ & $35.66 \pm 9.68$ \\
    F184 (hours) & $58.51$ & $34.26 \pm 22.55$ & $29.78 \pm 6.75$ \\
    \hline
    \end{tabular}
\end{table}

Table~\ref{tab:time_consump} compares number of core-hours consumed by the Cholesky and iterative kernels to \papone\ simulations (see its Appendix B for further details).
The Cholesky kernel is supposed to produce almost equivalent results as the eigendecomposition kernel, of which the precursor was adopted in \papone. Thanks to software improvements (see Section~\ref{ss:pyimcom} and \NewEdit{Appendices~\ref{app:pyimcom} and \ref{app:accel}}), the consumption of core-hours has been reduced by about an order of magnitude.
F184 band is an exception, as about $54.59\%$\footnote{The actual fraction is slightly smaller, as some double-counting is involved.} of the postage stamps encounter Cholesky decomposition failures, and {\sc Imcom} has to perform eigendecomposition to fix negative eigenvalues. In retrospect, $\kappa/C = 2\times 10^{-4}$ is probably not sufficient to suppress numerical errors, causing ${\mathbf A}$ matrices in F184 to be unstable. However, since the coaddition results are still valid, we choose not to rerun the simulations; the \NewEdit{OpenUniverse2024} simulations are coadded with larger values of $\kappa/C$ in the redder bands. % Open Universe 2024

Due to its significant overhead, although the iterative kernel uses a significantly smaller acceptance radius ($0.625 \,{\rm arcsec}$ versus $1.25 \,{\rm arcsec}$), it is a few times slower than the Cholesky kernel in general. Nevertheless, benefiting from the {\sc PyImcom} infrastructure, it is still faster than the \papone\ implementation and uses less cores.
The survey strategy adopted for \citet{2023MNRAS.522.2801T} image simulations was $2$ passes in each band, each with $3$ dithers in Y106 or F184 bands and $4$ dithers in J129 and H158 bands. Comparing the time consumption of the iterative and Cholesky kernels in the three bluer bands, we see that the former is slightly less sensitive to coverage (number of exposures; see \papone\ Fig.~1), as expected from its ${\cal O}(n^2)$ complexity (as opposed to ${\cal O}(n^3)$ of the latter).
In addition, based on its performance in F184, the iterative kernel is stable against positive-definiteness issue caused by numerical errors. However, these potential advantages are only useful if it can become faster and more accurate; such improvements are left for future work.

The performance of the empirical kernel is not included in Table~\ref{tab:time_consump}. For simulations in this work, its time consumption is slightly greater than that of the Cholesky kernel.
However, this does not manifest the true performance of this mock kernel: as noted in Section~\ref{ss:empir-kernel}, most of the time is spent to compute PSF leakage and noise amplification, both defined in Eq.~(\ref{eq:U_Sigma}); furthermore, since it does not solve linear systems, it cannot benefit from shortcuts provided by Eqs.~(\ref{eq:U_reduced}) and (\ref{eq:Sigma_reduced}).
In the ``no-quality control'' mode of the empirical kernel,\footnote{When the {\tt LAKERNEL} configuration entry is set to {\tt \textquotesingle Empirical\textquotesingle}, there are two alternative ways to enable this mode: i) exclude both ``U'' (for PSF leakage) and ``S'' (for noise amplification) from {\tt OUTMAPS}; ii) set {\tt EMPIRNQC} to {\tt true} in JSON or {\tt True} in Python. {\sc PyImcom} automatically updates {\tt OUTMAPS} and {\tt EMPIRNQC} if this mode is enabled in either way.} the time consumption of coadding a block is at the level of $10$ minutes.
In addition, the empirical kernel only requires one core per thread. Therefore, the affordability makes it a competitive ``quick look'' tool like {\sc Drizzle}.

\begin{figure*}
    \centering
    \includegraphics[width=0.85\textwidth]{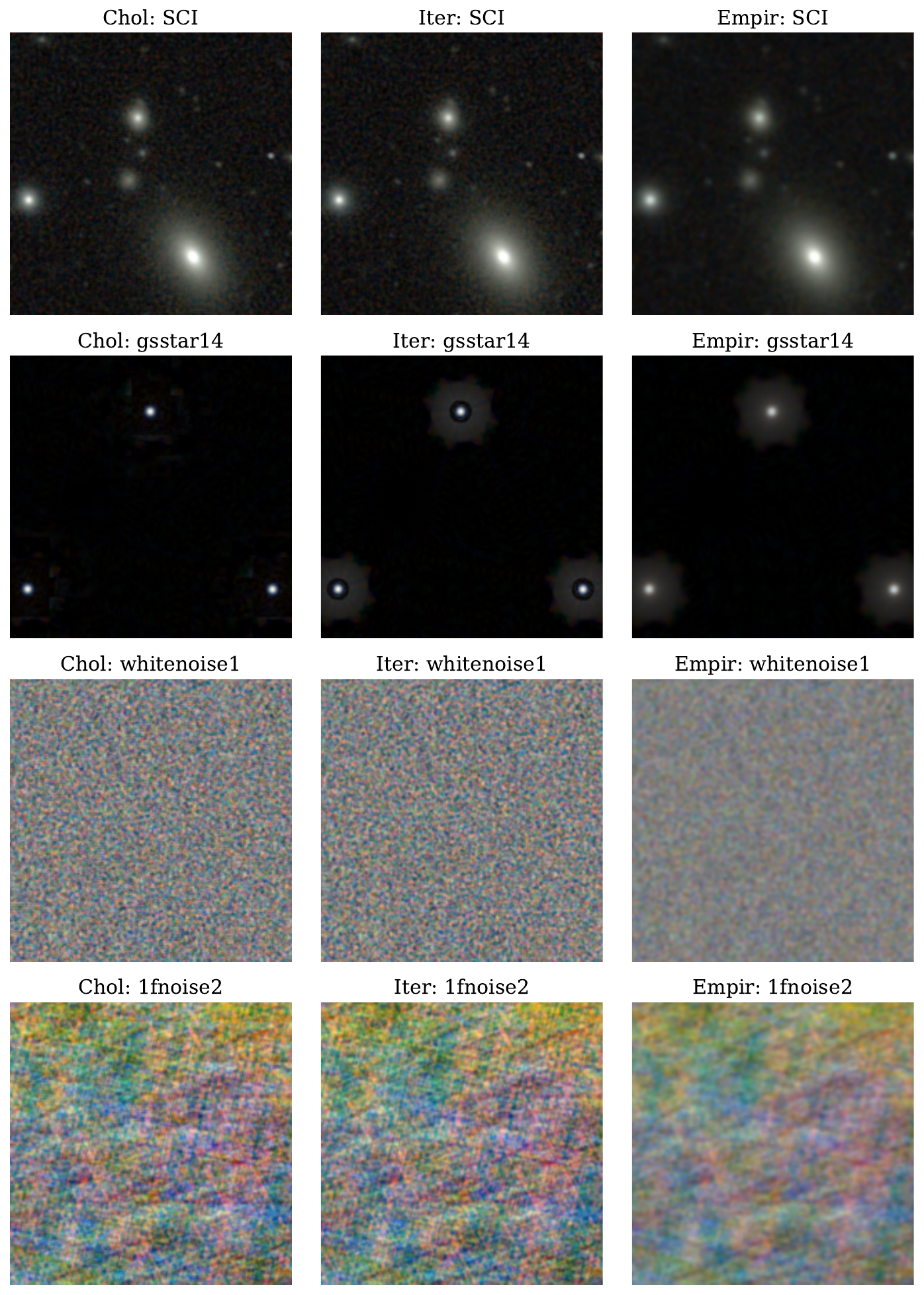}
    \caption{\label{fig:example_images}Four layers in a field of $17.5 \,{\rm arcsec}$ ($700$ output pixels) on a side, coadded by the Cholesky kernel ({\it left} column), the iterative kernel ({\it middle} column), and the empirical kernel ({\it right} column). Each panel is a Y106 ({\color{Y106} \#001AA6}) + J129 ({\color{J129} \#006659}) + H158 ({\color{H158} \#596600}) + F184 ({\color{F184} \#A61A00}) composite; note that these four colors have similar lightnesses and add up to white (\#FFFFFF).
    From {\it top} row to {\it bottom} row, the four layers are: simulated science images ({\tt \textquotesingle SCI\textquotesingle}), injected stars drawn by {\sc GalSim} ({\tt \textquotesingle gsstar14\textquotesingle}; see Section~\ref{sec:gsstar14}), simulated white noise frames ({\tt \textquotesingle whitenoise1\textquotesingle}; see Section~\ref{ss:whitenoise1}), and simulated $1/f$ noise frames ({\tt \textquotesingle 1fnoise2\textquotesingle}; see Section~\ref{ss:1fnoise2}). The scaling is set following \papone\ Fig.~8 for {\tt \textquotesingle SCI\textquotesingle} and following \paptwo\ Fig.~1 for the other three layers.}
\end{figure*}

Fig.~\ref{fig:example_images} compares four layers of a $17.5 \times 17.5 \,{\rm arcsec}^2$ field coadded by the Cholesky, iterative, and empirical kernels.
The first row presents simulated science images ({\tt \textquotesingle SCI\textquotesingle} layer), which are produced for ``sanity check'' purposes and are not analyzed in this work. \NewEdit{According to visual inspection, the three versions look quite similar to each other, indicating that all three kernels have successfully produced reasonable coaddition results; however, quantitative analyses should be able to tell the differences.} Although the background noise level is significantly lower in the empirical kernel results, we will show that this kernel is not suitable for measurements.
The second row presents injected stars drawn by {\sc GalSim} ({\tt \textquotesingle gsstar14\textquotesingle} layer), which are analyzed in Section~\ref{sec:gsstar14}. Images produced by the Cholesky kernel seem clean, yet careful scrutiny would reveal postage stamp boundary effects; the other two kernels have a halo surrounding each star, which is another artifact. The upper star will be re-plotted in a different scale in Fig.~\ref{fig:injected_star}.
The third and fourth rows present simulated white noise frames ({\tt \textquotesingle whitenoise1\textquotesingle}) and $1/f$ noise frames ({\tt \textquotesingle 1fnoise2\textquotesingle}), which are analyzed in the two subsections of Section~\ref{sec:noiseps}, respectively. The $1/f$ noise frame produced by the Cholesky kernel manifests a square grid (aligned with the panel) which again corresponds to postage stamp boundaries; otherwise the subtle differences between the first two kernels are hardly noticeable from this figure, while the empirical kernel results are considerably blurry.

\subsection{Statistics of {\sc Imcom} diagnostics} \label{ss:diagnostics}

\begin{figure*}
    \centering
    \includegraphics[width=0.95\textwidth]{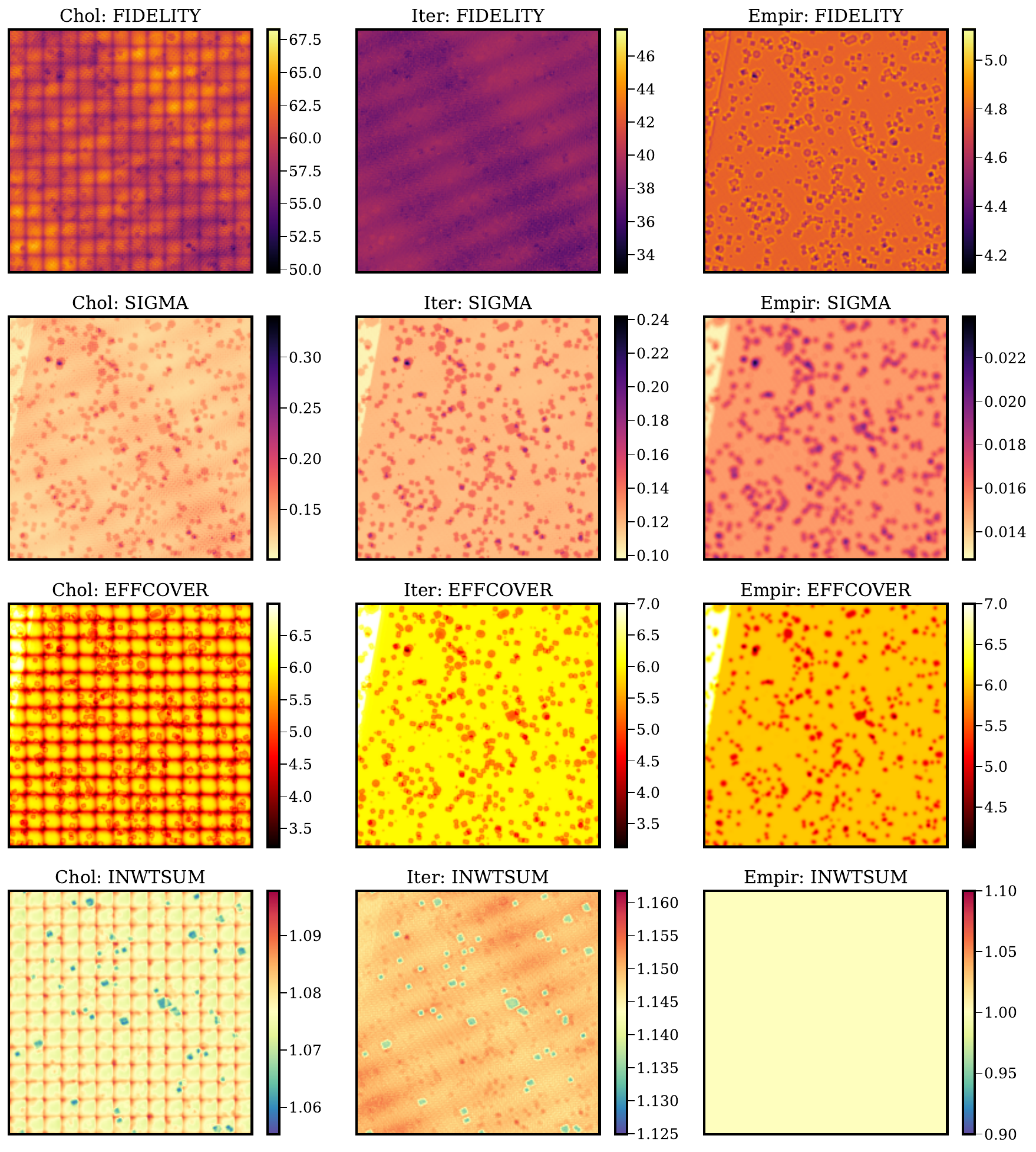}
    \caption{\label{fig:cf_outmaps}Output maps in the H158 band produced by the Cholesky kernel ({\it left} column), the iterative kernel ({\it middle} column), and the empirical kernel ({\it right} column).
    From {\it top} to {\it bottom}: fidelity (negative logarithmic PSF leakage in ${\rm dB}$, i.e., $-10 \log_{10} (U_\alpha/C)$), noise amplification (Eq.~\ref{eq:U_Sigma}), total weight (Eq.~\ref{eq:Tsum}), and effective coverage (Eq.~\ref{eq:Neff}). Note that we deliberately choose different color bar ranges to better display spatial structures.}
\end{figure*}

Fig.~\ref{fig:cf_outmaps} displays {\sc Imcom} output maps --- PSF leakage, noise amplification, effective coverage, and total weight --- in the same field as Fig.~\ref{fig:example_images} produced by the Cholesky, iterative, and empirical kernels in the H158 band. In the Cholesky kernel results, postage stamp boundaries can be clearly seen in all maps but noise amplification. Moir\'e patterns (see \papone\ Section~5.3 for discussion) show up to varying degrees in some maps produced by the first two kernels.
Detector defects and cosmic ray hits (injected as input pixel masks) are manifested as clusters of ``less good'' values in most maps, especially the effective coverage map produced by the iterative kernel or the empirical kernel. This demonstrates that, combined with simulations, {\sc Imcom} is suitable for high-precision depth variation studies, which are an important consideration in survey design.

In the following text, we discuss these four diagnostics one at a time: we present distributions of each quantity given by each band-kernel combination, and refer back to this figure for spatial structures if needed.
Since each mosaic in this work contains $[(n_{\rm block} n_1 + 2 {\tt PAD}) n_2]^2 = [16 \times 48 + 2 \times 2) \times 50]^2 \approx 1.49 \times 10^9$ pixels in total, some downsampling is necessary. Here we take $15 \times 15$ pixel cutouts centered at each of the $5517$ \NewEdit{stars injected onto a HEALPix\footnote{\url{https://healpix.sourceforge.io/}} grid \citep{2005ApJ...622..759G} with \NewEdit{${\tt NSIDE} = 14$} (see Section~\ref{sec:gsstar14})},\footnote{Practically, centers of an injected star and an output pixel never coincide; central pixels of such cutouts are those closest to expected centroids of injected stars.}\footnote{In \paptwo\ Section~5.1, it was stated that there were ``54\,597 unique simulated stars,'' although \papone\ simulations only covered an area $\sim 9$ times larger than those in this work. This is because \paptwo\ duplicated some of the injected stars by accident, see Section~\ref{ss:fixes}.}
compute either the average (for most quantities) or standard deviation (for total weight only) within each cutout, and take the logarithm afterwards.\footnote{In \paptwo, the ``mean fidelity'' was defined as $\langle -10 \log_{10} (U/C) \rangle$ (average of the logarithm); in this paper, we define it as $-10 \log_{10} \langle U/C \rangle$ (logarithm of the average). While the previous definition literally matches its name, taking the average in logarithmic space reduces the impact of pixels with large $U_\alpha$ values. Nevertheless, the differences are not very large --- usually up to $1$ in terms of fidelity.}
In addition, correlations between these diagnostics will be shown in the upper-left portion of each heatmap in Fig.~\ref{fig:heatmaps}.

\begin{figure}
    \centering
    \includegraphics[width=\columnwidth]{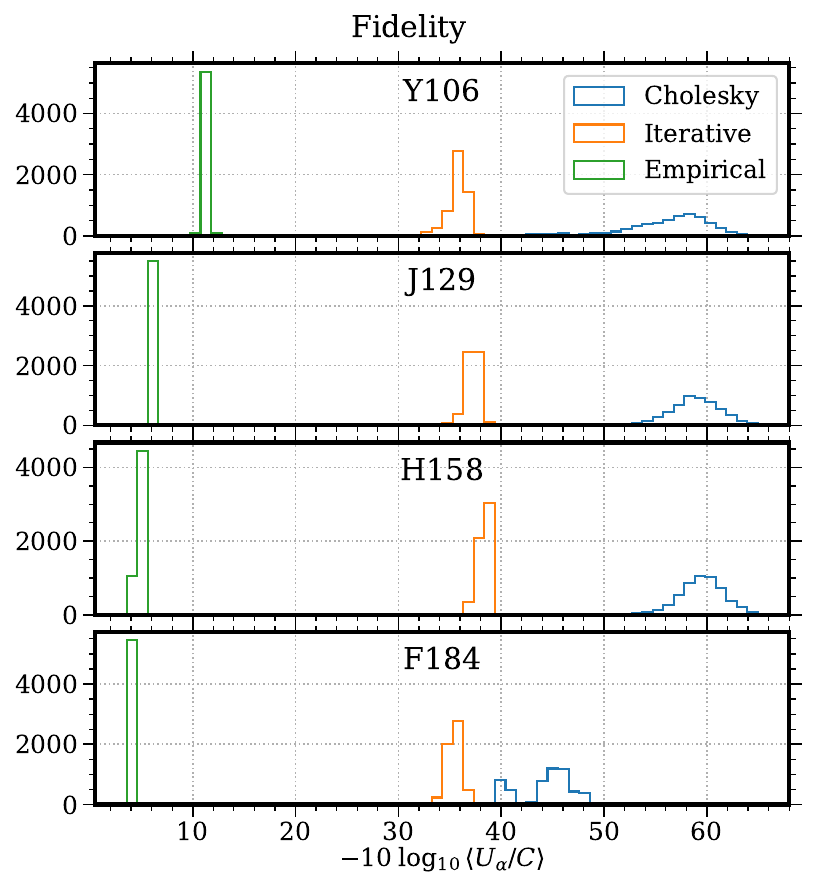}
    \caption{\label{fig:outmap_fidelity}Histograms of $5517$ mean fidelity values yielded by three linear algebra kernels in four bands.
    Mean fidelity is defined as $-10\log_{10} \langle U_\alpha/C\rangle$, where $U_\alpha$ is the PSF leakage metric defined in Eq.~(\ref{eq:U_Sigma}), and $\langle\cdot\rangle$ denotes an average over $15 \times 15$ pixels centered at a HEALPix node with \NewEdit{${\tt NSIDE} = 14$}.
    From {\it top} to {\it bottom}, the four panels present histograms in Y106, J129, H158, and F184 bands; results given by the Cholesky, iterative, and empirical kernels are shown in blue, orange, and green, respectively.}
\end{figure}

Fig.~\ref{fig:outmap_fidelity} presents the distribution of mean fidelity ($-10\log_{10} \langle U_\alpha/C\rangle$) values given by each band-kernel combination studied in this work. Fidelity directly mirrors how the as-realized coadded PSF in each pixel differs from the target output PSF; \papone\ aimed for a fidelity of $60$, corresponding to $0.1\%$ ``leakage'' in the root mean square sense.
As shown by the blue histograms, the Cholesky kernel, which is supposed to produce results similar to those given by eigendecomposition (adopted in \papone), has mean fidelity values in the vicinity of $60$ in the first three bands. In F184, the fidelity values are not as desirable, because the target output PSF is not sufficiently wider than input PSFs (see Section~\ref{ss:whitenoise1} for further evidence).
The obvious bimodal feature is probably due to the positive definiteness issue mentioned in Section~\ref{ss:preview} --- by fixing the negative eigenvalues, {\sc PyImcom} effectively adopts a larger Lagrange multiplier $\kappa_\alpha$, which increases the PSF leakage.

The mean fidelity values given by the iterative kernel are not as good, but are relatively consistent in all bands. This is a strong indication that, for this linear algebra kernel, PSF leakage is dominated by the random errors due to finite tolerance instead of nature of the system matrices.
As a reminder, we adopt a relative tolerance of ${\tt rtol} = 1.5 \times 10^{-3}$ for simulations in this work, which would lead to a fidelity value close to $-10\log_{10} (1.5 \times 10^{-3}) = 28.34$.\footnote{Intuitively, the relative tolerance ${\tt rtol}$ directly maps to the amplitude of errors in coaddition weights $T_{\alpha i}$. Eq.~(\ref{eq:Sigma_mat}) has three terms, a quadratic term and a linear term in $T_{\alpha i}$, and $C = \Vert \Gamma \Vert^2$ which does not depend on $T_{\alpha i}$. Since $|T_{\alpha i}| \ll 1$, we expect the linear term to dominate over the quadratic term when the weights have errors.} Fortunately, it turns out that errors in coaddition weights of different input pixels (for the same output pixel) are not (or only weakly) correlated to each other, hence the holistic effect is that the fidelity value is usually better than $28.34$.
As we will see in Section~\ref{sec:gsstar14}, poor fidelity strongly limits the accuracy of the iterative kernel results, motivating a more stringent tolerance parameter.

Unlike the previous two kernels, the empirical kernel is just a ``mock'' kernel, and is agnostic on the target output PSF. Therefore, it is not surprising that it has a much larger PSF leakage.
As mentioned in Section~\ref{ss:empir-kernel}, the acceptance radius $\rho_{\rm acc} = 0.5 \,{\rm arcsec}$ is chosen based on visual inspection of Fig.~\ref{fig:locality}, which is based on results in Y106 band; since target output PSFs are different in different bands, the fidelity is even worse in the other three bands.
Due to its enormous PSF leakage, it is unwise to perform shape measurements on the empirical kernel results and expect high precision. Nevertheless, source detection algorithms may be able to withstand this, and the empirical kernel remains a valid ``quick look'' tool.

\begin{figure}
    \centering
    \includegraphics[width=\columnwidth]{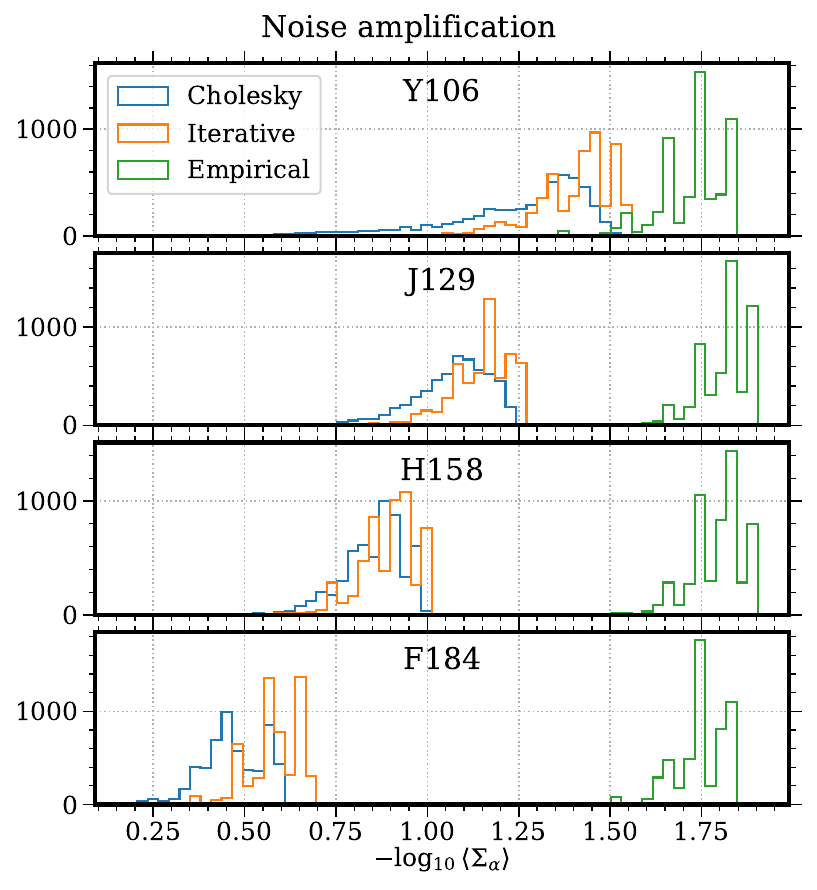}
    \caption{\label{fig:outmap_sigma}Histograms of $5517$ logarithmic mean noise amplification values yielded by three linear algebra kernels in four bands.
    Logarithmic mean noise amplification is defined as $\log_{10} \langle \Sigma_\alpha\rangle$, where $\Sigma_\alpha$ is the noise amplification metric defined in Eq.~(\ref{eq:U_Sigma}), and $\langle\cdot\rangle$ denotes an average over $15 \times 15$ pixels centered at a HEALPix node with \NewEdit{${\tt NSIDE} = 14$}.
    A minus sign is added so that ``better'' values are shown on the right; otherwise layout and format of the histograms are the same as in Fig.~\ref{fig:outmap_fidelity}.}
\end{figure}

Likewise, Fig.~\ref{fig:outmap_sigma} presents the distribution of logarithmic mean noise amplification ($\log_{10} \langle \Sigma_\alpha\rangle$) values; a minus sign is added so that ``better'' values are shown on the right, which is consistent with other histograms in this section.
Despite its poor fidelity, the empirical kernel has the best control over noise amplification among all three kernels considered in this work. This is because the way it assigns coaddition weights makes very good use of available input pixels if we want to minimize $\sum_{i=0}^{n-1} T_{\alpha i}^2$ while maintaining $T_{{\rm tot}, \alpha} = \sum_{i=0}^{n-1} T_{\alpha i} = 1$.
However, as mentioned in Section~\ref{ss:outmaps}, a sum of $1$ may not be the most reasonable normalization: if $T_{{\rm tot}, \alpha}$ is supposed to be a number larger than $1$, $\langle \Sigma_\alpha\rangle$ deteriorates by a factor of $T_{{\rm tot}, \alpha}^2$. This point will be made clearer in Section~\ref{ss:1pt-stats}.
The distribution shows several peaks, which correspond to different coverage values --- this statement is validated by the clearer peaks in Fig.~\ref{fig:outmap_neff} (see below) and the strong correlation between these two quantities in both Figs.~\ref{fig:cf_outmaps} and \ref{fig:heatmaps}.

For the iterative and Cholesky kernels, noise amplification shows a strong dependence on band: it is better in bluer bands and not as good in redder bands.
This is understandable: recall that \papone\ adopted wider target output PSFs for bluer bands, in which input PSFs are actually narrower; therefore, the level of concentration (i.e., the degree of heavily relying on neighboring input pixels) monotonically increases with wavelength.
Such observation indicates that the relative size between target and input PSFs matters a lot for noise control; choice of target PSF will be further explored in \papfour.
Besides, the iterative kernel consistently outperforms the Cholesky kernel in terms of noise control, which we will further investigate in Section~\ref{sec:noiseps}.

\begin{figure}
    \centering
    \includegraphics[width=\columnwidth]{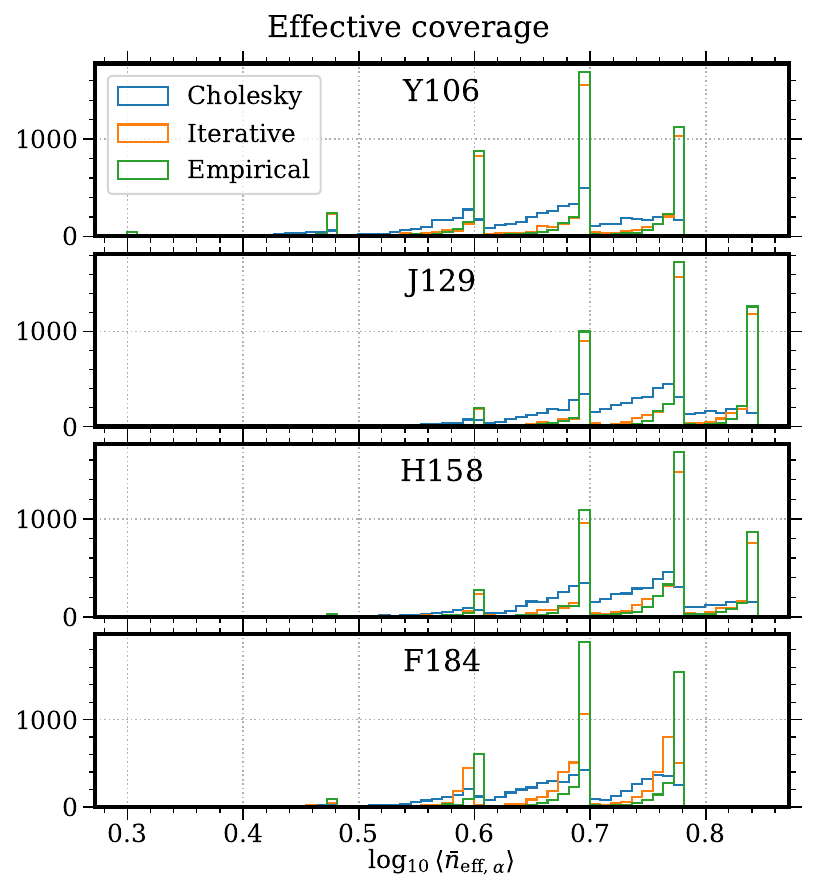}
    \caption{\label{fig:outmap_neff}Histograms of $5517$ logarithmic mean effective coverage values yielded by three linear algebra kernels in four bands.
    Logarithmic mean effective coverage is defined as $\log_{10} \langle \bar{n}_{{\rm eff}, \alpha}\rangle$, where $\bar{n}_{{\rm eff}, \alpha}$ is the effective coverage metric defined in Eq.~(\ref{eq:Neff}), and $\langle\cdot\rangle$ denotes an average over $15 \times 15$ pixels centered at a HEALPix node with \NewEdit{${\tt NSIDE} = 14$}.
    Layout and format of the histograms are the same as in Fig.~\ref{fig:outmap_fidelity}.}
\end{figure}

Fig.~\ref{fig:outmap_neff} presents the distribution of logarithmic mean effective coverage ($\log_{10} \langle \bar{n}_{{\rm eff}, \alpha}\rangle$) values.
It is perspicuous that this quantity is quantized: we can easily see peaks corresponding to $3$, $4$, $5$, and $6$ exposures in Y106 and F184 bands, as well as peaks corresponding to $4$, $5$, $6$, and $7$ exposures in J129 and H158 bands.
A closer look would reveal a tail on the left side of each peak, which manifests the existence of input pixel masks: {\sc Imcom} uses a set of permanent masks for bad pixels, and produces for each exposure a random mask (which is consistent in different runs) for ``cosmic ray hits.''
Before moving on to the other two kernels, we note that in \paptwo, mean coverage was defined in a coarse way: an input image is counted as one, as long as it has non-zero (strictly speaking, $> 0.01$) contribution to a specific postage stamp. Therefore, effective coverage is always preferred for finer purposes; however, for the purpose of binning blocks into mean coverage bins, this difference should be negligible, hence we use the old definition in Section~\ref{sec:noiseps} to facilitate comparisons with \paptwo\ results.

Since the empirical kernel makes ``very good'' use of available input pixels, its effective coverage results can be used as a benchmark for the other two kernels. Although the iterative kernel is a ``real'' LA kernel which solves linear systems, thanks to its circular window for input pixels (see Eq.~\ref{eq:n-circ} and surrounding text), it also makes ``quite good'' use of them, although F184 results are undermined by the insufficiently wide target output PSF.
Unlike the iterative kernel, the Cholesky kernel uses a rounded square window function (see the lower panel of Fig.~\ref{fig:inpix}, as well as Eq.~\ref{eq:n-rdsq} and surrounding text); if we consider the relative position between the output pixel and the window, we see that such a window is asymmetric, especially for output pixels close to postage stamp boundaries. This asymmetry biases the coaddition weights; therefore, the Cholesky kernel still has room for improvement despite its high fidelity.

\begin{figure}
    \centering
    \includegraphics[width=\columnwidth]{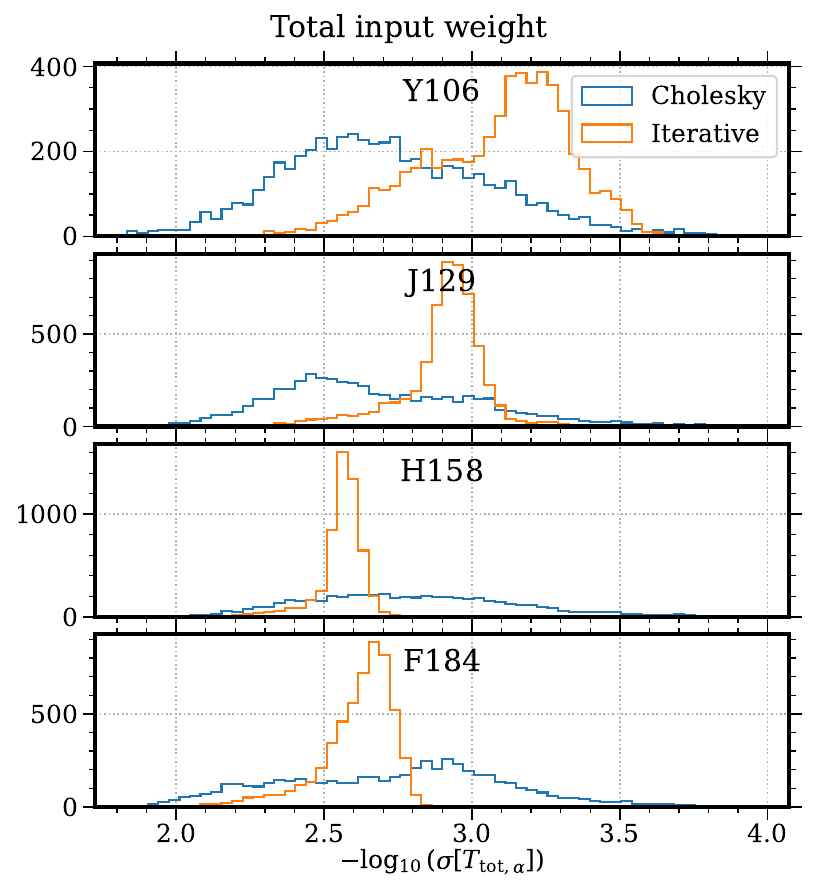}
    \caption{\label{fig:outmap_tsum}Histograms of $5517$ logarithmic standard deviation of total input weight values yielded by three linear algebra kernels in four bands.
    Logarithmic standard deviation of total weight is defined as $\log_{10} \sigma[T_{{\rm tot}, \alpha}]$, where $T_{{\rm tot}, \alpha}$ is the total input weight metric defined in Eq.~(\ref{eq:Tsum}), and $\sigma[\cdot]$ denotes a standard deviation within $15 \times 15$ pixels centered at a HEALPix node with \NewEdit{${\tt NSIDE} = 14$}.
    A minus sign is added so that ``better'' values are shown on the right; the empirical kernel results are not shown as the standard deviation is uniformly zero by definition; otherwise layout and format of the histograms are the same as in Fig.~\ref{fig:outmap_fidelity}.}
\end{figure}

Fig.~\ref{fig:outmap_tsum} presents the distribution of logarithmic standard deviation of total weight ($\log_{10} \sigma[T_{{\rm tot}, \alpha}]$) values; like in Fig.~\ref{fig:outmap_sigma}, a minus sign is added so that ``better'' values are shown on the right.
The empirical kernel results are not shown in this figure, as it has $T_{{\rm tot}, \alpha} = 1$ for all output pixels; for the other two kernels, distribution of this standard deviation is driven by different factors, which we discuss separately next.
The Cholesky kernel results are consistent across all four bands (note that the $y$-axis scale is different). As can be seen from Fig.~\ref{fig:cf_outmaps}, the total weight $T_{{\rm tot}, \alpha}$ is a strong function of output pixel position relative to postage stamp boundaries. Since the positions of injected stars are the same, the consistency among bands indicates that the level of total weight variation is similar.
The iterative kernel has a different story: the spread in $T_{{\rm tot}, \alpha}$ is again due to random errors caused by finite tolerance. As mentioned in the discussion of noise amplification, the redder the band, the higher the level of concentration, i.e., the less the number of input pixels dominating the fluctuation of total weight. With a smaller tolerance, the iterative kernel would outperform the Cholesky kernel, since it is not subject to postage stamp boundary effects.

\section{Noise power spectra} \label{sec:noiseps}

In this section, we examine power spectra of coadded noise frames following \paptwo. As mentioned in Section~\ref{ss:config}, simulations in this work include two types of simulated noise, white noise ({\tt \textquotesingle whitenoise1\textquotesingle}) and $1/f$ noise ({\tt \textquotesingle 1fnoise2\textquotesingle}); for a detailed analysis of real-world noise data from laboratory tests, see \NewEdit{\citet{Laliotis2024PASP}}. % Laliotis et al. (in prep)

We refer interested readers to \papone\ Section~3.4 for how {\sc Imcom} generates input noise frames, and to \paptwo\ Section~3 for technical details regarding the analysis. Below is a brief summary:
\begin{itemize}
    \item The white noise input is a realization of the uncorrelated component of readout noise. For each exposure, {\sc Imcom} generates a $4088\times 4088$ Gaussian random field with mean $0$ and variance $1$.
    \item The $1/f$ noise input is a realization of the principal correlated component of readout noise. For each channel ($4096 \times 128$ pixels, including reference pixels) of a {\slshape Roman} detector \citep{2020JATIS...6d6001M}, it is scale-invariant in the time stream and has unit variance per logarithmic range in frequency.

    \item For noise signal $S$, the 2D power spectra are computed as
    \begin{equation}
    P_{\rm 2D}(u, v) = \frac{s_{\rm out}^2}{N^2} \left| \sum_{j_x, j_y} S_{j_x, j_y} e^{-2\pi i s_{\rm out} (uj_x+vj_y)} \right|^2,
    \label{eq:P2D}
    \end{equation}
    where $u$ and $v$ are sampled at integer multiples of $1/(Ns_{\rm out})$; here $N = n_1 n_2 = 2400$ is the number of pixels on each side of a block, and $s_{\rm out} \equiv \Delta\theta = 1.25 \,{\rm arcsec}$ is the output pixel scale.
    Note that we have fixed the undue repetition of padding postage stamps in \paptwo.

    \item The azimuthally averaged power spectra are computed using the method from \citet{2023MNRAS.520.4715C}. We take the azimuthal average of 2D power spectra over $150$ thin annuli of equal width; to condense the results, $16 \times 16$ blocks of each mosaic are binned into $5$ equally sized bins according to their mean coverage (see Section~\ref{ss:diagnostics} for some discussion about the coverage).
\end{itemize}

We first investigate white noise in Section~\ref{ss:whitenoise1}, and then $1/f$ noise in Section~\ref{ss:1fnoise2}.

\subsection{Simulated white noise} \label{ss:whitenoise1}

\begin{figure*}
    \centering
    \includegraphics[width=0.95\textwidth]{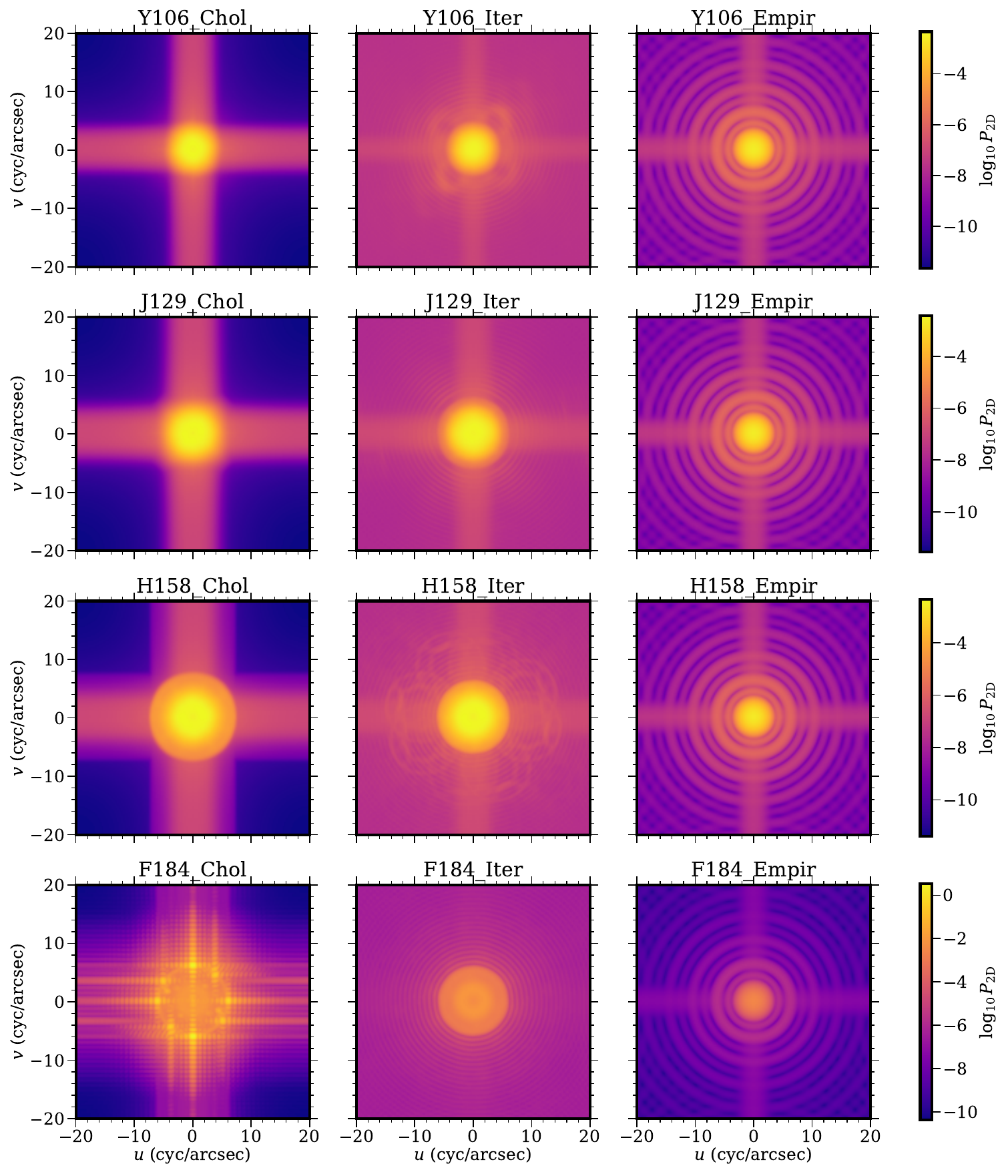}
    \caption{\label{fig:whitenoise1_2d}2D power spectra of simulated white noise frames ({\tt \textquotesingle whitenoise1\textquotesingle}), averaged over $16 \times 16$ blocks of each band-kernel combination and binned by $8 \times 8$ modes, plotted on a logarithmic scale. From {\it left} column to {\it right} column: results given by the Cholesky, iterative, and empirical kernels; from {\it top} row to {\it bottom} row: results in Y106, J129, H158, and F184 bands.
    Following \paptwo\ Fig.~2, the horizontal and vertical axes show wave vector components ($u$ and $v$ respectively) ranging from $-20$ to $+20$ cycles arcsec$^{-1}$. The color scale shows the power $P(u, v)$ in units of ${\rm arcsec}^2$ (Eq.~\ref{eq:P2D}).}
\end{figure*}

\begin{table}[]
    \centering
    \caption{\label{tab:extrama_white}Minimum and maximum values in 2D white noise power spectra shown in Fig.~\ref{fig:whitenoise1_2d}.}
    \begin{tabular}{cccc}
    \hline
        LA kernel & Cholesky & Iterative & Empirical \\
    \hline
        Y106 min & $2.270 \times 10^{-12}$ & $1.970 \times 10^{-8}$ & $6.935 \times 10^{-11}$ \\
        Y106 max & $4.487 \times 10^{-3}$ & $3.446 \times 10^{-3}$ & $2.491 \times 10^{-3}$ \\
        J129 min & $2.753 \times 10^{-12}$ & $1.077 \times 10^{-8}$ & $5.059 \times 10^{-11}$ \\
        J129 max & $3.683 \times 10^{-3}$ & $3.273 \times 10^{-3}$ & $2.024 \times 10^{-3}$ \\
        H158 min & $4.098 \times 10^{-12}$ & $2.088 \times 10^{-8}$ & $5.835 \times 10^{-11}$ \\
        H158 max & $4.532 \times 10^{-3}$ & $4.483 \times 10^{-3}$ & $2.115 \times 10^{-3}$ \\
        F184 min & $4.105 \times 10^{-11}$ & $2.471 \times 10^{-7}$ & $5.885 \times 10^{-11}$ \\
        F184 max & $3.503 \times 10^{0}$ & $7.505 \times 10^{-3}$ & $2.376 \times 10^{-3}$ \\
    \hline
    \end{tabular}
\end{table}

Fig.~\ref{fig:whitenoise1_2d} compares the three linear algebra kernels in terms of 2D white noise power spectra; minimum and maximum values are tabulated in Table~\ref{tab:extrama_white}. The left column shows results given by the Cholesky kernel, which is supposed to basically replicate what we had in \papone\ and \paptwo.
Such expectation is largely met in Y106, J129, H158 bands, in terms of the central bright regions and large $+$ signs --- the former are basically the square of the quotient of output and input module translation functions (MTFs; magnitude of Fourier transform of PSFs), while the later are artifacts, partially caused by selection of input pixels (see below for further discussion in the context of the other two kernels).
Because of different forms of target output PSFs -- obscured Airy disk convolved with Gaussian in \papone\ and simple Gaussian in this work --- the dynamic range of the power is significantly extended, causing the large $+$ signs to be more eminent.

\begin{figure}
    \centering
    \includegraphics[width=\columnwidth]{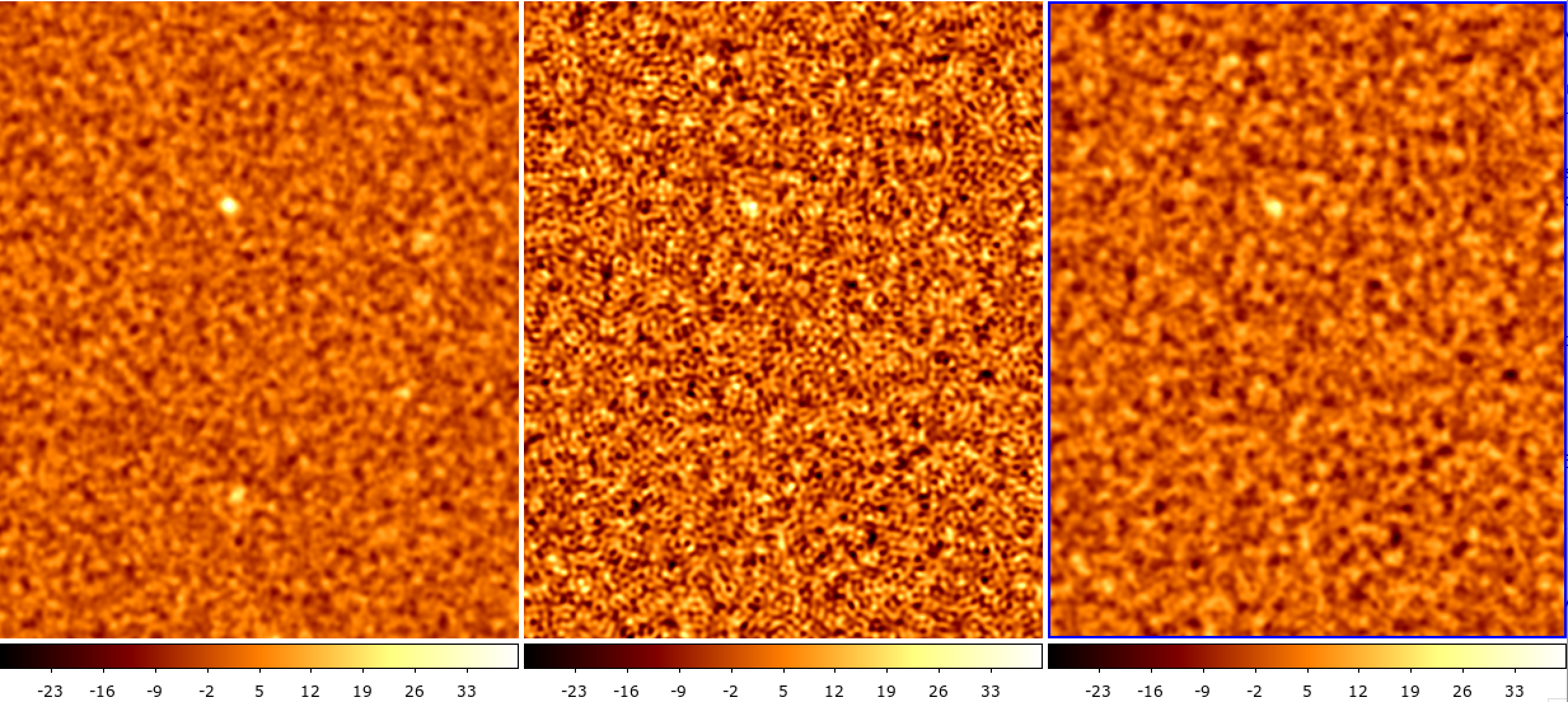}
    \caption{\label{fig:whitenoise1_F184}The screenshot above is the outputs from the Cholesky kernel, center of block $(0,0)$: Left: H158; Middle: F184; Right: F184, smoothed with Gaussian of sigma = 2 output pixels. The right panel can be thought of as having the output PSF widened in postprocessing. See text for further explanation.}
\end{figure}

The Cholesky kernel results in F184 band are evidently worth more attention: the maximum power is a few orders of magnitude larger than those in the other bands, and significant horizontal and vertical fringes occur in outer regions.
This is because the target output PSF chosen for F184 is not sufficiently wider than the input PSFs --- although it has the same FWHM as the one used in \papone, its different form leads to non-zero MTF values as input MTFs almost vanish. If it was sufficiently wide (in real space), its MTF would be narrower (in Fourier space), and thus this division by ``zero'' issue would be mitigated.
In Fig.~\ref{fig:whitenoise1_F184}, we smooth a sample of the coadded white noise frame in F184 and compare the smoothed version with its H158 counterpart. There is excess noise at very high wavenumber with this kernel, but that phenomenon can be suppressed with some output smoothing and the lower wavenumbers are unaffected. Our power spectra make this explicit. While the output smoothing is successful in suppressing the high spatial frequency noise pattern, in the future we plan to directly use a wider output PSF in F184 if the Gaussian option is used (since we prefer to suppress this noise in {\sc Imcom} rather than in postprocessing).
Since other aspects of the Cholesky kernel results in F184 are still valid (see Section~\ref{ss:diagnostics} and Section~\ref{sec:gsstar14}), we choose not to rerun the simulations with a more reasonable target output PSF, but warn {\sc Imcom} users to be careful regarding the choice of target PSFs. A systematic investigation of such choice will be presented in \papfour.

Comparing results given by the three LA kernels, it is clear that the shape of 2D power spectra mirrors the shape of the window function for input pixels. For the Cholesky kernel, it is a weighted average (over relative positions of output pixels) of rounded squares (see the lower panel of Fig.~\ref{fig:inpix}).
For the iterative and empirical kernels, we see ring structures due to uniform circular windows with different radii: relatively larger acceptance radius $\rho_{\rm acc}$ (in real space) for the former leads to finer rings (in Fourier space), while smaller $\rho_{\rm acc}$ for the latter leads to thicker rings.
A more important difference between these two kernels is that, the empirical kernel assigns coaddition weights in exactly the same way in different bands, while the iterative kernel solves linear systems and provides answers close to what one would expected from input and output PSFs.
The main drawback of the iterative kernel, though, is that the noise power spectra it yields contain a significant ``white'' component --- this is again due to the finite tolerance, and has nothing to do with the input, hence we may refer to it as the ``output white noise.'' How this output noise scales with the tolerance will be explored in \papfour.
In addition, the iterative kernel results display some noticeable patterns in Y106 and H158 bands. These are likely due to some ``resonances'' between the two roll angles of the survey in each band, instead of {\sc Imcom} artifacts. We will further investigate these patterns if needed in our future work.

Interestingly --- contradicting our speculation in \paptwo\ Section~3.2 --- postage stamp boundaries effects are not solely responsible for the large $+$ signs: neither the iterative kernel nor the empirical kernel is subject to such boundaries effects by definition, yet the large $+$ signs are still omnipresent.
If we only had results from the Cholesky and iterative kernels, one would ascribe them to our input PSF sampling grid, which is a square grid of $2\times 2$ groups of postage stamps (see \NewEdit{Appendix~\ref{app:pyimcom}}). % Section~\ref{ss:pyimcom}
However, the empirical kernel does not involve input PSF sampling at all. Therefore, our current speculation is that the large $+$ signs are due to the incorrect assumption about periodicity in Eq.~(\ref{eq:P2D}).\footnote{Discrete Fourier transform (DFT) itself does not assume periodicity, but squaring DFT results does. Specifically, $P_{\rm 2D}(u, v)$ in Eq.~(\ref{eq:P2D}) is invariant under the ``rotation'' $j_x \to (j_x + \Delta j_x) \operatorname{mod} N$, $j_y \to (j_y + \Delta j_y) \operatorname{mod} N$, where $\Delta j_x$ and $\Delta j_y$ are arbitrary integers. \NewEdit{\citet{Laliotis2024PASP}} break this symmetry using apodization \citep[see, e.g.,][]{1978IEEEP..66...51H} so that large $+$ signs do not appear in 2D noise power spectra.} % Laliotis et al. (in prep)

\begin{figure*}
    \centering
    \includegraphics[width=0.9\textwidth]{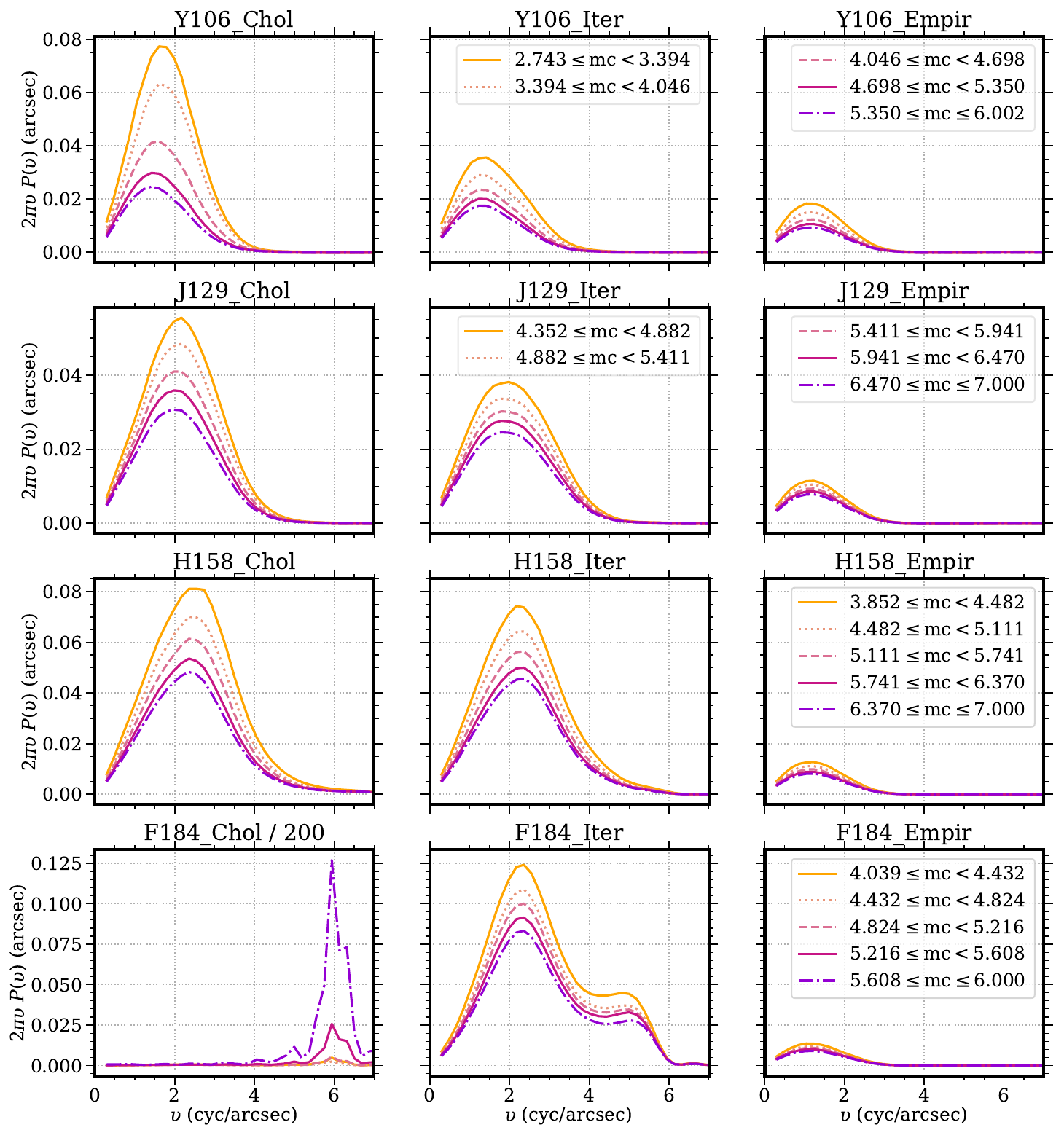}
    \caption{\label{fig:whitenoise1_1d}Azimuthally averaged power spectra of simulated white noise frames, averaged over modes within each of the $150$ radial bins and blocks in each mean coverage (``mc'' in short) bin for each band-kernel combination.
    The arrangement is the same as Fig.~\ref{fig:whitenoise1_2d}. Curves given by the Cholesky kernel in F184 band are divided by a factor of $200$.}
\end{figure*}

Fig.~\ref{fig:whitenoise1_1d} compares the three LA kernels in terms of the azimuthally averaged version of white noise power spectra.\footnote{The analytical expectation for 1D noise power spectra (see \paptwo\ Appendix A for derivation) depends on the choice of target output PSF and is not included in this work.}
Comparing the left and middle columns, it is clear that the iterative kernel outperforms the Cholesky kernel, especially in Y106 and J126 bands and at limited mean coverage; the contrast becomes less significant in H158, possibly because of the ``output white noise.'' Therefore, an upgraded version of the iterative kernel has the potential of reducing the needed number of exposures while maintaining the signal-to-noise ratio.
In F184, the division by ``zero'' features in the Cholesky kernel results flip the relationship between power and mean coverage; a reasonable explanation would be: the larger the coverage, the closer {\sc Imcom} gets to the (poorly chosen) target output PSF, and the larger the quotient becomes.
These features show up as second bump in the iterative kernel results, indicating that the iterative kernel may be more stable against defects in linear systems.
The Empirical kernel results are only sensitive to the mean coverage but not directly to the band, as expected. However, as mentioned in Section~\ref{ss:outmaps}, these curves need to be rescaled if $T_{{\rm tot}, \alpha} = 1$ turns out to be an inappropriate normalization for coaddition weights.

\subsection{Simulated $1/f$ noise} \label{ss:1fnoise2}

\begin{figure*}
    \centering
    \includegraphics[width=0.95\textwidth]{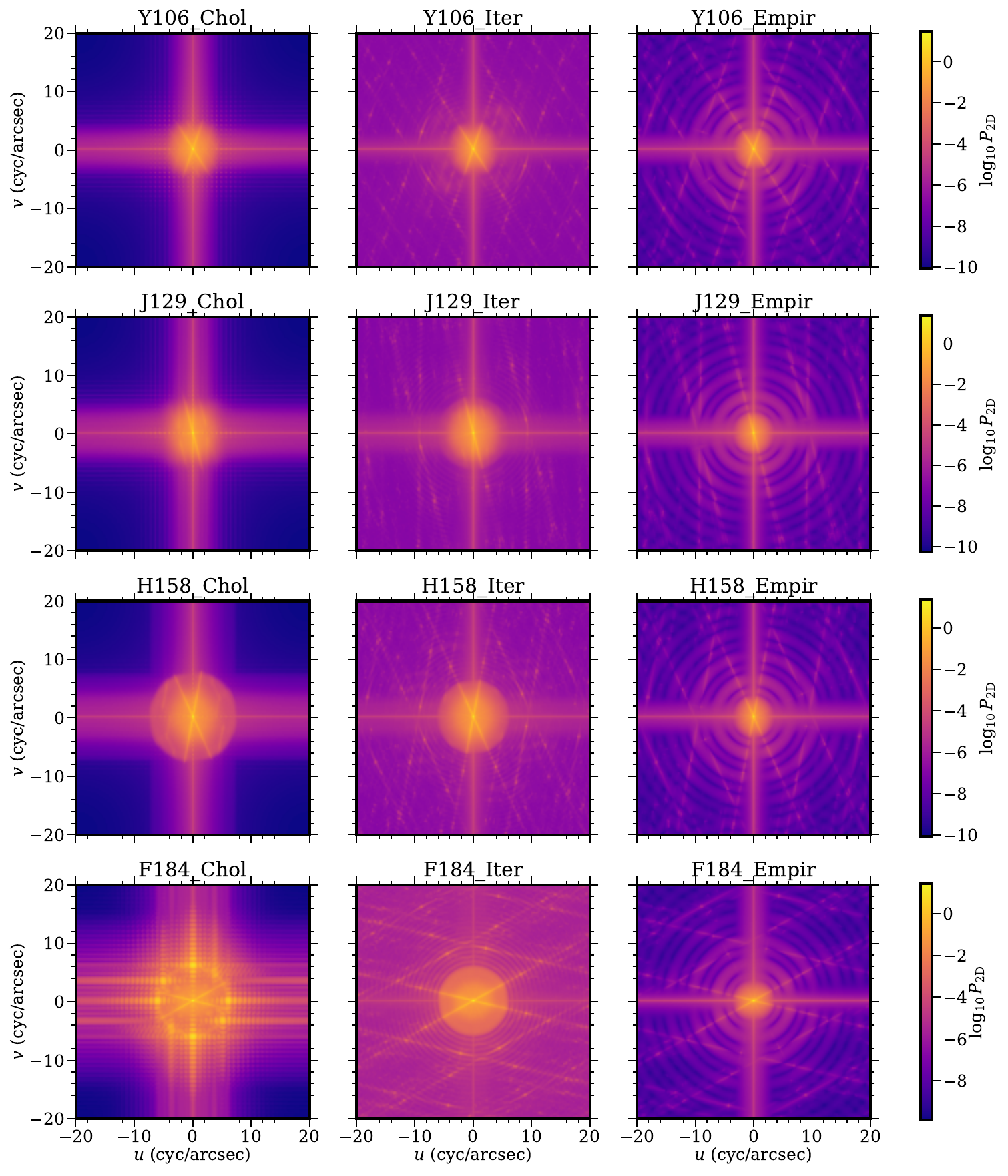}
    \caption{\label{fig:1fnoise2_2d}Same as Fig.~\ref{fig:whitenoise1_2d}, but for simulated $1/f$ noise frames ({\tt \textquotesingle 1fnoise2\textquotesingle}).}
\end{figure*}

\begin{table}[]
    \centering
    \caption{\label{tab:extrama_1f}Minimum and maximum values in 2D $1/f$ noise power spectra shown in Fig.~\ref{fig:1fnoise2_2d}.} % \NewEdit{\st{Caption}}
    \begin{tabular}{cccc}
    \hline
        LA kernel & Cholesky & Iterative & Empirical \\
    \hline
        Y106 min & $8.811 \times 10^{-11}$ & $1.600 \times 10^{-7}$ & $5.842 \times 10^{-10}$ \\
        Y106 max & $2.647 \times 10^{1}$ & $2.926 \times 10^{1}$ & $2.201 \times 10^{1}$ \\
        J129 min & $5.195 \times 10^{-11}$ & $6.921 \times 10^{-8}$ & $3.487 \times 10^{-10}$ \\
        J129 max & $2.266 \times 10^{1}$ & $2.538 \times 10^{1}$ & $1.898 \times 10^{1}$ \\
        H158 min & $8.928 \times 10^{-11}$ & $1.232 \times 10^{-7}$ & $3.989 \times 10^{-10}$ \\
        H158 max & $2.168 \times 10^{1}$ & $2.431 \times 10^{1}$ & $1.780 \times 10^{1}$ \\
        F184 min & $1.267 \times 10^{-10}$ & $1.482 \times 10^{-6}$ & $3.579 \times 10^{-10}$ \\
        F184 max & $2.449 \times 10^{1}$ & $2.817 \times 10^{1}$ & $1.974 \times 10^{1}$ \\
    \hline
    \end{tabular}
\end{table}

Next, we switch gears from uncorrelated noise to correlated noise. Fig.~\ref{fig:1fnoise2_2d} is similar to Fig.~\ref{fig:whitenoise1_2d}, showing 2D $1/f$ noise power spectra; minimum and maximum values are tabulated in Table~\ref{tab:extrama_1f}.
We can see many features we have seen before: central bright regions \NewEdit{as} quotients of output and input MTFs \NewEdit{squared}, X shapes due to the two roll angles in each band (see \paptwo\ Section~2.2), $+$ signs \NewEdit{owing to} artifacts caused by {\sc Imcom} {\slshape per se} and analysis method used here (see Section~\ref{ss:whitenoise1} for discussion), etc. % at ... caused by
Nonetheless, there are also some features not seen in \paptwo\ or Section~\ref{ss:whitenoise1}. For the Cholesky kernel, alternating bright and dark fringes now appear in both directions, most clearly in Y106 and J129 bands. Such fringes do not appear in results given by other kernels, and the spacing between them are still $0.8 \,{\rm cyc} \,{\rm arcsec}^{-1}$, hence we still believe they are artifacts caused by postage stamp boundaries.
For both the iterative and empirical kernels, the X shapes are extended and reoccurring at large wavenumbers. A closer look at the Cholesky kernel results, especially those in H158 band, would reveal similar reoccurring X shapes in central bright regions. Therefore, we tentatively conclude that the extended X shapes are again artifacts caused by the incorrect assumption about periodicity in Eq.~(\ref{eq:P2D}); they are not seen in the Cholesky kernel results, \NewEdit{as they are} averaged out due to non-uniform window functions for input pixels. % because

\begin{figure*}
    \centering
    \includegraphics[width=0.9\textwidth]{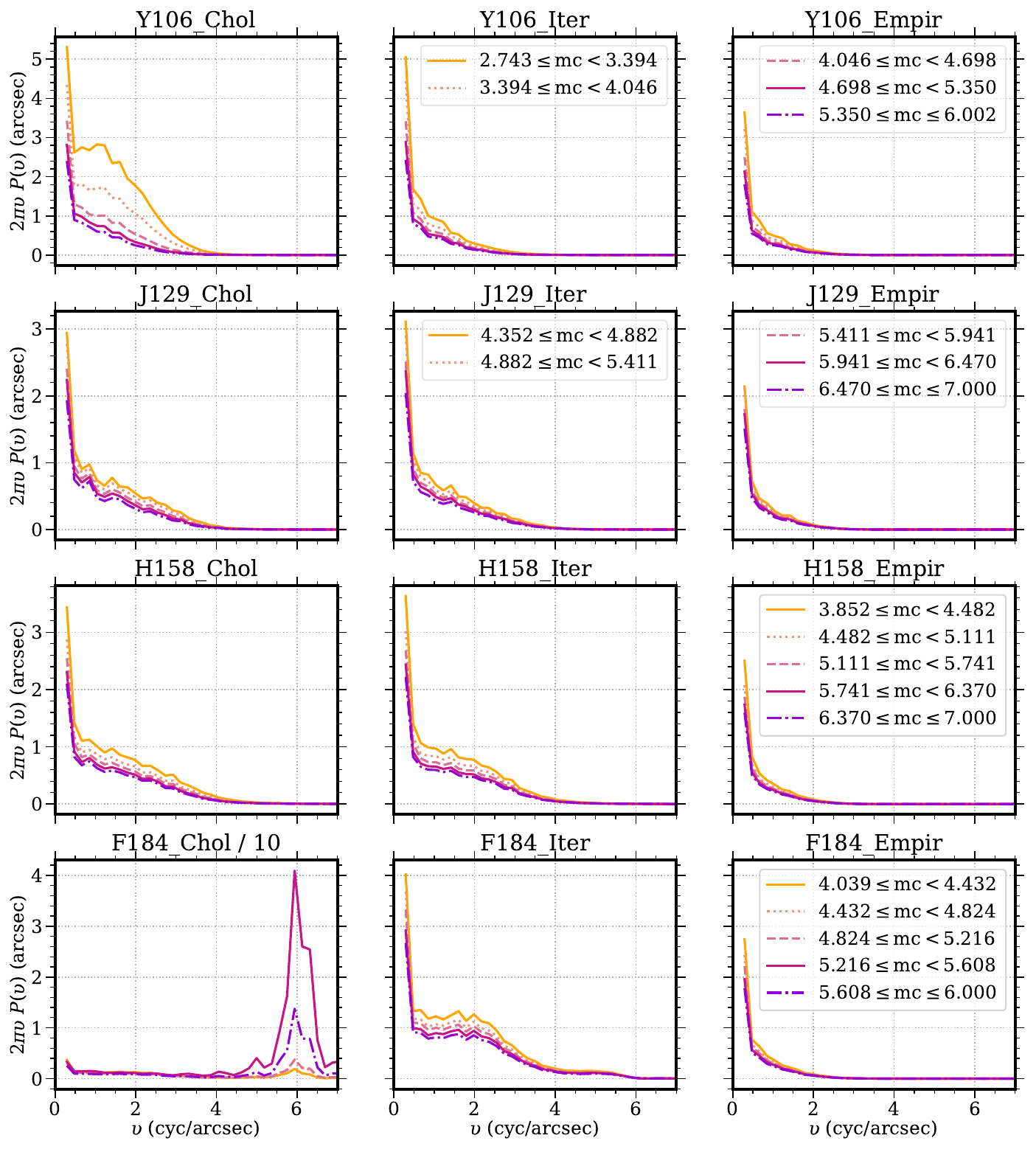}
    \caption{\label{fig:1fnoise2_1d}Same as Fig.~\ref{fig:whitenoise1_1d}, but for simulated $1/f$ noise frames. Curves given by the Cholesky kernel in F184 band are divided by a factor of $10$.}
\end{figure*}

Fig.~\ref{fig:1fnoise2_1d} is similar to Fig.~\ref{fig:whitenoise1_1d}, showing 2D $1/f$ noise power spectra. Since the extended and reoccurring X shapes are also averaged out in curves given by the iterative and empirical kernels, it seems reasonable not to worry too much about them.
The advantage of the iterative kernel over the Cholesky kernel is only considerable in Y106 band and at small mean coverage, i.e., not in J129 and H158 bands or at decent mean coverage. Interestingly, the peaks in the Cholesky kernel results in F184 band are $\sim 20$ times less significant than in white noise power spectra, and the relationship between power and mean coverage does not seem monotonic. Both could be explained by the different nature of $1/f$ noise; we refer interested readers to \paptwo\ Appendix~A for the expected behaviors of these two types of simulated noise.

To conclude this section, we comment that better noise control would allow us to shorten the exposure time and survey a larger area of the sky within a given time.

\section{Measuring point sources} \label{sec:gsstar14}

In this section, we perform shape measurements on coadded stars. Specifically, we measure injected point sources drawn by {\sc GalSim} (i.e., those in {\tt \textquotesingle gsstar14\textquotesingle} layer) instead of more realistic stars (i.e., those in {\tt \textquotesingle SCI\textquotesingle} or {\tt \textquotesingle truth\textquotesingle} images), as the former have completely known properties and are better for diagnostic purposes.
Ultimately, shape measurements will be performed on galaxies (extended sources); as mentioned in \paptwo\ Section~1, stars constitute a ``stress test,'' since they are narrower in real space and wider in Fourier space, maximizing the impact of undersampling.

\begin{figure*}
    \centering
    \includegraphics[width=0.9\textwidth]{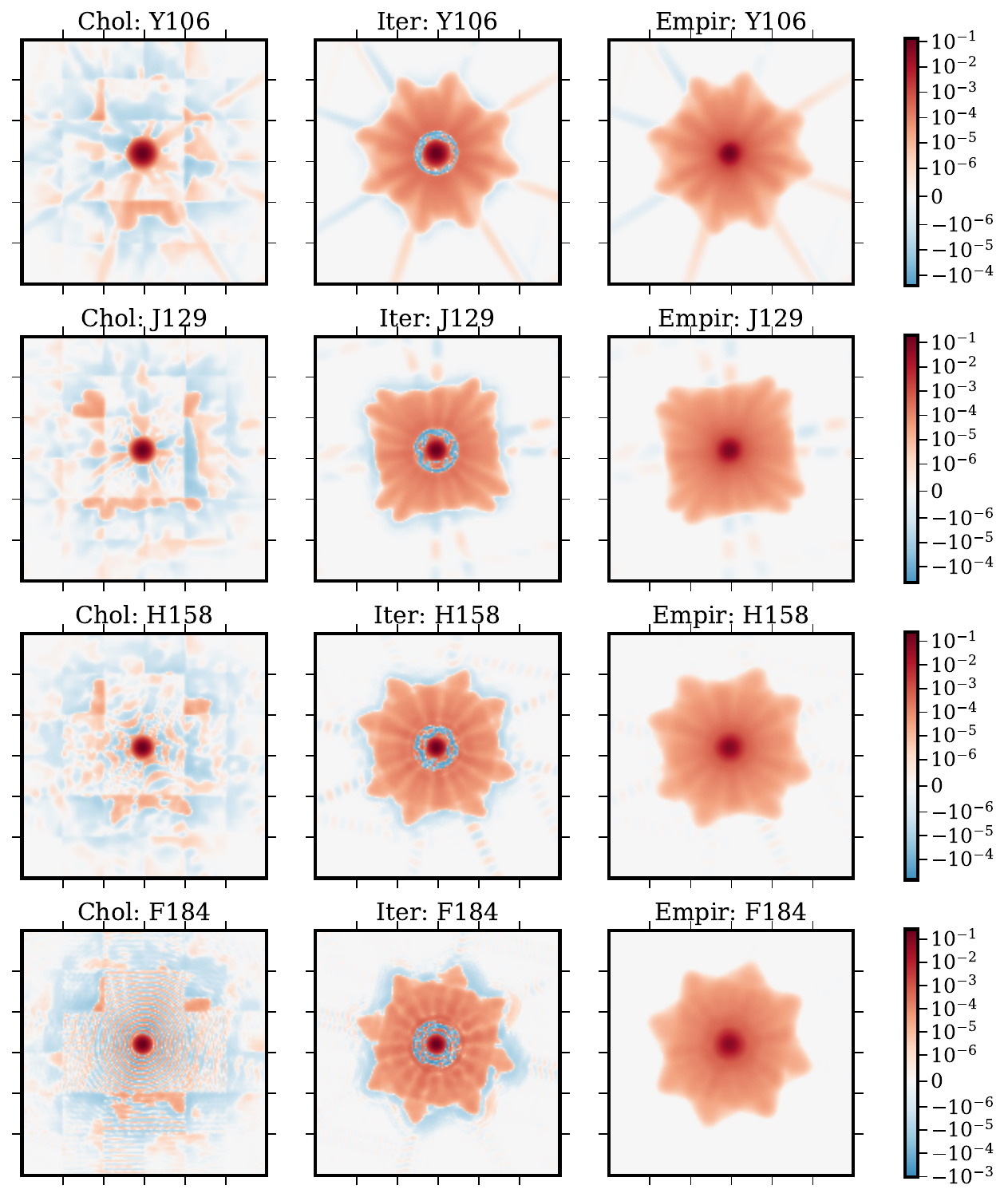}
    \caption{\label{fig:injected_star}An example injected {\sc GalSim} star (upper one in the second row of Fig.~\ref{fig:example_images}) coadded by the three linear algebra kernels (from {\it left} to {\it right}: Cholesky, iterative, and empirical) in the four bands (from {\it top} to {\it bottom}: Y106, J129, H158, and F184). Each panel is a cutout of $7.5 \,{\rm arcsec}$ ($300$ output pixels) on a side. Note that: i) \NewEdit{the} star is not located at the center of this cutout; ii) \NewEdit{measurements} in Section~\ref{sec:gsstar14} are performed on considerably smaller cutouts of $1.975 \,{\rm arcsec}$ ($79$ output pixels) on a side and centered at pixels closest to expected centroids. The symmetrical logarithmic scale ({\tt matplotlib.colors.SymLogNorm}) is used to better display the outer regions.} % The ... Measurements
\end{figure*}

Fig.~\ref{fig:injected_star} shows an example injected star coadded by all three LA kernels studied in this work in all four bands. These plots have a different scaling than Fig.~\ref{fig:example_images}, in order to better display outer regions of the coadded images.
From the empirical kernel column, we can clearly see how the injected stars are drawn: square input PSF patches of size $32 \times 32$ native \NewEdit{pixels} are ``injected'' onto the {\tt \textquotesingle gsstar14\textquotesingle} input layer. However, due to the inaccuracy of this kernel, the resulting coadds are not well-suited for measurements. % pixel
The iterative kernel column is similar to the empirical kernel column, in terms of clearly seen square patches. Although it is supposed to produce higher-quality results, we see significant ``output white noise'' acting on negative annuli surrounding the star and thus undermining shape measurements.
The story of the Cholesky kernel column is two-fold: on the one hand, thanks to its much larger windows for input pixels, it is able to partially remove the diffraction spikes; on the other hand, due to the irregularity and non-uniformity of its windows, it also distorts the diffraction spikes and is subject to significant postage stamp boundary effects.
In the F184 band, alternating negative and positive rings, probably due to insufficient target output PSF width (see Section~\ref{ss:whitenoise1} for discussion), can be clearly seen in the Cholesky kernel image.

Outside of the inner regions (roughly corresponding to the input PSF patches), some ``large diffraction spikes'' can be clearly seen in cutout produced by the iterative and empirical kernels. In the images yielded by the iterative kernel, we can even see such spikes from adjacent injected stars in the grid (not shown here, but some are shown in Fig.~\ref{fig:example_images}).
From Fig.~\ref{fig:injected_star} alone, it may not be clear why these extended ``diffraction spikes'' only appear in several directions, and are not always aligned with those in the input PSF patches. However, if we put these cutouts and Fig.~\ref{fig:1fnoise2_2d} side by side, it becomes obvious that the extended spikes coincide with the X shapes in the 2D power spectra of correlated noise. Therefore, we think such ``large diffraction spikes'' are also related to the two roll angles of the survey in each band, and leave investigation of the specific mechanism to future work.
By comparing these ``diffraction spikes'' in different bands, we notice that they are usually oscillating, and the ``wavelength'' of such oscillation is negatively correlated with the wavelength of the band --- therefore in Y106, the bluest band, cutouts shown here are not large enough to display a complete wavelength. Since shapes of these patterns are the same in the iterative kernel and empirical kernel results, we get the clue that they are mainly determined by the input PSFs rather than the target output PSF. Meanwhile, it is unclear why they become dimmer in redder bands with the empirical kernel, but basically remain the same amplitude with the iterative kernel.
Before concluding this discussion on artificial ``diffraction spikes,'' it is worth noting that work is underway on the removal of physical diffraction spikes of very bright stars (Macbeth et al. in prep), which would interfere with measurements of other objects.

With the above observations in mind, if we take another look at the Cholesky kernel results, we see broken versions of both square patches and ``large diffraction spikes,'' mirroring commonality among different linear algebra strategies and brokenness caused by non-uniform window functions for input pixels. Another feature shared between the Cholesky and iterative kernels is that there is always a negative region surrounding the input PSF patches; we think this limitation comes from the finiteness of PSFs in image simulations, rather than the coaddition process.

This work focuses on $1$-point statistics of star shapes for two reasons: first, the quality of $2$-point statistics are based on that of $1$-point measurements; second, we have only coadded a footprint of $0.071 \,{\rm deg}^2 \sim 0.253$ {\slshape Roman} fields of view, which is not sufficient to produce meaningful $2$-point statistics.
In Section~\ref{ss:1pt-stats}, we investigate five properties of injected stars, all of which are defined based on moments and measured using the {\sc HSM} module \citep{2003MNRAS.343..459H, 2005MNRAS.361.1287M} of {\sc GalSim}. To perform such measurements, a $1.975 \times 1.975 \,{\rm arcsec}^2$ ($79 \times 79$ output pixels) cutout is made for each of the $5517$ injected point sources in each band.
In Section~\ref{ss:heatmaps}, we go one step further and study correlations between {\sc Imcom} diagnostics (see Sections~\ref{ss:outmaps} and \ref{ss:diagnostics}) and errors in these stellar properties.

\subsection{$1$-point statistics} \label{ss:1pt-stats}

\begin{figure}
    \centering
    \includegraphics[width=\columnwidth]{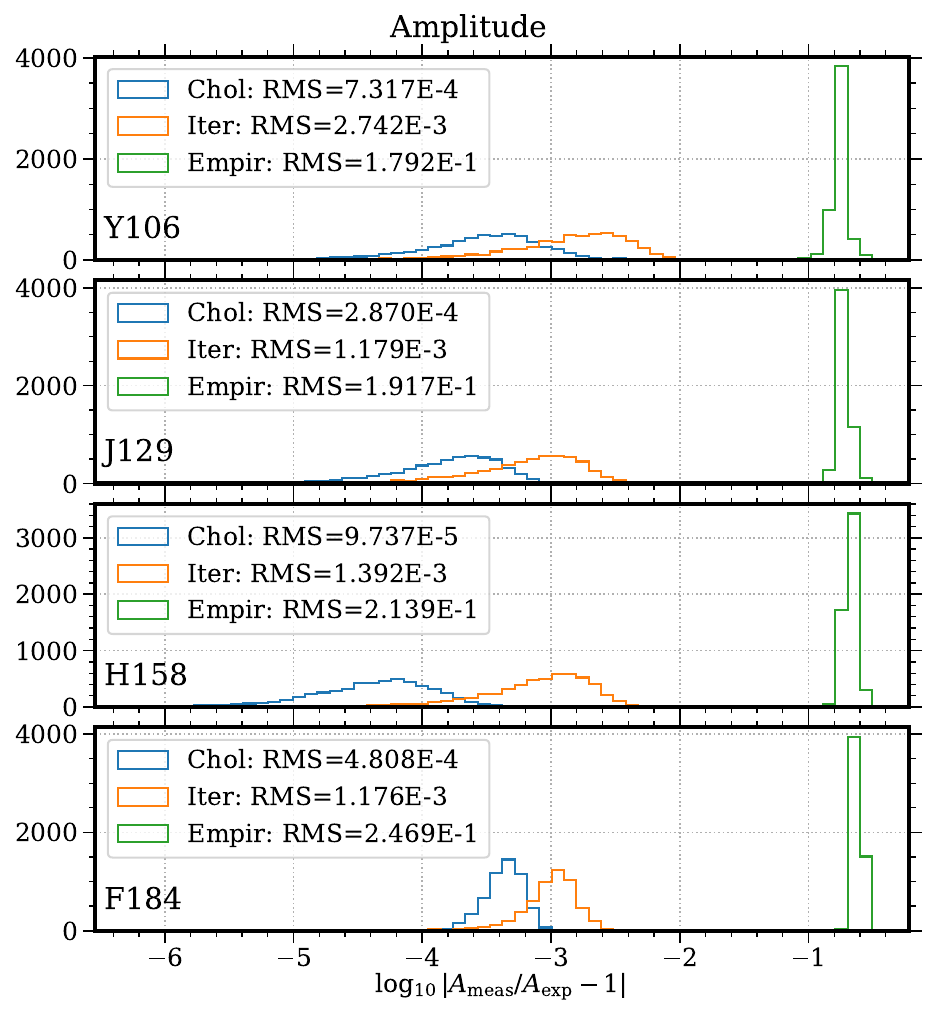}
    \caption{\label{fig:star_amplitude}Histograms of logarithmic absolute amplitude errors of $5517$ injected stars coadded by three linear algebra kernels in four bands.
    Logarithmic absolute amplitude error is computed as $\log_{10} |A_{\rm meas}/A_{\rm exp}-1|$, where $A_{\rm meas}$ and $A_{\rm exp}$ are measured and expected amplitudes of an injected star, respectively.
    Root-mean-square (RMS) errors are annotated in the legends; otherwise layout and format of the histograms are the same as in Fig.~\ref{fig:outmap_fidelity}.}
\end{figure}

Amplitude is a simple measure of the total flux of a star; in terms of moments, it can be comprehended as the $0^{\rm th}$ moment.
Fig.~\ref{fig:star_amplitude} presents the distribution of logarithmic absolute amplitude errors ($\log_{10} |A_{\rm meas}/A_{\rm exp}-1|$)\NewEdit{\footnote{Throughout this section, we use logarithm to better compare the three linear algebra kernels. Modulo the $A_{\rm meas}$ values given by the empirical kernel, we do not think there are significant biases regarding the signs of discrepancies.}} given by each band-kernel combination studied in this work. Note that unlike those in Section~\ref{ss:diagnostics}, histograms in this section always have ``better'' values shown on the left.
In \paptwo\ Fig.~12, amplitudes of {\sc Imcom} coadded stars were consistently biased in all bands; specifically, most of the {\tt \textquotesingle SCI\textquotesingle} stars had $A_{\rm meas} < A_{\rm exp}$. That was because although the target output PSFs were Airy disks convolved with Gaussians, {\sc HSM} tried to fit a 2D Gaussian to each cutout, and thus fluxes in the rings were largely ignored.
In simulations for this work, target output PSFs are chosen to be simple 2D Gaussians, hence such consistent bias is not observed in neither the Cholesky kernel nor the iterative kernel results, showcasing an important advantage of Gaussian output PSFs.
In general, the Cholesky kernel outperforms the iterative kernel by half to one order of magnitude. For both kernels, the spread is smaller in F184, probably because the alternating negative and positive rings surrounding each star (see the last row of Fig.~\ref{fig:injected_star}, especially the left panel) consistently bias the Gaussian fit, indicating again that the target output PSF needs to be sufficiently wide.

All the stars coadded by the empirical kernel also have $A_{\rm meas} < A_{\rm exp}$, but the cause is completely different from that in \paptwo: the problem has nothing to do with the target output PSF form (recall that this kernel does not know about the target PSFs), and can be ascribed to the normalization of coaddition weights.
Since the median of measured amplitudes are $15.93$, $15.73$, $15.24$, and $14.66$ in \NewEdit{Y106, J129, H158, and F184} bands, respectively, and the expected amplitude is always $(s_{\rm in} / \Delta\theta)^2 = (0.11 / 0.025)^2 = 19.36$ (since the total flux is normalized in the input layer), it is advisable to rescale the empirical kernel results by $1.216$, $1.231$, $1.270$, and $1.321$, respectively. If we do so, the noise amplification metric (see Section~\ref{ss:diagnostics}) and all the noise power spectra (see Section~\ref{sec:noiseps}) should be multiplied by those factors squared. % the four
However, we choose not to do so, as these numbers are not derived from first principles, and it is impossible to ``correct'' fidelity values after {\sc Imcom} runs in an easy way. Note that both effective coverage and shape measurements are unaffected by a global scaling of coaddition weights. Therefore, we leave the derivation of such factors for future work if needed.

\begin{figure}
    \centering
    \includegraphics[width=\columnwidth]{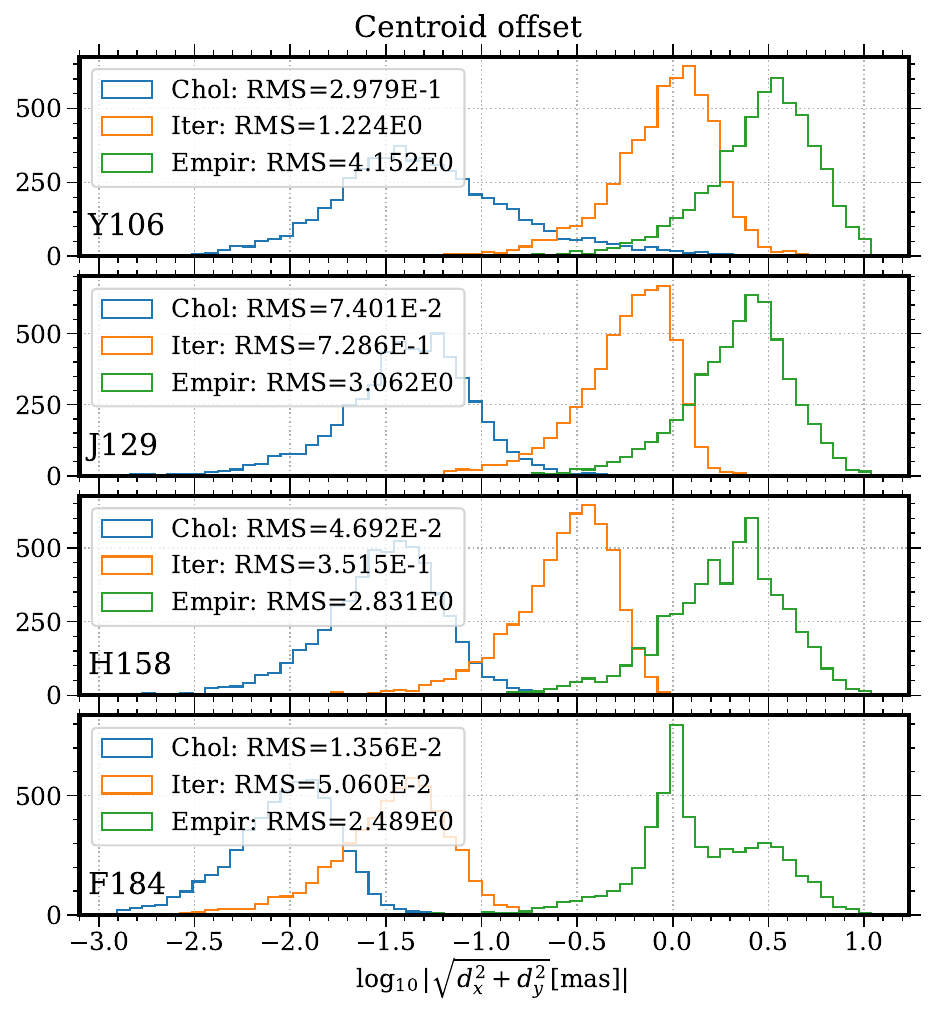}
    \caption{\label{fig:star_offset}Histograms of logarithmic centroid offsets (in milliarcseconds) of $5517$ injected stars coadded by three linear algebra kernels in four bands.
    Logarithmic centroid offset is computed as $\log_{10} |\sqrt{d_x^2 + d_y^2}|$, where $d_x$ and $d_y$ are $x$ and $y$ components of the centroid offset (discrepancy between measured and expected centroids; in milliarcseconds) of an injected star, respectively.
    Layout and format of the histograms are the same as in Fig.~\ref{fig:star_amplitude}.}
\end{figure}

As indicated by its name, {\tt \textquotesingle gsstar14\textquotesingle} point sources are injected onto a {\sc HEALPix} grid with \NewEdit{${\tt NSIDE} = 14$}. \NewEdit{Thence}, we known exactly where they are supposed to be, and centroid offset is the discrepancy between the measured and expected centroids ($1^{\rm st}$ moments) of a star. % Thenceforce
Fig.~\ref{fig:star_offset} presents the distribution of logarithmic absolute centroid offsets ($\log_{10} |\sqrt{d_x^2 + d_y^2}|$). In all four bands, the Cholesky kernel provides the most accurate results, the empirical kernel provides the least accurate results, and the iterative kernel is somewhere in between.
Interestingly, results given by both the Cholesky and iterative kernels get better as the wavelength increases, and this trend is more significant for the latter. It appears that, for astrometry purposes, a narrower target output PSF leads to smaller centroid offsets, and the narrowness benefits the iterative kernel to a larger degree.

\begin{figure}
    \centering
    \includegraphics[width=\columnwidth]{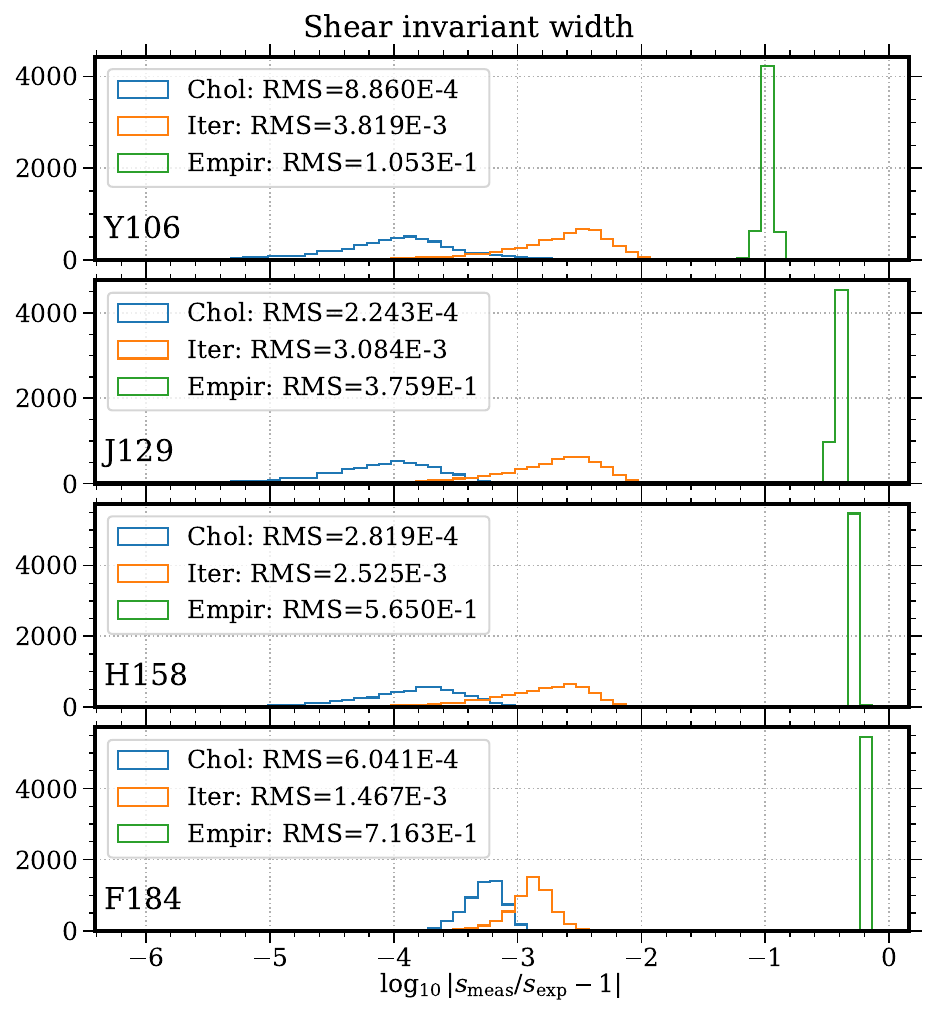}
    \caption{\label{fig:star_width}Histograms of logarithmic absolute size errors of $5517$ injected stars coadded by three linear algebra kernels in four bands.
    Logarithmic absolute size error is computed as $\log_{10} |s_{\rm meas}/s_{\rm exp}-1|$, where $s_{\rm meas}$ and $s_{\rm exp}$ are measured and expected shear-invariant widths of an injected star, respectively.
    Layout and format of the histograms are the same as in Fig.~\ref{fig:star_amplitude}.}
\end{figure}

Following \paptwo, we then examine the $2^{\rm nd}$ moments, which can be written as a $2\times 2$ symmetric matrix, or equivalently the combination of shear-invariant width (here we use the symbol $s$ to avoid confusion with standard deviation)
\begin{equation}
\label{eq:size_definition}
s = \sqrt[4]{M_{xx}M_{yy}-M_{xy}^2},
\end{equation}
and ellipticity components
\begin{equation}
\label{eq:shape_definition}
(g_1,g_2) = \frac{(M_{xx}-M_{yy}, 2M_{xy})}
{M_{xx} + M_{yy} + 2\sqrt{M_{xx}M_{yy} - M_{xy}^2}}.
\end{equation}

Fig.~\ref{fig:star_width} presents the distribution of logarithmic absolute size errors ($\log_{10} |s_{\rm meas}/s_{\rm exp}-1|$). For both the Cholesky and iterative kernels, the situation is very similar to that of logarithmic absolute amplitude errors, indicating that the shear-invariant width is also somehow based on a Gaussian fit, validating our choice of target output PSF (see Section~\ref{ss:config}). The iterative kernel results are worse than the Cholesky kernel results to a larger degree, as the negative annuli surrounding stars (see Fig.~\ref{fig:injected_star}) due to ``output white noise'' have a larger impact on $2^{\rm nd}$ moments than on $0^{\rm th}$ moments.
The Empirical kernel results are again orders of magnitude worse than the other two kernels; however, unlike amplitude errors, size errors are larger in redder bands. This is understandable: this kernel uses the same acceptance radius $\rho_{\rm acc}$ to assign coaddition weights according to Eq.~(\ref{eq:empir}), consequently the measured sizes ($s_{\rm meas}$) are set by input PSF widths, which increase at larger wavelength; meanwhile, the ``true'' sizes ($s_{\rm exp}$) are measured from target output PSFs, of which the widths decreases at larger wavelength following \papone.
In a sense, our definition of size error is unfair for the empirical kernel; nevertheless, predictable output PSFs are preferred for shape measurements, as better shown by ellipticity.

\begin{figure}
    \centering
    \includegraphics[width=\columnwidth]{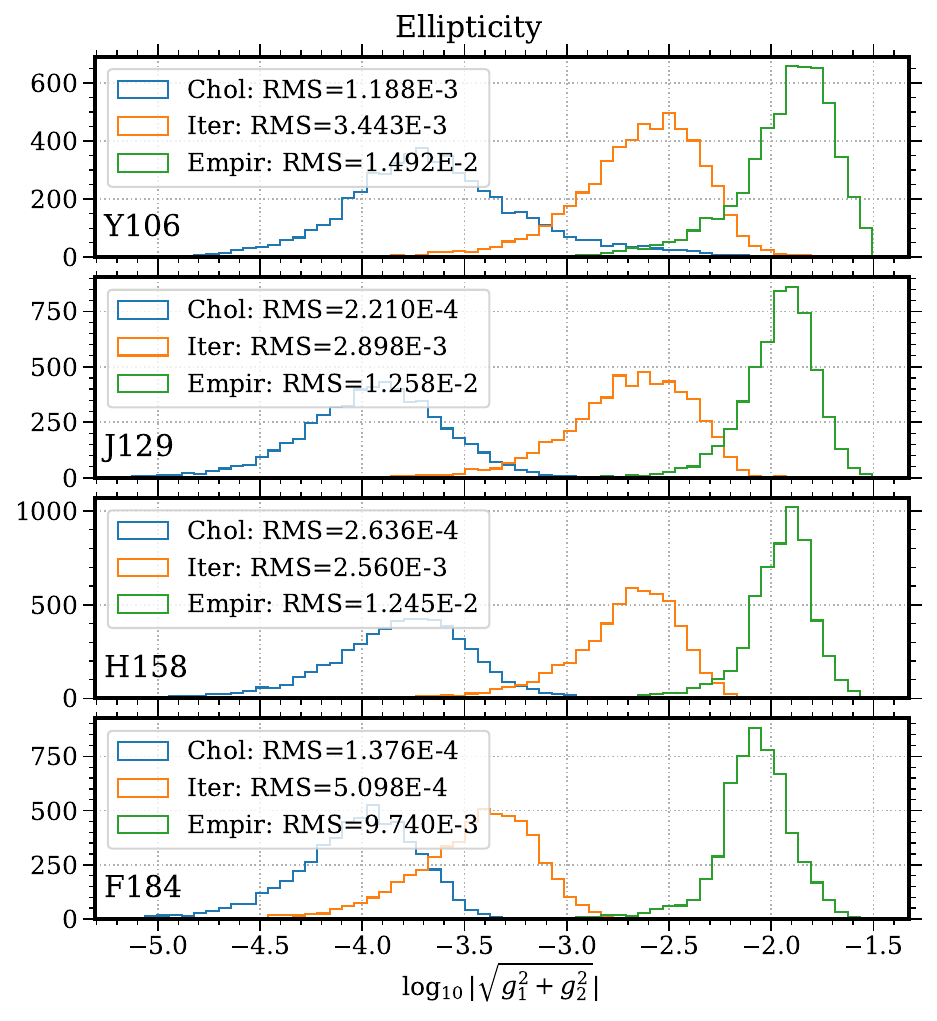}
    \caption{\label{fig:star_ellipticity}Histograms of logarithmic ellipticities of $5517$ injected stars coadded by three linear algebra kernels in four bands.
    Logarithmic ellipticity is computed as $\log_{10} \sqrt{g_1^2 + g_2^2}$, where $g_1$ and $g_2$ are the two components of measured ellipticity of an injected star; note that the quantity is expected to be zero for ideal, circular sources. Layout and format of the histograms are the same as in Fig.~\ref{fig:star_amplitude}.}
\end{figure}

Injected stars are ideal point sources, and our target output PSFs are perfectly circular, hence both components of ellipticity are expected to be zero. Fig.~\ref{fig:star_ellipticity} presents the distribution of logarithmic ellipticities ($\log_{10} \sqrt{g_1^2 + g_2^2}$).
For all the three linear algebra kernels studied in this work, the situation is similar to that of centroid offset. Note that centroid offset is measured in milliarcseconds, while ellipticity is dimensionless, hence the numerical values of the errors are not comparable.
The outperformance of the Cholesky kernel over the iterative kernel again makes sense, as both $1^{\rm st}$ and $2^{\rm nd}$ moments focus on the central regions of star cutouts; the Cholesky kernel results are mainly biased near postage stamp boundaries, while the iterative kernel results are subject to ``output white noise'' everywhere.

\begin{figure}
    \centering
    \includegraphics[width=\columnwidth]{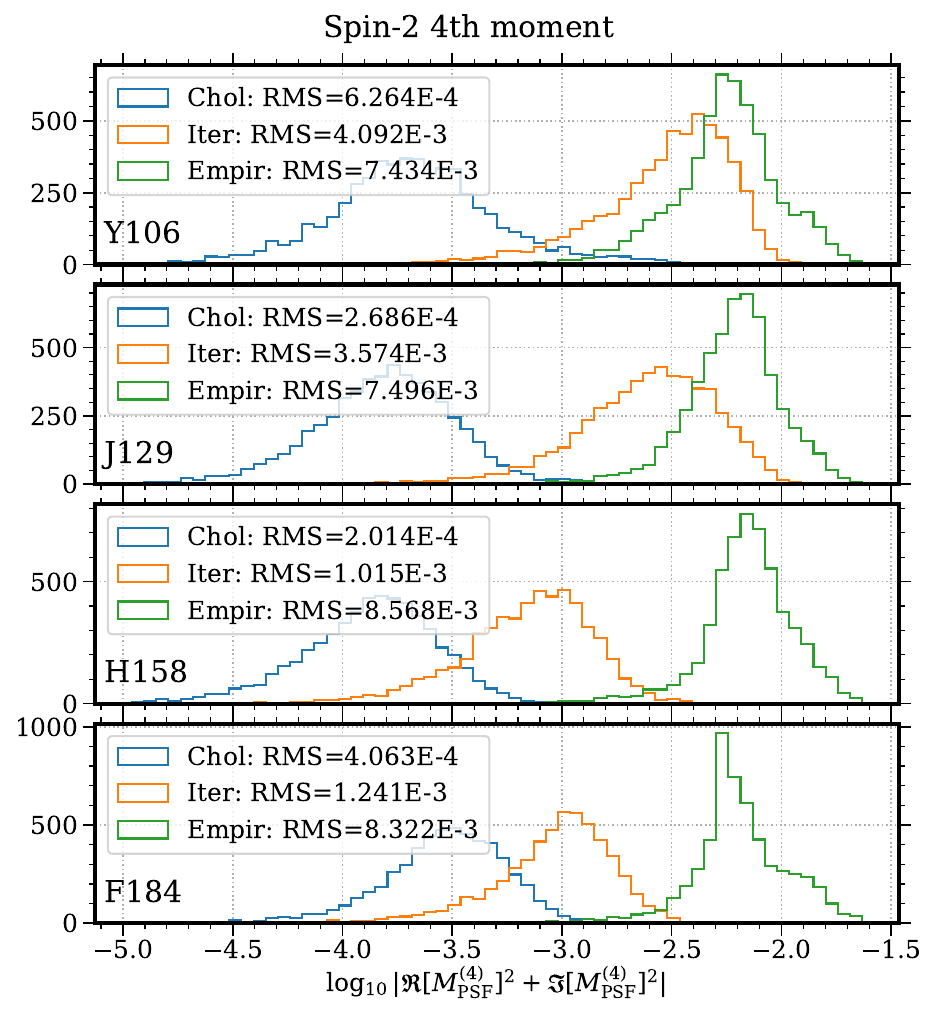}
    \caption{\label{fig:star_4th_moment}Histograms of logarithmic spin-$2$ $4^{\rm th}$ moments of $5517$ injected stars coadded by three linear algebra kernels in four bands.
    Logarithmic spin-$2$ $4^{\rm th}$ moment is computed as $\log_{10} (\Re[M^{\rm (4)}_{\rm PSF}]^2 + \Im[M^{\rm (4)}_{\rm PSF}]^2)$, where $\Re[M^{\rm (4)}_{\rm PSF}]$ and $\Im[M^{\rm (4)}_{\rm PSF}]$ are real and imaginary components of measured spin-$2$ $4^{\rm th}$ moment of an injected star, respectively; note that the quantity is expected to be zero for ideal, circular sources.
    Layout and format of the histograms are the same as in Fig.~\ref{fig:star_amplitude}.}
\end{figure}

What if we consider higher moments of the stars, which are more concerned about outer regions of the cutouts? Following \cite{2023MNRAS.520.2328Z}, we define standardized higher moments as
\begin{equation}
\label{eq:moment_define}
M_{pq} = \frac{\int {\rm d}x \, {\rm d}y \, u^p \, v^q \, \omega(x,y)
\, I(x,y)}{\int {\rm d}x \, {\rm d}y \, \omega(x,y) \, I(x,y) },
\end{equation}
where the transformed coordinates $(u,v)$ are chosen for each star cutout, such that the second moment shapes in Eq.~(\ref{eq:shape_definition}) vanish, and the second moment size in Eq.~(\ref{eq:size_definition}) is normalized to $1$; $p$ and $q$ are integer indices, $\omega(x,y)$ is the adaptive weight function, and $I(x,y)$ is the image.
Specifically, the complex spin-2 $4^{\rm th}$ moment is defined as
\begin{equation}
\label{eq:spin-2-fourth-moment}
    M^{\rm (4)}_{\rm PSF} = M_{40} - M_{04} + 2i (M_{31}+M_{13}).
\end{equation}
\citet{2023MNRAS.525.2441Z} demonstrated that this quantity contributes substantially to additive cosmic shear systematics in two-point correlation function; for ideal, circular sources like our injected stars, it is also expected to be zero. Fig.~\ref{fig:star_4th_moment} presents the distribution of logarithmic spin-$2$ $4^{\rm th}$ moments ($\log_{10} (\Re[M^{\rm (4)}_{\rm PSF}]^2 + \Im[M^{\rm (4)}_{\rm PSF}]^2)$).
The situation is still similar to that of centroid offset or ellipticity, except in F184 band, where the Cholesky kernel results are still better than the iterative kernel results, but break the trend in wavelength. We tentatively conclude that the Cholesky kernel yields desirable $4^{\rm th}$ moments as long as the target output PSF is sufficiently wide, and the potential of the iterative kernel can only be revealed if we have enough control over its ``output'' white noise.
As for the empirical kernel, spin-2 $4^{\rm th}$ moment results basically have the same quality as ellipticity results, possibly because $M^{\rm (4)}_{\rm PSF}$ and $(g_1,g_2)$ are both spin-2 quantities and are equally affected by diffraction spikes.

\subsection{What causes measurement errors?} \label{ss:heatmaps}

\begin{figure*}
    \centering
    \includegraphics[width=0.85\textwidth]{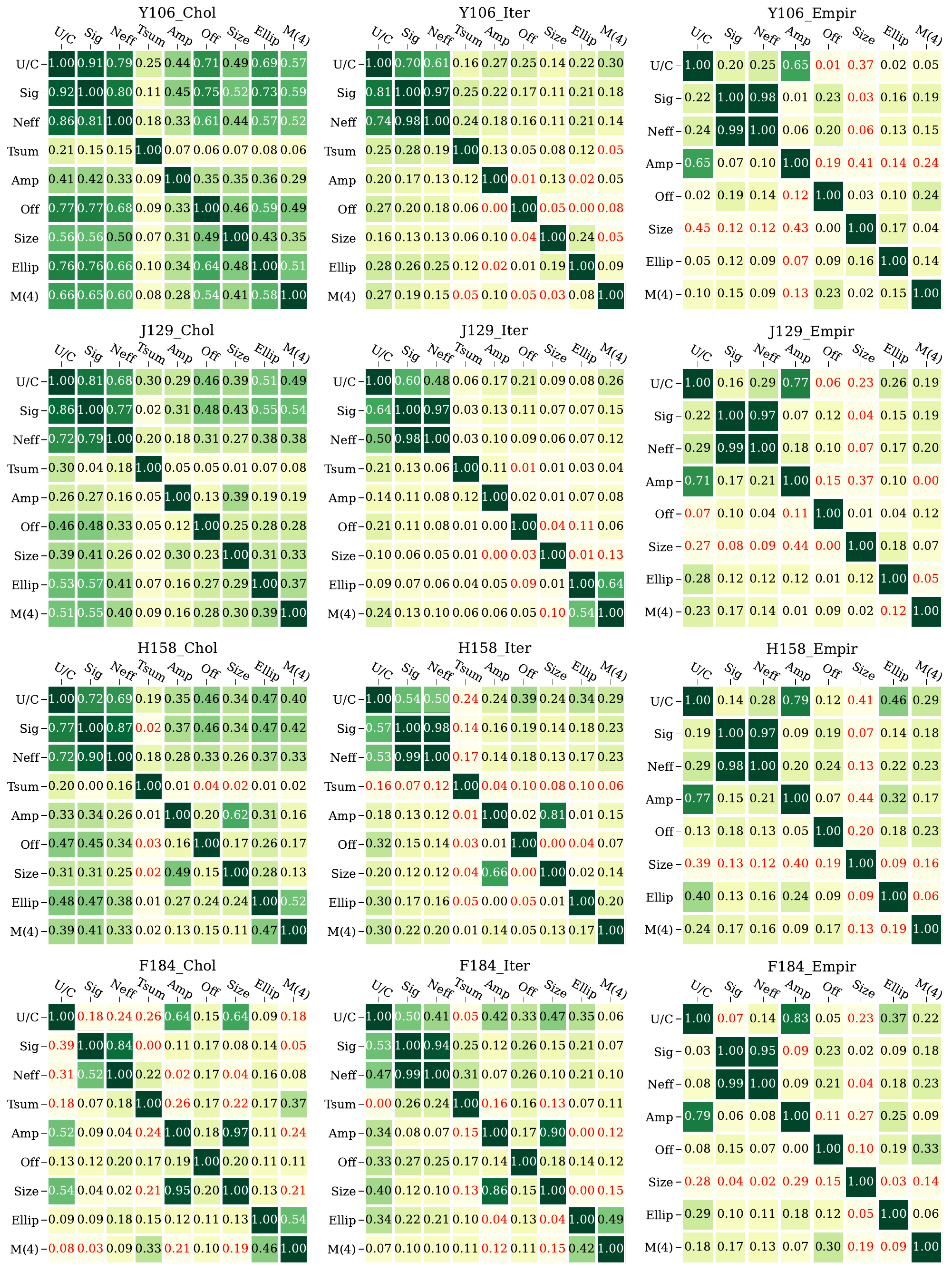}
    \caption{\label{fig:heatmaps}Correlations between the four {\sc Imcom} diagnostics (discussed in Section~\ref{ss:diagnostics}; ``Tsum'' given by the empirical kernel is not included as it is constantly zero) and five $1$-point statistics of injected stars (discussed in Section~\ref{ss:1pt-stats}), yielded by three linear algebra kernels in four bands.
    Minus signs are added to all five $1$-point statistics, so that all correlations shown here are expected to be positive.
    Pearson correlation coefficients (Spearman's rank correlation coefficients) are shown in below-diagonal (above-diagonal) cells of each heatmap; correlation coefficients are annotated in corresponding cells, and negative ones are shown in red.
    Abbreviated labels are used for the nine quantities: ``U/C'' --- see Fig.~\ref{fig:outmap_fidelity}, ``Sig'' --- see Fig.~\ref{fig:outmap_sigma}, ``Neff'' --- see Fig.~\ref{fig:outmap_neff}, ``Tsum'' --- see Fig.~\ref{fig:outmap_tsum}, ``Amp'' --- see Fig.~\ref{fig:star_amplitude}, ``Off'' --- see Fig.~\ref{fig:star_offset}, ``Size'' --- see Fig.~\ref{fig:star_width}, ``Ellip'' --- see Fig.~\ref{fig:star_ellipticity}, and ``M(4)'' --- see Fig.~\ref{fig:star_4th_moment}.}
\end{figure*}

In this section, we investigate correlations between {\sc Imcom} diagnostics discussed in Section~\ref{ss:diagnostics} and $1$-point statistics of injected stars discussed in Section~\ref{ss:1pt-stats}. Intuitively, with higher fidelity, smaller noise amplification, larger effective coverage, and smaller spread in total weights, errors in all measurements should be reduced.
To test such intuition, Fig.~\ref{fig:heatmaps} visualizes Pearson correlation coefficients and Spearman's rank correlation coefficients between these quantities. Note that the former mirrors linear correlation between vectors, while the \NewEdit{latter} do not assume linearity and thus supplement the picture. % later

The Cholesky kernel displays significant correlations between three of the {\sc Imcom} diagnostics and most of the measurement errors in Y106 bands; these correlations become weaker in J129 and H158, and almost vanish in F184.
In general, results in the three bluer bands support our intuition: a larger coverage leads to better fidelity and noise control, and thus better shape measurements. The standard deviation of total weight is an exception, as its correlations with both other diagnostics and measurement errors are weak, indicating that this quantity provides little information about the quality of output images.
In F184 band, most results are disrupted by the poorly chosen target output PSF (see Section~\ref{ss:whitenoise1}). However, the fidelity remains correlated with errors in amplitude and size --- these two errors are almost perfectly correlated themselves --- consolidating our suspicion about their dependence on a Gaussian fit (see Section~\ref{ss:1pt-stats}).

In most cases, the iterative and empirical kernels do not show as significant correlations between {\sc Imcom} diagnostics and measurement errors, but for different reasons.
As mentioned in previous sections, the iterative kernel results are subject to ``output white noise;'' since quantities in the heatmaps are defined in different ways, it is not unexpected that the correlations are wiped out. Nevertheless, for both these two kernels, the correlation between noise amplification and effective coverage are remarkable and consistent, showcasing an advantage of uniform and circular windows for input pixels.
By definition (see Section~\ref{ss:empir-kernel}), the empirical kernel is agnostic on target output PSFs, hence the fidelity is not expected to indicate well the quality of output images; although it is significantly correlated with amplitude error in all bands, such correlation does not mean causation, as both depend on the reasonableness of the normalization $T_{{\rm tot}, \alpha} = \sum_\alpha T_{\alpha i} = 1$.
By design (see Section~\ref{ss:empir-kernel}), the empirical kernel somehow has these two diagnostics as ``goals,'' hence its noise amplification and effective coverage results are desirable. However, due to the simplicity of the way it assigns coaddition weights, good noise control and usage of input images do not transfer to good measurements.

\section{Summary} \label{sec:summary}

{\sc Imcom} \citep{2011ApJ...741...46R} is an image coaddition algorithm with control over resulting point spread functions in output images, designed for the weak gravitational lensing program of the {\slshape Nancy Grace Roman Space Telescope}.
In this work, we have refactored our previous implementation of {\sc Imcom}, {\sc furry-parakeet} and {\sc fluffy-garbanzo} \citep[][also known as \papone]{2024MNRAS.528.2533H}, into a fully object-oriented new package with better data structures, {\sc PyImcom} (Section~\ref{ss:pyimcom} \NewEdit{and Appendix~\ref{app:pyimcom}}).
We have deployed multiple approaches to enhance the software performance, including reusing system submatrices, searching for optimal Lagrange multipliers with nodes rather than bisection, solving linear systems using Cholesky composition or iterative method, etc.
To produce almost equivalent coadded images, the consumption of core-hours has been reduced by about an order of magnitude: from ${\cal O}(10^2)$ in \papone\ to ${\cal O}(10^1)$ in this work for each $1.0 \times 1.0 \,{\rm arcmin}^2$ block (Section~\ref{ss:preview}).
In addition, the new implementation is supposed to facilitate further improvements and extensions, and to have better readability for both developers and users.

We have integrated or designed several diagnostics into {\sc Imcom} (Section~\ref{ss:outmaps}), and have developed several new strategies to assign coaddition weights (Section~\ref{sec:lakernel}). To compare these linear algebra (LA) kernels, we have re-coadded a $16 \times 16 \,{\rm arcmin}^2$ region of synthetic {\slshape Roman} images \citep{2023MNRAS.522.2801T} used in \papone.
We have configured simulations in this works using ``benchmark'' configurations for all configurations of bands and LA kernels (Section~\ref{ss:config}); fine-tuning of {\sc Imcom} hyperparameters will be explored in a companion paper (referred to as \papfour).
We have examined these simulations in terms of {\sc Imcom} diagnostics (Section~\ref{ss:diagnostics}); following \citet[][also known as \paptwo]{2024MNRAS.528.6680Y}, we have investigated power spectra of simulated noise frames (Section~\ref{sec:noiseps}) and measurements of ideal injected stars drawn by {\sc GalSim} (Section~\ref{sec:gsstar14}).

The Cholesky kernel (introduced in Section~\ref{ss:kappa-search}) is designed based on the observation that the search for an optimal Lagrange multiplier can be simplified using a series of nodes.
It is efficient and relatively accurate (in terms of measurements of injected stars), yet the irregular and non-uniform window functions for input pixels it requires slightly bias coaddition results, leading to postage stamp boundary effects which may confuse subsequent analyses.

The iterative kernel (introduced in Section~\ref{ss:iter-kernel}) avoids these issues by using a circular window function for each output pixel. Such uniformity and regularity lead to better noise control, especially when the coverage (i.e., number of exposures covering a given location on the sky) is limited.
But this kernel is slow due to significant overhead, and the results are subject to random errors caused by finite tolerance. Until these difficulties are overcome, using this kernel is largely experimental.

The empirical kernel (introduced in Section~\ref{ss:empir-kernel}) coadds images using an empirical relation based on geometry. Although such relation is informed by results from the Cholesky kernel, it is not able to capture nuances in linear systems, and the normalization of coaddition weights is hard to derive from first principles.
Therefore, it does not produce as good results, but is a valid option for ``quick look'' purposes.

Since different linear algebra strategies have different advantages and disadvantages, it would be ideal to combine their strengths in the future; for now, the Cholesky kernel is the most recommended choice in general.
Our ultimate goal is to conduct high-precision shape measurements on galaxies observed by {\slshape Roman} and map the mass distribution in the Universe.
To this end, for future work, we will prioritize the application of shear measurement pipelines, e.g. {\sc Metacalibration} \citep{2017arXiv170202600H, 2017ApJ...841...24S, 2020ApJ...902..138S} or AnaCal \citep[also known as FPFS;][]{2023MNRAS.521.4904L, 2024MNRAS.52710388L}, to our coadded images.

\section*{Acknowledgements}

K.C., C.H., K.L., and E.M. received support from the National Aeronautics and Space Administration, under subaward AWP-10019534 from the Jet Propulsion Laboratory. C.H. additionally received support from the David \& Lucile Packard Foundation award 2021-72096. M.Y. and M.T. were supported by NASA under JPL Contract Task 70-711320, “Maximizing Science Exploitation of Simulated Cosmological Survey Data Across Surveys.” M.T. was supported by the “Maximizing Cosmological Science with the Roman High Latitude Imaging Survey” Roman Project Infrastructure Team (NASA grant 22-ROMAN11-0011).

The detector mask files used in the {\slshape Roman} image simulations are based on data acquired in the Detector Characterization Laboratory (DCL) at the NASA Goddard Space Flight Center. We thank the personnel at the DCL for making the data available for this project.

Computations for this project used the Pitzer cluster at the Ohio Supercomputer Center \citep{Pitzer2018} and the Duke Compute Cluster.

This project made use of the {\sc NumPy} \citep{2020Natur.585..357H}, {\sc Astropy} \citep{2013A&A...558A..33A, 2018AJ....156..123A, 2022ApJ...935..167A}, and {\sc SciPy} \cite{Virtanen2020NatMe} packages.
Most of the figures were made using {\sc Matplotlib} \citep{Hunter:2007}; {\sc SAOImageDS9} \citep{2003ASPC..295..489J} played an important role as a preview tool.
Some of the results in this paper have been derived using the {\sc healpy} and {\sc HEALPix} package \citep{2005ApJ...622..759G, 2019JOSS....4.1298Z}.

\section*{Data Availability}

The codes for this project, along with sample configuration files and setup instructions, are available in the three GitHub repositories:
\begin{itemize}
\item \url{https://github.com/hirata10/furry-parakeet.git} (\papone\ implementation, postage stamp coaddition)
\item \url{https://github.com/hirata10/fluffy-garbanzo.git} (\papone\ implementation, mosaic driver)
\item \url{https://github.com/kailicao/pyimcom.git} (new implementation, introduced in this work)
\end{itemize}
This project used {\sc PyImcom} v1.0.1 and C routines in {\sc furry-parakeet} v0.1.1 for simulations, and {\sc PyImcom} v1.0.2 for postprocessing and analysis; {\sc fluffy-garbanzo} v0.1.1 was used during development but is now obsolete.

The following changes were made to the previous code (but not the underlying algorithm) to improve performance or maintainability:
\begin{list}{$\bullet$}{}
\item Options were added to the coaddition kernel ({\sc furry-parakeet}) to save only the outputs requested by the user. This reduces memory usage.
\end{list}

\appendix

\section{{\sc PyImcom}: in-depth description} \label{app:pyimcom}

\NewEdit{This appendix supplements Section~\ref{ss:pyimcom} by explaining} the ``physical meaning'' of the {\sc PyImcom} procedure; we refer readers interested in the implementation details to the docstrings and comments in the new repository. % explains
Here we emphasize that we have prioritized conceptual comprehensibility over strict encapsulation, and have put a lot of effort into enhancing code readability.
\NewEdit{Section~\ref{ss:app-config} describes how {\sc PyImcom} parses the configuration and prepares the input data, and Section~\ref{ss:app-linsys} presents how it builds and solves linear systems.}
We note that \NewEdit{this appendix} is rather technical, \NewEdit{and} skipping footnotes \NewEdit{herein} does not affect the understanding of the rest of this paper. % Section~\ref{ss:pyimcom} ... but is included here for logical coherence; ... therein % \NewEdit{\st{(}}, \NewEdit{\st{)}}

\subsection{Configuration and input data} \label{ss:app-config}

To coadd a block with {\sc PyImcom}, one needs to build a {\tt Config} instance, and use it as an argument to construct a {\tt Block} instance. (See our example Python driver scripts in the {\tt examples/} subdirectory; note that driver scripts are supposed to be run outside of the package directory.)
A {\tt Config} instance can be built either from a JSON configuration file or from scratch using our command-line interface.\footnote{A more common situation in practice might be that the user customizes a configuration built from JSON file before using it by simply changing instance attributes, without the need to make another configuration file. This should be useful if people want to test series or grids of alternative parameters. {\sc PyImcom} will record the actually used configuration in a header of the output FITS file.} It contains instructions about what sky area to coadd as well as how to coadd it.
An incomplete list of configuration entries, particularly relevant to the initialization of a {\tt Block} instance, includes:
\begin{itemize}
    \item {\tt OBSFILE} and {\tt INDATA}: Catalog, directory, and format of input images.
    \item {\tt INPSF}: Directory, format, and resolution \NewEdit{(oversampling rate)} of input PSFs.
    \item {\tt EXTRAINPUT}: Additional input layers to coadd using the same ${\mathbf T}$ matrices. These include several types of noise frames and injected sources; see \papone\ Table~1.
    \item {\tt CTR}: Center of the mosaic ($n_{\rm block} \times n_{\rm block}$ array of blocks) in terms of equatorial coordinates (right ascension and declination, or RA and Dec in short). Combined with $n_{\rm block}$ (the {\tt BLOCK} entry, $48$ in \papone\ and $16$ in this paper), the block index (the {\tt this\_sub} argument of the {\tt Block} constructor), and {\tt OUTSIZE} (see below), the block location is fully specified.
    \item {\tt OUTSIZE}: This entry has $3$ components, $n_1$, $n_2$, and $\Delta\theta$. Each block is an $n_1 \times n_1$ array of postage stamps, each (output) postage stamp is an $n_2 \times n_2$ array of output pixels, and each output pixel has scale $\Delta\theta$. These are $48$, $50$, and $0.025 \,{\rm arcsec}$, respectively, in both \papone\ and this paper.
\end{itemize}
Each {\sc Imcom} output pixel grid (of size $[(n_{\rm block} n_1 + 2 {\tt PAD}) n_2]^2$; see below for {\tt PAD}) comes from a stereographic projection centered at the specified mosaic center and aligned with the line of longtitude in the vertical direction; see \papone\ Fig.~4 for an illustration of {\tt CTR}, {\tt BLOCK}, {\tt OUTSIZE}, and some other parameters.
Note that the extra postage stamps for overlap, for which the number of rows or columns on each side of a block is set by the {\tt PAD} entry (usually $2$, corresponding to $2.5 \,{\rm arcsec}$ when $n_2 \Delta\theta = 1.25 \,{\rm arcsec}$), are useful for studying objects lying between blocks. Yet unless they lie on the mosaic boundaries, these padding postage stamps can be reused from neighboring blocks if they are coadded. This trick is a new feature of {\sc PyImcom}.
Neglecting the finiteness of the mosaic size ($n_{\rm block}$), given $n_1 = 48$ and ${\tt PAD} = 2$, this reuse reduces the number of postage stamps being coadded by a fraction of $1 - (48/52)^2 = 14.8\%$; the gain is not dramatic, but easily achievable via postprocessing.\footnote{Such postprocessing can be performed with the {\tt share\_padding\_stamps} method of the class {\tt Mosaic}. After the postprocessing for simulations in this work (see Section~\ref{sec:sims}) has been done, we have identified a subtle glitch in the method mentioned above: If an input image is only used for padding postage stamps of a block, i.e., not for its central $n_1 \times n_1$ postage stamps, its contribution is not added to the {\tt \textquotesingle INWEIGHT\textquotesingle} HDU, which documents the total weight of each input image for each postage stamp (note that this involves a summation over output pixels, while the ``total weight'' defined in Section~\ref{ss:outmaps} is a summation over input pixels), and its flatten version, the {\tt \textquotesingle INWTFLAT\textquotesingle} HDU. Even if such a situation does exist, it is expected to be very rare, and its impact is insignificant as these two HDUs are only used for deriving the ``mean coverage'' (see Section~\ref{sec:noiseps}), which is the average count of $> 0.01$ values among postage stamps. Other components of the postprocessing results, including all coadded layers and all {\sc Imcom} diagnostics, are not affected by this glitch. In conclusion, this glitch does not grant re-performing the postprocessing.}
Therefore, {\sc PyImcom} only coadds extra postage stamps on mosaic boundaries by default,\footnote{The default value of the {\tt PADSIDES} configuration entry is {\tt \textquotesingle auto\textquotesingle}, i.e., {\sc PyImcom} determines whether to pad on each side automatically. Alternatively, it can be set to {\tt \textquotesingle all\textquotesingle} ({\tt \textquotesingle none\textquotesingle}), so that all four sides (none of the sides) are padded on. If it is none of the above, {\sc PyImcom} searches for capital letters {\tt \textquotesingle B\textquotesingle} (bottom), {\tt \textquotesingle T\textquotesingle} (top), {\tt \textquotesingle L\textquotesingle} (left), and {\tt \textquotesingle R\textquotesingle} (right) in the string, and pads on the corresponding side(s).} and we only consider the $n_1 \times n_1$ array in this section.
Some other entries will be introduced in the following text as needed, mostly in footnotes; for a comprehensive list, see example configuration files in the {\tt configs/} subdirectory or source code in the {\tt config.py} module of {\sc PyImcom}.

After being informed about the input and expected output by parsing its {\tt Config}, the {\tt Block} instance prepares input data for the coaddition.
It searches for relevant input images in the catalog,\footnote{The first half of the second known issue reported in \papone\ Section~4.4, that the search radius for input images does not account for plate scale variations, has not been addressed as of production runs for this paper. When only corners of input and output images overlap, the corresponding input image may not be used for that output image in some rare cases. This issue does not affect main conclusions of this paper, but will be addressed in the future.} and constructs a list of {\tt InImage} instances.
Each {\tt InImage} checks whether the corresponding FITS files exist, and gets all required layers by calling the {\tt get\_all\_data} function of the {\tt layer.py} module; some of the layers are read from input files, while others are made during runtime according to user-specified parameters.
The above steps are similar to our previous implementation.
However, {\sc fluffy-garbanzo} stores input data in a {\tt numpy.ndarray} of shape $(n_{\rm layer}, n_{\rm image}, 4088, 4088)$, where $4088$ is the number of native pixels on each side of a {\slshape Roman} SCA; it sometimes resorts to {\tt numpy.memmap} to reduce memory usage since this array is enormous: $4088^2 \times 4 \,{\rm B} = 63.8 \,{\rm MB}$ per layer per image in {\tt numpy.float32}.
Provided that a block ($1.0 \times 1.0 \,{\rm arcmin}^2$ by default) is much smaller than an SCA ($7.5 \times 7.5 \,{\rm arcmin}^2$), only about $(1.0/7.5)^2 = 1.78\%$ of these input pixels are truly relevant to a block;\footnote{Or $\sim 2.25\%$, if we include ${\tt PAD} = 2$ padding postage stamps on each side and take into account the fact that {\sc Imcom} needs some of the input pixels outside boundaries of the output region, but these do not change the order of magnitude.} ergo in {\sc PyImcom}, an {\tt InImage} partitions its pixels into postage stamps, and only stores positions and signals of those needed by the {\tt Block}. Permanent and cosmic ray masks are applied here, i.e., masked input pixels are not selected in this step.
Note that {\sc fluffy-garbanzo} does not need to store input pixel positions,\footnote{Instead, for each postage stamp, {\sc fluffy-garbanzo} uses its center as the pivot point, and approximates input pixel positions on the output map with distortion matrices $\operatorname{d}[(X, Y)_{\rm perfect}] / \operatorname{d}[(X, Y)_{\rm native}]$, where $(X, Y)$ are pixel indices. {\sc PyImcom} performs accurate mapping using both input and output world coordinate systems (WCSes), although the improvement in precision is not significant.} yet storing them is only a small price to pay for big gains.

\NewEdit{The upper panel of Fig.~\ref{fig:inpix} illustrates this partitioning process.} Once the input pixels are partitioned, the {\tt Block} instance reorganizes the input data into an $(n_1+2) \times (n_1+2)$ array of {\tt InStamp} instances.
Upon completion of the reorganization, input data are removed from {\tt InImage} instances, which are however kept as interfaces to input PSFs and plate distortions; an $n_1 \times n_1$ array of {\tt OutStamp} instances are then initialized to coadd the block stamp by stamp.\footnote{To support extra postage stamps and use the same indices, {\sc PyImcom} always makes $(n_1 + 2 {\tt PAD} + 2) \times (n_1 + 2 {\tt PAD} + 2)$ arrays (implemented as Python nested lists) for both {\tt InImage} and {\tt OutStamp} instances, using the Python {\tt None} object as placeholders.}
\NewEdit{For the selection of input pixels for each output postage stamp, see the lower panel of Fig.~\ref{fig:inpix} and the description in Section~\ref{ss:pyimcom}.}

\subsection{Building linear systems} \label{ss:app-linsys}

\NewEdit{To build system matrices, {\sc PyImcom} computes PSF overlaps (see Eq.~\ref{eq:Aij_Bi}) and performs interpolations to obtain individual elements.}
{\sc Imcom} samples input PSFs for every $2 \times 2$ group of postage stamps, i.e., every $2.5 \times 2.5 \,{\rm arcsec}^2$ region in \papone\ and this paper.
In {\sc PyImcom}, groups of PSFs are implemented as {\tt PSFGrp} instances, attached to {\tt InStamp} instances with only even or only odd (depending on the parity of {\tt PAD}) indices, in the case of input PSFs, and the {\tt Block} instance, in the case of target PSF (modeled using {\tt OutPSF} static methods).
{\tt PSFGrp} instances rotate each input PSF by an appropriate angle mainly set by the roll angle of the corresponding exposure and corrected using its distortion at the specific sampling point.
To avoid wrapping artifacts, all PSF arrays are zero-padded before {\tt PSFGrp} instances perform forward real FFT operations on them; see Section~\ref{ss:accel-fft} for how we accelerate FFT operations in this case.
{\tt PSFOvl} instances are built on the basis of {\tt PSFGrp} instances to compute PSF overlaps (correlations): they multiply a transformed PSF array and the conjugate of another or itself, and perform inverse real FFT operations to attain the overlap (correlation) between a pair of PSFs or a PSF and itself.
There are three types of {\tt PSFOvl} instances:
\begin{itemize}
    \item Input-input: overlap between a pair of input {\tt PSFGrp} instances (input-input cross-overlap), or an input {\tt PSFGrp} instance and itself (input self-overlap); {\sc PyImcom} computes ${\mathbf A}$ matrix elements with them.
    \item Input-output: overlap between an input {\tt PSFGrp} instance and the output {\tt PSFGrp} instance of the {\tt Block}; {\sc PyImcom} computes ${\mathbf B}$ matrix elements with them.\footnote{For convenience, {\sc PyImcom} directly computes and stores elements of the $-{\mathbf B}/2$ matrix, which we sometimes also refer to as the ${\mathbf B}$ matrix in the text when they are practically equivalent (in terms of amount of computation, memory usage, etc.).}
    \item Output-output: overlap between the output {\tt PSFGrp} instance and itself; {\sc PyImcom} computes $C = \Vert \Gamma \Vert^2$ with it.
\end{itemize}
{\tt PSFOvl} is designed to be a callable class: its {\tt \_\_call\_\_} method takes postage stamps ({\tt InStamp} or {\tt OutStamp} instances, depending on the nature of each {\tt PSFOvl} instance) as arguments, and performs interpolations to produce system submatrices.

\NewEdit{Fig.~\ref{fig:matrix} presents a pair of ${\mathbf A}$ and ${\mathbf B}$ matrices and the resulting ${\mathbf T}$ matrix. Here we continue the discussion in Section~\ref{ss:pyimcom}.}
Simple counting tells us that each {\tt OutStamp} makes use of $45$ ${\mathbf A}$ submatrices and $9$ ${\mathbf B}$ submatrices.
We have noticed that, although none of the ${\mathbf B}$ submatrices are shared by adjacent postage stamps, an ${\mathbf A}$ submatrix is relevant to up to $9$ {\tt OutStamp} instances, and thus can be reused.
Neglecting fewer uses near block boundaries, and assuming $\rho_{\rm acc} = n_2\Delta\theta$ following \papone, reusing ${\mathbf A}$ submatrices can reduce interpolations to a fraction of $(9/2/9 + 12/6 + 8/4 + 6/3 + 8/2 + 2/1) / ((5+\pi)^2/2) = 37.7\%$.\footnote{Among a $3 \times 3$ group of {\tt InStamp} instances relevant to an {\tt OutStamp} instance, a total of $45$ ${\mathbf A}$ submatrices need to be computed. $9$ of them correspond to $(0, 0)$ displacement in terms of postage stamp indices, can be used $9$ times, and we only need to calculate half of the elements; $12$ of them correspond to $(0, +1)$ or $(+1, 0)$ displacement and can be used $6$ times; $8$ of them, $(+1, \pm 1)$ displacement, $4$ times; $6$ of them, $(0, +2)$ or $(+2, 0)$ displacement, $3$ times; $8$ of them, $(+2, \pm 1)$ or $(\pm 1, +2)$ displacement, $2$ times; $2$ of them, $(+2, \pm 2)$ displacement, $1$ time. Note that in the new framework, even if some of the input pixels are not used for a specific output postage stamp, all the possible ${\mathbf A}$ matrix elements need to be computed as they may be relevant for other stamps, and thus there is no $\pi$ in the numerator of the expression in the text.}
The reuse of ${\mathbf A}$ submatrices causes a slight difference between {\sc PyImcom} and our previous implementation.
In {\sc furry-parakeet}, a {\tt PSF\_Overlap} instance uses the same group of input PSFs for all selected input pixels; in other words, {\tt PSF\_Overlap} is equivalent to the input self-overlap case of {\tt PSFOvl} in {\sc PyImcom}. When a pair of input pixels is relevant to multiple output postage stamps, it is possible for the corresponding ${\mathbf A}$ matrix element to have different values in different {\tt PSF\_Overlap} instances.
In the new framework, all ${\mathbf A}$ matrix elements for a given input pixel are computed using the PSF sampled at the sampling point closest to that pixel. When a pair of input pixels belongs to different $2 \times 2$ groups of {\tt InStamp} instances, {\sc PyImcom} resorts to the input-input cross-overlap case of {\tt PSFOvl}.
This multiplies the number of inverse FFT operations by $\sim 8$,\footnote{A typical input {\tt PSFGrp} instance is involved in $8$ input-input cross-overlap {\tt PSFOvl} instances, each requires $\bar{n}_{\rm image}^2$ suites of inverse FFT operations; meanwhile, a self-overlap {\tt PSFOvl} instance only requires $\bar{n}_{\rm image}(\bar{n}_{\rm image}+1)/2$ suites. Therefore, the factor is $(8\bar{n}^2/2 + \bar{n}(\bar{n}+1)/2) / (\bar{n}(\bar{n}+1)/2) = 9 - 8/(\bar{n}+1)$, where we have omitted the subscript ``image'' for simplicity. This evaluates to $7.86$ for $\bar{n}_{\rm image}=6$, and $8.11$ for $\bar{n}_{\rm image}=8$, hence $\sim 8$ for our purposes.} yet we believe that it is more accurate to use cross-overlaps, especially when we include input PSF variation over SCAs (see \papfour\ Section~2).

Each {\tt InStamp} instance contains $\bar{n}_{\rm image} (n_2\Delta\theta/s_{\rm in})^2 \approx 129 \;\bar{n}_{\rm image}$ input pixels on average, hence the size of each ${\mathbf A}$ submatrix (in {\tt numpy.float64}) is $(129 \;\bar{n}_{\rm image})^2 \times 8 \,{\rm B} = 0.127 \;\bar{n}_{\rm image}^2 \,{\rm MB}$.
Due to the large number of submatrices, we need to manage the memory usage dynamically.
This motivates the class {\tt SysMatA}, which can be envisioned as an interface to a huge ${\mathbf A}$ matrix for all the input pixels relevant to the block being coadded, which only produces and returns submatrices as needed.
As the size of input-input {\tt PSFOvl} instances is also proportional to $\bar{n}_{\rm image}^2$, instead of storing any of them, {\tt SysMatA} constructs each of them only when a dependent ${\mathbf A}$ submatrix is requested by {\tt OutStamp} instances, use it to produce all dependent ${\mathbf A}$ submatrices, and then destroys it immediately.
The strategy of {\tt SysMatB}, its sibling managing an imaginary huge ${\mathbf B}$ matrix, is different.
Given that each {\tt OutStamp} instance contains $m = n_2^2 = 2500$ output pixels,\footnote{The number of output pixels of each postage stamp is $m = (n_2+3\times2)^2 = 3136$ if we include transition pixels to mitigate boundary effects (see \papone\ Eq.~4), where $3$ is the default value of the {\tt fade\_kernel} parameter (the {\tt FADE} configuration entry), number of rows or columns on each side.} submatrices of ${\mathbf B}$ are larger than those of ${\mathbf A}$; meanwhile, the size of input-output {\tt PSFOvl} instances is only proportional to $\bar{n}_{\rm image}$.
Therefore, instead of ${\mathbf B}$ submatrices, {\tt SysMatB} stores {\tt PSFOvl} instances.

To free up the memory occupied by ${\mathbf A}$ submatrices or input-output {\tt PSFOvl} instances as soon as possible, {\tt SysMatA} and {\tt SysMatB} keep track of the remaining reference count to each of them, using the 3D array {\tt iisubmats\_ref} and the 2D array {\tt iopsfovl\_ref}, respectively.
Likewise, each {\tt InStamp} instance harboring an input {\tt PSFGrp} keeps track of its remaining reference count using the integer attribute {\tt inpsfgrp\_ref}.
To get the total reference counts to all these arrays or instances, {\sc PyImcom} loops over {\tt OutStamp} instances in the simulation mode ({\tt sim\_mode}) before actually performing FFT operations and interpolations or solving linear systems.
Based on its reference counting mechanism, we have deployed virtual memory to {\tt SysMatA}; see Section~\ref{ss:loop-virmem} for details.

After the {\tt sim\_mode} loop is done, \NewEdit{{\sc PyImcom} does its job by looping over over output postage stamps, again as $2 \times 2$ groups. It builds and solves linear systems, performs the actual coaddition (Eq.~\ref{eq:coadd}), and reports the results (see Section~\ref{ss:pyimcom}).}

\section{Acceleration measures} \label{app:accel}

Multiple techniques and choices described in Section~\ref{sec:framework} speed up {\sc PyImcom} to varying degrees. This appendix presents some additional acceleration measures, independently applied to interpolation, fast Fourier transform (FFT), and memory management, respectively.

\subsection{Interpolation from a regular grid} \label{ss:reg-interp}

In order to achieve very high accuracy in interpolation (especially for interpolating the PSF overlap, $G_i\otimes G_j$), \papone\ Appendix A used 10-point interpolation routines:
\begin{equation}
\hat f(x) = \sum_{\mu=-4}^5 w_\mu(x-\lfloor x\rfloor) f(\lfloor x\rfloor + \mu),
\end{equation}
where $\lfloor \cdot\rfloor$ is the floor function, and the interpolation coefficients $w_\mu$ were written as
\begin{equation}
w_\mu(\xi) = \sum_{l=1}^5 \left\{
H^{\rm c}_{\mu,l} \cos [\zeta_l(\xi-\tfrac12)]
+ H^{\rm s}_{\mu,l} \sin [\zeta_l(\xi-\tfrac12)]
\right\},
\end{equation}
where $\zeta_l$ and $H^{\rm c,s}_{\mu,l}$ are coefficients given in \papone\ Table~A3. This D5,5,$\frac1{12}$ scheme is the ``optimal'' interpolation method in the sense of minimizing least-square errors as defined in \papone, and it achieves relative errors of $<10^{-9}$ for any function that is $\ge 6\times$Nyquist sampled. However, profiling showed that the trigonometric functions contributed significantly to the computation time.

An alternative is to expand the trigonometric functions using the Bessel functions $J_m$ and the Chebyshev polynomials $T_m$ \citep[e.g.][9.1.45,46]{1972hmfw.book.....A}. This leads to
\begin{equation}
w_\mu(\xi) = \sum_{m=0}^\infty a_{\mu, m} T_m(2\xi-1),
~~~0\le\xi<1
\label{eq:wmu}
\end{equation}
where $-1\le 2\xi-1<1$ and the coefficients are
\begin{equation}
a_{\mu , m} = \left\{\begin{array}{lll}
\sum_{l=1}^5 H^{\rm c}_{\mu,l} J_0( \frac{\zeta_l}2) & & m=0 \\
2(-1)^{m/2} \sum_{l=1}^5 H^{\rm c}_{\mu,l} J_m( \frac{\zeta_l}2) & & m\ge 2, ~~m~{\rm even} \\
2(-1)^{(m-1)/2} \sum_{l=1}^5 H^{\rm s}_{\mu,l} J_m( \frac{\zeta_l}2) & & m~{\rm odd}. \\
\end{array}
\right.
\end{equation}
While the series is formally infinite, 5 terms (8th or 9th order) suffices to reach $<10^{-9}$ accuracy. The inversion symmetry of the interpolation problem $\xi\rightarrow 1-\xi$ guarantees $a_{1-\mu,m} = (-1)^m a_{\mu,m}$: therefore, by separately saving the even-$m$ and odd-$m$ contributions to Eq.~(\ref{eq:wmu}) we can compute only the 5 values $w_{-4}...w_0$, and then obtain $w_1...w_5$ by subtracting rather than adding the odd-$m$ terms.

We evaluate the sum in Eq.~(\ref{eq:wmu}) through $m=9$ by writing the even-$m$ contribution as a single 4th order polynomial in $(\xi-\frac12)^2$ and the odd-$m$ contribution as a single 4th order polynomial in $(\xi-\frac12)^2$ multiplied by $\xi-\frac12$, and using Horner's method. This requires 46 multiplies in total, as opposed to 10 trigonometric functions and 105 multiplications required for the implementation in \papone.

\subsection{PSF sampling and FFT operations} \label{ss:accel-fft}

Theoretically, point spread functions are real-valued bivariate functions (${\mathbb R}^2 \mapsto {\mathbb R}^+$); practically, these must be sampled as discrete and finite 2D arrays.
In our image simulations \NewEdit{\citep{2023MNRAS.522.2801T, OpenUniverse2025arXiv}}, PSFs of {\slshape Roman} exposures are modeled as arrays of shape $(256, 256)$, spanning regions of $32 \times 32$ native pixels with an oversampling rate of $8$. [For different physical sizes and/or different oversampling rates, users need to set the {\tt npixpsf} parameter correspondingly (see below) and/or tell {\sc PyImcom} the oversampling rate (see the \NewEdit{first} paragraph of \NewEdit{Section~\ref{ss:app-config}}).] % \citep[][in prep]{2023MNRAS.522.2801T} ... second, Section~\ref{ss:pyimcom}
{\sc Imcom} resamples each of these PSF arrays so that the resulting arrays are aligned with the output pixel grid; to first order, this rotates PSF arrays by appropriate angles. These resampled arrays have shape $(n_{\rm samp}, n_{\rm samp})$, where $n_{\rm samp}$ corresponds to local variable {\tt ns2} of {\tt PSF\_Overlap.\_\_init\_\_} in {\sc furry-parakeet} and class attribute {\tt PSFGrp.nsamp} in {\sc PyImcom}.
The physical span of PSF sampling arrays is {\tt npixpsf} (in native pixels), which is a global variable of {\sc fluffy-garbanzo} script {\tt run\_coadd.py} and an instance attribute of {\sc PyImcom} class {\tt Config} (set by the configuration entry {\tt NPIXPSF}). {\tt npixpsf} needs to be at least $\sqrt{2} n_{\rm samp}$ so that entire input PSF are kept regardless of the rotation angle (odd multiples of $\pi/4$ are the most demanding cases).
Since FFT-based correlation computation assumes periodicity, {\sc Imcom} zero-pads the PSF arrays to avoid wrapping artifacts and achieves the shape $(n_{\rm FFT}, n_{\rm FFT})$, where $n_{\rm FFT} \geq 2 n_{\rm samp}$ and should be a nice number for FFT purposes. Note that when the prime factorization of $n_{\rm FFT}$ is $\prod_{i} p_i^{k_i}$, FFT of a 1D array of length $n_{\rm FFT}$ has complexity $n_{\rm FFT} \sum_{i} k_i p_i$, hence it is preferable for $n_{\rm FFT}$ to have a large $\sum_{i} k_i$.

\papone\ further enhanced the oversampling rate by a factor of $2$ while resampling PSFs, padded the sampling arrays by $5$ rows or columns on each side, and forced $n_{\rm FFT}$ to be a multiple of $2^{\lceil\log_2 n_{\rm samp}\rceil-2}$. With ${\tt npixpsf} = 64$, these lead to $n_{\rm samp} = 1033$ (we subtracted $1$ for better performance) and $n_{\rm FFT} = 2560$.
During the development of {\sc PyImcom}, we have realized that ${\tt npixpsf} = 48$ should be sufficient, and neither the factor of $2$ nor the padding leads to noticeable improvements. {\sc PyImcom} multiplies {\tt npixpsf} by the oversampling rate to obtain $n_{\rm samp} = 383$ (we subtract $1$ following the \papone\ convention), and doubles the product to obtain $n_{\rm FFT} = 768$. In consequence, the complexity of each suite of FFT operations has been reduced to a fraction $[768^2 \times (8 \times 2 + 1 \times 3)] / [2560^2 \times (9 \times 2 + 1 \times 5)] = 7.43\%$; note that FFT operations are performed on 2D arrays, hence an additional factor of $n_{\rm FFT}$ needs to be included.
Recall that we have introduced input-input cross-overlaps in {\sc PyImcom}, and the number of inverse FFT operations (which is proportional to $n_{\rm image}^2$) has been multiplied by $\sim 8$ (see \NewEdit{Section~\ref{ss:app-linsys}}). Since the number of forward FFT operations is only proportional to $n_{\rm image}$, inverse FFT operations dominate. Combining all these factors, at this point, the total complexity of FFT operations has been reduced by about half. % Section~\ref{ss:pyimcom}
[Furthermore, the physical span of PSF overlap arrays does not need to exceed the largest distance in one direction between pixels, which is determined by the linear algebra kernel and the acceptance radius $\rho_{\rm acc}$: $n_2\Delta\theta + 2\rho_{\rm acc}$ for the Cholesky kernel, and $2\rho_{\rm acc}$ for iterative kernel. ${\tt npixpsf}$ should be at least twice this distance; however, we choose not to further reduce this parameter to avoid large discrepancies in central regions of PSF overlaps.]

Both sampling and overlap arrays are real-valued, thus real FFT operations can be used. The computation of PSF overlaps can be broken down into the following steps:
\begin{enumerate}
    \item Zero-pad a sampling array of shape $(n_{\rm samp}, n_{\rm samp})$ to the shape $(n_{\rm FFT}, n_{\rm FFT})$.
    \item Perform forward real 1D FFT in one direction to get a complex 2D array of shape $(n_{\rm FFT}, n_{\rm FFT}/2+1)$ (since our $n_{\rm FFT}$ is always even).
    \item Perform forward complex 1D FFT in the other direction to get another complex 2D array of the same shape.
    \item Perform element-wise multiplication to an transformed 2D array and the complex conjugate of another such array (cross-overlap) or itself (self-overlap).
    \item Perform inverse complex 1D FFT in the second direction to get a complex 2D array of shape $(n_{\rm FFT}, n_{\rm FFT}/2+1)$.
    \item Perform inverse real 1D FFT in the first direction to get a real 2D array of shape $(n_{\rm FFT}, n_{\rm FFT})$.
    \item Perform 2D inverse zero-frequency shift and extract an overlap array of shape $(n_{\rm samp}, n_{\rm samp})$.
\end{enumerate}
Examining the above procedure, we have noticed that in step (ii), $(n_{\rm FFT} - n_{\rm samp})$ FFT operations (out of $n_{\rm FFT}$) are performed on all-zero arrays; likewise, in step (vi), results of $(n_{\rm FFT} - n_{\rm samp})$ FFT operations (again out of $n_{\rm FFT}$) are to be discarded.
Thenceforth, we have developed special tools to handle such situations economically, namely static methods {\tt PSFGrp.accel\_pad\_and\_rfft2} and {\tt PSFOvl.accel\_irfft2\_and\_extract}.
These functions further break down steps (i) or (vii) and intertwine the sub-steps with the corresponding FFT operations to avoid unnecessary computations (the latter performs inverse zero-frequency shifts manually). See their docstrings for ASCII art illustrations.
They further reduce the total complexity of FFT operations by an additional factor of $\sim 2$; the reduction in time consumption is not as large, because some additional overhead is involved, and transforming all-zero arrays is faster than transforming general ones.

Before concluding this section on FFT, we mention that {\sc PyImcom} automatically detects {\tt mkl\_fft.\_numpy\_fft}, interface to Intel (R) MKL FFT functionality,\footnote{\url{https://pypi.org/project/mkl-fft/}} if it is available, and uses it instead of {\tt numpy.fft}.
The gain in performance fluctuates significantly, roughly between as fast (no gain) and twice as fast. We encourage {\sc PyImcom} users using Intel machines to try this facility.

\subsection{Looping order and virtual memory} \label{ss:loop-virmem}

As mentioned in \NewEdit{Section~\ref{ss:app-linsys}}, the {\tt Block} instance loops over output postage stamps as $2 \times 2$ groups. Technically, this looping order can be illustrated by the following Python-style pseudocode: % Section~\ref{ss:pyimcom}
\begin{minted}[
 breaklines,
 ]{python}
for j in range(j_min, j_max+1, 2):
    for i in range(i_min, i_max+1, 2):
        coadd_outstamp((j  , i  ))
        coadd_outstamp((j  , i+1))
        coadd_outstamp((j+1, i  ))
        coadd_outstamp((j+1, i+1))
\end{minted}
where {\tt j} is the row index and {\tt i} is the column index.
{\tt j\_min} is $1$ (we remind the readers that ${\tt j} = 0$ is for input stamps only, and {\sc PyImcom} uses a unified indexing) if the block is padded on the bottom, and ${\tt PAD}+1$ if not; likewise, {\tt j\_max} is $n_1+{\tt PAD}\times 2$ if the block is padded on the top, and $n_1+{\tt PAD}$ if not. The story of {\tt i\_min} and {\tt i\_max} is basically the same, except for padding directions.

Assuming we want to coadd an entire block with all sides padded; for simplicity, we abbreviate the postage stamp index to $(j, i)$ in the following discussion.
The first {\tt OutStamp} to coadd is $(1, 1)$, and it depends on $9$ {\tt InStamp} instances, namely those with $j = 0, 1, 2$ and $i = 0, 1, 2$.
To obtain the corresponding system submatrices, $4$ {\tt PSFGrp} instances need to be sampled; they are attached to {\tt InStamp} instances $(0, 0)$, $(0, 2)$, $(2, 0)$, and $(2, 2)$, respectively, and each is sampled at the upper-right corner of its harboring {\tt InStamp} (i.e., the center of the $2 \times 2$ group).
Based on these $4$ {\tt PSFGrp} instances, $10$ input-input {\tt PSFOvl} instances are constructed, including:
\begin{itemize}
    \item $4$ self-overlaps, each producing ${4 \choose 1} + {4 \choose 2} = 10$ ${\mathbf A}$ submatrices;
    \item $4$ cross-overlaps with $(0, +2)$ or $(+2, 0)$ displacement, each producing $4^2 - 2^2 = 12$ ${\mathbf A}$ submatrices;
    \item $2$ cross-overlaps with $(+2, \pm 2)$ displacement, each producing $1 + 2 \times 2 + 4 = 9$ ${\mathbf A}$ submatrices.
\end{itemize}
We have used the fact that \NewEdit{an} ${\mathbf A}$ submatrix does not need to be computed if its two {\tt InStamp} instances have $\max(\{|\Delta j|, |\Delta i|\}) \geq 3$. % a
Because of the symmetry of this $2 \times 2$ collection of {\tt PSFOvl} instances, once all the $45$ ${\mathbf A}$ submatrices for {\tt OutStamp} $(1, 1)$ are ready, those for {\tt OutStamp} instances $(1, 2)$, $(2, 1)$, and $(2, 2)$ are also all ready (recall that {\sc PyImcom} produces all the dependent ${\mathbf A}$ submatrices once an input-input {\tt PSFOvl} instance is constructed).
This validates our looping order, which is somewhat counter-intuitive as the grid of $2 \times 2$ groups of output postage stamps and that for input postage stamps are misaligned in both directions.

Now we move on to the reuse of ${\mathbf A}$ submatrices. Neglecting block boundaries, the ${\mathbf A}$ submatrix computed for {\tt InStamp} instances $(j_1, i_1)$ and $(j_2, i_2)$ is used by all {\tt OutStamp} instances with $(j, i)$ satisfying $\min(\{j_1, j_2\}) \leq j \leq \max(\{j_1, j_2\})$ and $\min(\{i_1, i_2\}) \leq i \leq \max(\{i_1, i_2\})$, and the total number of {\tt OutStamp} instances is $|\Delta j| \cdot |\Delta i|$.
The key question is whether an ${\mathbf A}$ submatrix is reused by the next row of $2 \times 2$ {\tt OutStamp} groups, as if so, it stays in the dictionary {\tt SysMatA.iisubmats} during the coaddition of $\sim 2n_1$ postage stamps, which is completely a waste of memory.
We can divide ${\mathbf A}$ submatrices into $3$ categories according to their $|\Delta j|$ values:
\begin{itemize}
    \item $|\Delta j| = 0$: The above situation never happens.
    \item $|\Delta j| = 1$: The above situation happens half the time; specifically, it happens when $\min(\{j_1, j_2\})$ is odd, and does not happen when it is even.
    \item $|\Delta j| = 2$: The above situation always happens.
\end{itemize}

The universality of the waste mentioned above motives the deployment of virtual memory: if an ${\mathbf A}$ submatrix is used in two rows of $2 \times 2$ {\tt OutStamp} groups, we can save it in a temporary file, and load it when it is needed again. Obviously each ${\mathbf A}$ submatrix needs to be saved and loaded at most once.
If virtual memory is used (determined by the {\tt VIRMEM} configuration entry), it slows down {\sc Imcom} slightly, but allows users to request fewer CPUs and thus complete same tasks with less computing resources.
In other words, virtual memory decelerates {\sc PyImcom} in terms of wall time for a specific computational job; however, it accelerates our program in the sense that, given a fixed number of CPUs and a fixed number of blocks to coadd, users can finish all their jobs in less time.

\bibliography{mainbib}{}
\bibliographystyle{aasjournal}

\end{document}